\documentclass[journal,10pt]{IEEEtran}
\usepackage{graphicx}
\usepackage{epstopdf}
\usepackage{amsthm}

\usepackage{times,amsmath,epsfig}
\usepackage{amssymb}
\usepackage{resizegather}
\usepackage{color}
\usepackage{cite}
\usepackage{makecell}
\usepackage{subcaption}
\usepackage{color}
\usepackage[export]{adjustbox}
\graphicspath{{folder/}{Figures/}}
\usepackage{bm}
\usepackage[linesnumbered,ruled]{algorithm2e}
\usepackage{tabularx}
\usepackage{array}
\newcommand{\ttvar}{\begingroup\@makeother\#\@ttvar}
\usepackage{tablefootnote}
\newcolumntype{P}[1]{>{\centering\arraybackslash}p{#1}}
\usepackage{pifont}
\usepackage{longtable,tabularx,ltxtable,ragged2e}
\newcolumntype{Y}{>{\RaggedRight\arraybackslash}X}
\usepackage{placeins}
\usepackage[symbol]{footmisc}
\usepackage{threeparttable}
\usepackage{url}

\usepackage{breakurl}
\usepackage[breaklinks]{hyperref}
\usepackage{mathtools}
\usepackage{mathtools}
\usepackage{footnote}
\usepackage{float}

\usepackage[capitalise,noabbrev]{cleveref}

\DeclarePairedDelimiter\abs{\lvert}{\rvert}%
\DeclarePairedDelimiter\norm{\lVert}{\rVert}%

\makeatletter
\let\oldabs\abs
\def\abs{\@ifstar{\oldabs}{\oldabs*}}

\let\oldnorm\norm
\def\norm{\@ifstar{\oldnorm}{\oldnorm*}}
\makeatother

\newcolumntype{M}[1]{>{\centering\arraybackslash}m{#1}}

\begin{document}
		\title{Towards Addressing Training Data Scarcity Challenge in Emerging Radio Access Networks: A Survey and Framework}
		
	\author{{Haneya Naeem Qureshi}\IEEEauthorrefmark{1}, 
	{Usama Masood} \IEEEauthorrefmark{1}, 
		{Marvin Manalastas} \IEEEauthorrefmark{1}, {Syed Muhammad Asad Zaidi}\IEEEauthorrefmark{1},
	{Hasan Farooq}\IEEEauthorrefmark{2}, 
	{Julien Forgeat}\IEEEauthorrefmark{2}, 
	{Maxime Bouton}\IEEEauthorrefmark{2},
    {Shruti Bothe}\IEEEauthorrefmark{2}, 
    {Per Karlsson}\IEEEauthorrefmark{2},
    {Ali Rizwan} \IEEEauthorrefmark{3},
	and {Ali Imran}\IEEEauthorrefmark{1}\IEEEauthorrefmark{4}
    \\ \hspace{-2.2mm} \IEEEauthorrefmark{1}AI4Networks Research Center, School of Electrical \& Computer Engineering, University of Oklahoma, OK, USA
     \\ \IEEEauthorrefmark{4}James Watt School of Engineering, University of Glasgow, UK
    \\ \IEEEauthorrefmark{3} Department of Electrical Engineering, Qatar University, Doha, Qatar
	\\ \IEEEauthorrefmark{2} Ericsson Research, Santa Clara, CA, USA \\
	Corresponding author email: haneya@ou.edu (Haneya Naeem Qureshi)}

\maketitle
			\begin{abstract}

The future of cellular networks is contingent on artificial intelligence (AI) based automation, particularly for radio access network (RAN) operation, optimization, and troubleshooting. To achieve such zero-touch automation, a myriad of AI-based solutions are being proposed in literature to leverage AI for modeling and optimizing network behavior to achieve the zero-touch automation goal. However, to work reliably, AI based automation, requires a deluge of training data. Consequently, the success of the proposed AI solutions is limited by a fundamental challenge faced by cellular network research community: scarcity of the training data. In this paper, we present an extensive review of classic and emerging techniques to address this challenge.  We first identify the common data types in RAN and their known use-cases. We then present a taxonomized survey of techniques used in literature to address training data scarcity for various data types. This is followed by a framework to address the training data scarcity. The proposed framework builds on available information and combination of techniques including interpolation, domain-knowledge based, generative adversarial neural networks, transfer learning, autoencoders, few-shot learning, simulators and testbeds. Potential new techniques to enrich scarce data in cellular networks are also proposed, such as by matrix completion theory, and domain knowledge-based techniques leveraging different types of network geometries and network parameters. In addition, an overview of state-of-the art simulators and testbeds is also presented to make readers aware of current and emerging platforms to access real data in order to overcome the data scarcity challenge. The extensive survey of training data scarcity addressing techniques combined with proposed framework  to select a suitable technique for given type of data, can assist researchers and network operators in choosing the appropriate methods to overcome the data scarcity challenge in leveraging AI to radio access network automation.
\end{abstract}
			

\section{Introduction}

Future cellular networks are envisioned to have big data enabled network automation capabilities \cite{bson}. This includes functionalities of self-optimization, self-healing and self-configuration \cite{survey_self_organization}-\cite{survey_self_healing} that are essential to ensure the viability and sustainability of future cellular networks amid challenges, such as amalgam of new technologies, growing complexity, resource inefficiency and shrinking profit margins.  
In order to enable these automation capabilities in next generation cellular networks, the process of heterogeneous base station (BS) deployment, implementing existing and newly proposed network features and tuning the associated network parameters has to be meticulous. This is because the process of selecting an optimal network configuration that can maximize the vital key performance indicators, like coverage, capacity, reliability or energy efficiency is a rather challenging task. Identifying the optimal network configuration is necessary for network operators to fulfill the promises made by much anticipated 5G and beyond networks and to realize the efficacy of several new use cases. 

\begin{figure*}
	\begin{subfigure}{0.6\textwidth}
		\includegraphics[width=\textwidth]{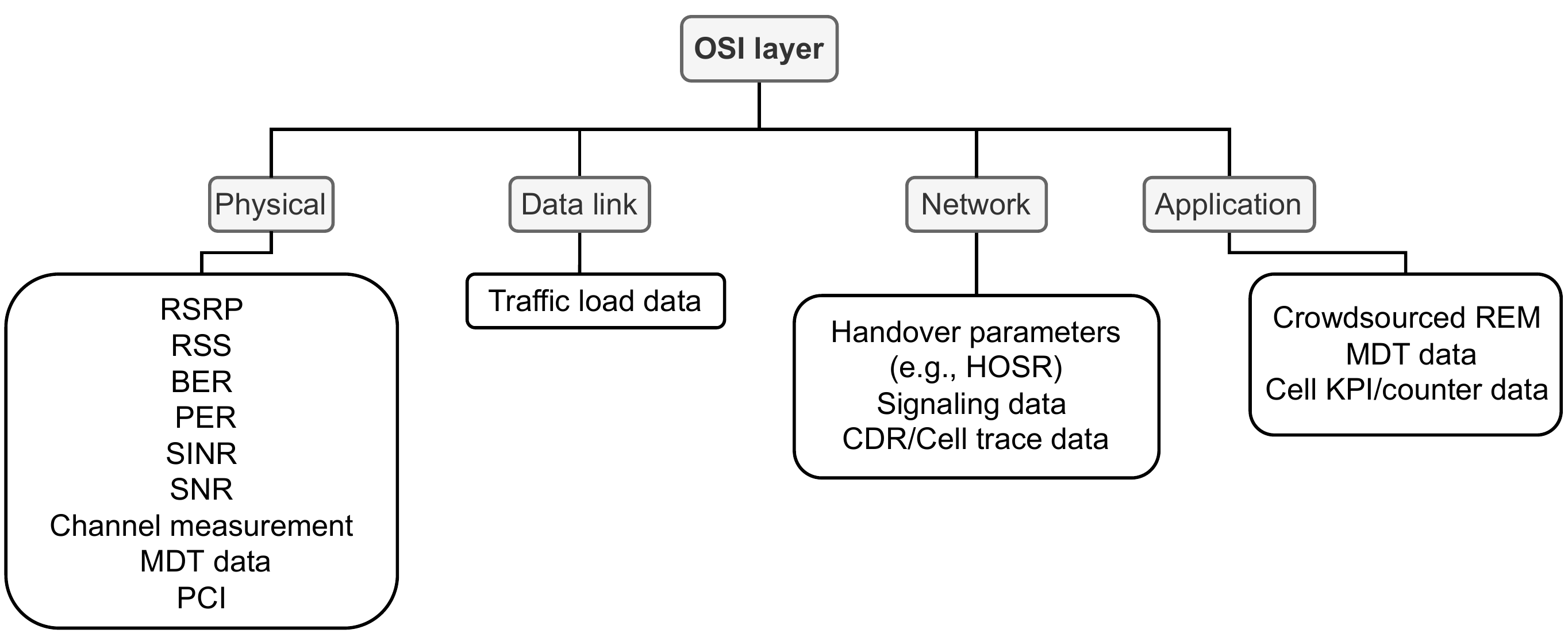}
		\caption{}
	\end{subfigure}
	\begin{subfigure}{0.4\textwidth}
		\includegraphics[width=\textwidth]{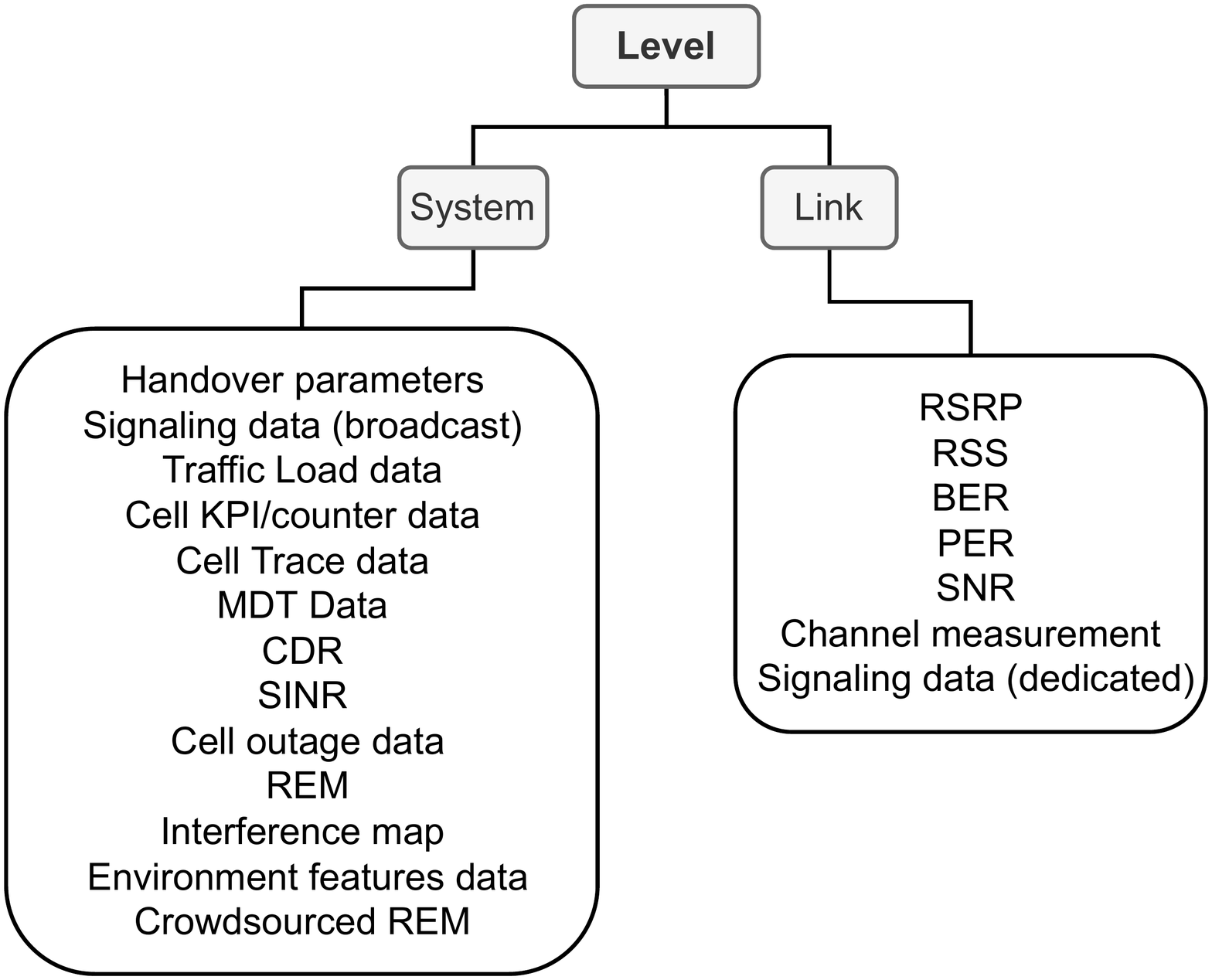}
				\caption{}
	\end{subfigure}
		\caption{Types of data on which techniques to address data scarcity have been applied in literature according to (a) OSI-layer based categories  (b) system/link level categories (figure is based on Table \ref{tab:sparsity})}
			\label{data_CAT}
\end{figure*}

Research community heavily rely on mathematical yet tractable analytical models \cite{analy1}\nocite{my_pimrc}\nocite{analy2}\nocite{my_massive_mimo}\nocite{analy3}-\cite{analy4} to propose
 planning, operation and optimization of different aspects of network. They, however, are based on restrictive assumptions and simplifications with respect to transceiver architecture, base station and user distributions and propagation characteristics, to name a few. Furthermore, stochastic geometry-based models are unable to capture the network dynamics which include mobility management and transmission latency. Therefore, several machine learning (ML) based techniques are proposed in current literature that leverage training and tuning of ML based models to determine the behavior of different configuration and optimization parameters (COPs), such as antenna tilt, transmit power, cell load in relation to different key performance indicators (KPIs), like coverage, capacity or energy efficiency \cite{survey_machine_learning}-\nocite{jeff_online_tuning}\cite{data_driven}. These COP-KPI relationships can then be used for COP-KPI optimization. Moreover, in cellular networks context, awareness about radio environment in a wireless system is crucial given that the radio spectrum is a limited resource \cite{yilmaz_location_2015}. Ample data is required for constructing radio environment maps (REMs) which can be used for operations such as spectrum management, to construct interference maps, to make decisions about spectrum availability for enabling dynamic spectrum access, for assessing/monitoring network health, minimizing signalling, interference management, optimization of radio resources allocation, dynamic spectrum allocation, identify bad-signal areas, automatic neighbor relation, minimize drive tests, handovers optimization and coexistence of various technologies \cite{comparison_in_cognitive}-\cite{colorado}. However, all such techniques face a common key challenge that undermine their utility: scarcity/sparsity of the training data. 
This fundamental problem has two facets: 
(i) \underline {Data scarcity}: Obtaining large amounts of pertinent training data from the operators is not a trivial task. Furthermore, as most of the data remain trapped in silos, even if willing, a single operator may not be able to provide the deluge of real data needed for developing models e.g., user (traffic, mobility pattern, QoE expectations) and network behavior (spatio-temporally robust COP-KPI) models. 
(ii) \underline {Data sparsity}: Network operators only try a limited range of COPs in live networks due to high probability of significant network performance impairment of live mobile network during the trial phase. Therefore, only a limited range of COP-KPI data can be obtained. Given that operators only try a limited range of COPs in live networks, despite sourcing from multiple operators, even when not scarce, the real data are expected to be sparse or unevenly distributed. In other words, term scarcity refers to problem when data is too little to train a model. Sparsity on the other hand refers to problem when there is some data, but it is thinly or unevenly distributed making reliable training of AI difficult. 
For sake of clarity, in rest of the manuscript we use only one term, scarcity to represent this problem irrespective of the reason behind data being not enough to train AI.

To illustrate the type of data in cellular networks which is scarce, Fig. \ref{data_CAT} shows the data on which data augmentation techniques have been applied in literature according to OSI layers and system/link level categorization.  Link level data corresponds to the point-to-point communication link, for example RSRP, and system level data takes the notion of data involving a large number of network elements including several links, for example REM.
The use cases of these data are elaborate in later sections and are summarized in last two columns of Table IV.

To address the data scarcity challenge, one solution can be to obtain data from field trials. However, conducting independent field trials on a large scale is costly and time-consuming, especially in dynamic scenarios, where the number and locations of measurements change, and it is infeasible to measure the radio frequency field strength values at every point of interest. 
Another way to obtain data is through mathematical models. However, they are based on too many assumptions and simplifications, that fail to depict real world scenarios. 
Moreover, in  ultra-dense deployments, small cells contain far fewer users compared to macro cells.  This makes user measurements at the base station of small cells scarce, which particularly poses a problem for automation solutions that leverage minimization of drive test (MDT) \cite{mdt1}\nocite{mdt2}-\cite{my_letter}. This problem is further aggravated if smaller bin size is used to reduce quantization error, attributing to the fact that many bins might not be visited by even a single user during the reporting period 	\cite{my_letter}.

Deploying the new 5G and beyond network functionalities in a real world cannot be done arbitrarily. If the training data is poorly distributed or scarce, it might not represent the actual network scenario very well, which could lead to over-fitting during the model training stage.
In order to develop accurate models, machine learning algorithms require large amounts of true training data since a model based on scarce data would rely on assumptions and weak correlations \cite{hughes_generative_2019}. 
In turn, unscrupulous network design and sub-optimal parameter configuration will hamper not only the capability of future networks that will impact the user experience negatively but will also increment the capital and operational expenditure (CAPEX/OPEX) of mobile operators \cite{survey_planning}.

\subsection{Related Work}
Data scarcity challenge has been addressed in the domain of environment sciences field, such as ecology, marine, agriculture, soil science, elevation, precipitation, and chemical concentrations, through review papers in \cite{survey_environmental_sciences}-\nocite{chemical_conc}\nocite{environment1}\cite{liheap}.
However, to the best of authors' knowledge, a survey paper on addressing the training data scarcity challenge in cellular networks is not present.

In cellular networks context, the closest survey papers to this work are  \cite{survey_REM}, \cite{survey_interference_maps} and \cite{zhang2019deep}. Authors in \cite{survey_REM} focus on the task of radio environment map (REM) construction techniques. Advantages, disadvantages, and asymptotic complexity comparison of seven  interpolation techniques (inverse distance weighted, nearest neighbor, spline, natural neighbor, modified Shepard’s method, gradient plus inverse distance squared method and Kriging). They also discuss some indirect construction methods that combine interpolation with  transmitter parameter information.
However, since work in \cite{survey_REM} is from 2014, many indirect methods developed after 2014 are not covered in it. Moreover, \cite{survey_REM} is limited to the task of REM construction only. Several methods that have gained popularity in past recent years to enrich scarce data, like advanced machine learning techniques and synthetic data generation, that are a part of this survey, are also not included in \cite{survey_REM}.

The other relevant study to this work is the study in \cite{survey_interference_maps}, where authors survey the use of interference maps. However,  the study in \cite{survey_interference_maps} focuses on spectrum occupancy measurement data only while reviewing studies till 2016. In contrast, in this survey, we cover variety of RAN data. Like \cite{survey_REM}, popular methods in recent years to augment scarce data, like advanced machine learning techniques and synthetic data generation are also not included in \cite{survey_interference_maps} as addressing data scarcity problem is not the  focus of the work in  \cite{survey_interference_maps}.

Simulators are another promising way to address the data scarcity challenge.  Two existing surveys on simulators include \cite{simulator_survey1} and \cite{simulator_survey2}. Authors in \cite{simulator_survey1}  compare 4G and 5G simulators and authors in \cite{simulator_survey2} provide a summary of the most significant 5G simulators. However, these works are restricted only to simulators as a tool for generating data.

Testbeds can also be used to generate real data to augment available scarce data. The work in \cite{turboran_access} compares key testbeds around the world in terms of location, scale of deployment, type of access,  key features, and supported experiments. However, these works are restricted to testbeds only, whereas this survey aims to address data sparsity challenge by considering additional techniques as identified in Fig. 2.

{
A more recent study from 2019 \cite{zhang2019deep} surveyed the applications of deep learning-based techniques, like transfer learning, autoencoders, generative adversarial networks techniques for wireless networks. The authors introduce the basics of deep learning and then identify wireless applications where those techniques can be used, for instance, mobile data analysis, mobility analysis, wireless sensor network, network control, network security, signal processing, and other emerging wireless applications. While some of the techniques discussed in \cite{zhang2019deep} can also be exploited to address data scarcity challenge in RANs to some extent for limited data types, the work in \cite{zhang2019deep} is not focused on addressing the training data scarcity challenge in RAN. In contrast, this survey not only provides a comprehensive review of techniques that can address  training data sparsity for a variety of RAN data but also it provides the first of its kind systematic framework to select the most suitable techniques for given data types.

}

To the best of authors' knowledge, there is no existing work that presents a consolidated survey and framework that aims to solve the training data scarcity challenge in cellular networks. This article presents the techniques in literature to address the training data scarcity problem over the period of 1991 to 2021 as they apply to radio access networks in wireless communications.

\subsection{Contributions and Organization}	
The key contributions in this paper can be summarized as follows:

\begin{figure*}[h]
        \centering
        \includegraphics[width=\linewidth]{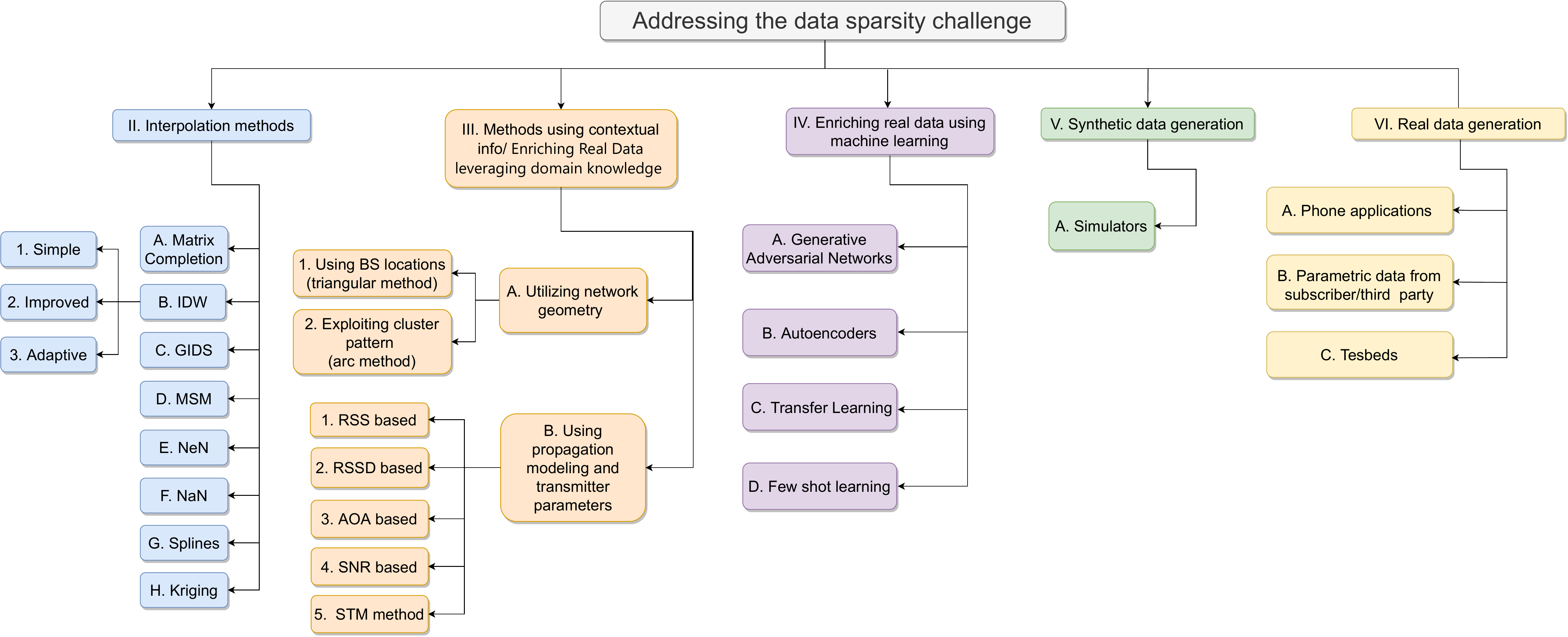}
        \caption{This figure presents one possible taxonomy for classifying the techniques to address data scarcity in RAN. 
        }
        \label{fig:taxonomy}
\end{figure*}

\begin{itemize}

\item 
To address the training data scarcity challenge, we present an overview of existing techniques, and potential new and emerging techniques, such as matrix completion theory (Section \ref{sec:matrix_completion}) leveraging different types of network geometries (Section \ref{sec:network_geometry}), and advanced machine learning techniques such as the use of generative adversarial networks (GANs) (Section \ref{sec:gans}), autoencoders (Section \ref{sec:autoencoders}), transfer learning (Section \ref{sec:transfer_learning}) and few-shot learning (Section \ref{sec:few shot}) to enrich scarce data in cellular networks. 
We also highlight the pros and cons of these approaches analyzed in context of different RAN focused use cases. A taxonomy of training data enrichment techniques is developed by grouping these techniques into various categories as shown in Fig. \ref{fig:taxonomy}.

\item We present a comparison of existing and emerging simulators (Section \ref{sec:simulators}) as tools for generating synthetic data to overcome the data scarcity issue which can greatly benefit researchers
    as the characterization and comparison among features of different simulators will enable them to identify publicly accessible simulators and use them for their specific problems.
    
\item An overview of state-of-the-art current and emerging testbeds for next generation cellular networks is presented in Section \ref{sec:testbeds} that will  make readers aware of current and emerging platforms to access real data in order to overcome data scarcity challenge. Most of these testbeds are available to external experiments, which will foster collaboration among different academic  institutions as well as with industry. This will in turn enable the utilization of these existing facilities to the fullest and accelerate quality research in the field of cellular networks.

 \item We propose a decision tree diagram, that will enable researchers and operators to choose appropriate methods to solve the training data scarcity challenge, based on the available information and network scenario. 
\end{itemize}

 It should be noted that measured data can be scarce and still be representative. On the other hand, data can be big but not representative. We begin  by presenting an overview of techniques that will work best in the first case.   In the case when data is scarce and representative, but the only information known are the measured data points and their location, interpolation methods in Section \ref{chap:interpolation} are likely to perform best. 

Moving forward, when some additional information beyond the data points and their locations is known, we can utilize the methods using contextual information or domain knowledge in Section \ref{chap:contextual_info}.
Several machine learning techniques can also be leveraged to address the data scarcity challenge. These  include generative adversarial networks, autoencoders, transfer learning and few-shot learning techniques (Section \ref{chap:ml}).

On the contrary, when the available data is big and non-representative or scarce and non-representative, the solution lies in either resorting to generate synthetic data (Section \ref{chap:synthetic}) or get real data (Section \ref{chap:real}).
In addition, for scenarios with no starting real data, for example, for new or anticipated scenarios which are not yet deployed in a real network, simulators, and testbeds to generate real data are most likely going to be the best option for wireless communications community.

Other classifications of data augmentation techniques, such as those based on OSI layer based, or system and link level grouping of the data streams are also possible. However, many data scarcity techniques can be applied to the data corresponding to multiple layers and levels. Therefore,  the rest of the paper is structured by organizing the techniques based on their technical grouping as shown in this tree diagram.  i.e., each branch represents a section, and each leaf represents a subsection of the paper.
Moreover, while it is intuitive to assume that data from different layers may require different generation techniques, but the suitability of a technique depends mainly on the characteristics of the data e.g., availability of latent distribution, completeness, representativeness, temporal or spatial nature and context and so on.  For example, traffic variation at base station data at the application level can be modelled as time series data, and same can be done for the packet error data at link level, and bit error data at physical layer. Similarly, data on traffic variation in space (system level data) bears similarity with, for instance, RSRP/SINR-based REM data (physical layer) and thus same techniques such as kriging, inverse distance weighted, nearest neighbor interpolation can be used.  While in most cases, the characteristics and contexts of the data may suffice to choose the best technique, in some cases, additional knowledge that can be extracted from knowing which layer the data belongs to may be helpful in improving the data augmentation.  However, so far in literature there does not exist examples of where knowledge of layer level mapping is exploited for data augmentation.

\section{Interpolation Methods}
\label{chap:interpolation}
When the only information required from cellular network are the measurement values (location-value pair) in order to recover the missing values,
we classify such methods as `interpolation methods', which assume that the data are spatially dependent and continuous over space \cite{Venezuela, rahman_creating_2019, chen_online_2018}. 

Interpolation methods are widely used in literature for radio environment map (REM) augmentation. REM for a coverage area consists of radio information, such as signal strength, signal quality or interference \cite{survey_REM}. Constructing REMs is done through manual drive tests, which leads to collection of data from scarce locations due to time and cost constraints. REM supports a variety of use cases, such as spectrum access management, identification of poor signal areas, automatic neighbor relation, power management, interference mitigation and management, optimization of radio resources allocation, radio resource management, dynamic spectrum allocation, handovers optimization, automated networks planning, maintenance and optimization of network parameters \cite{survey_REM}. Therefore, complete REMs from the available scarce REMs are required to support these use cases. 

\begin{figure*}
	\vspace{-0.2in}
	\centering
	\begin{subfigure}{0.41\textwidth}
		\includegraphics[width=\textwidth]{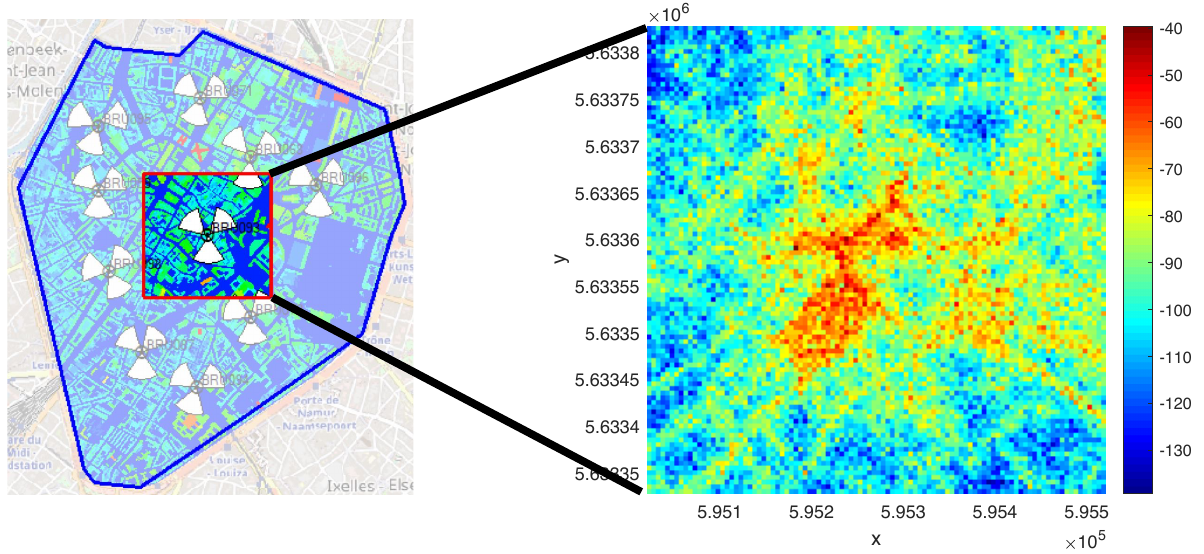}
		\caption{Full coverage map}
		\label{full_with_area}
	\end{subfigure}
	\begin{subfigure}{0.23\textwidth}
		\includegraphics[width=\textwidth]{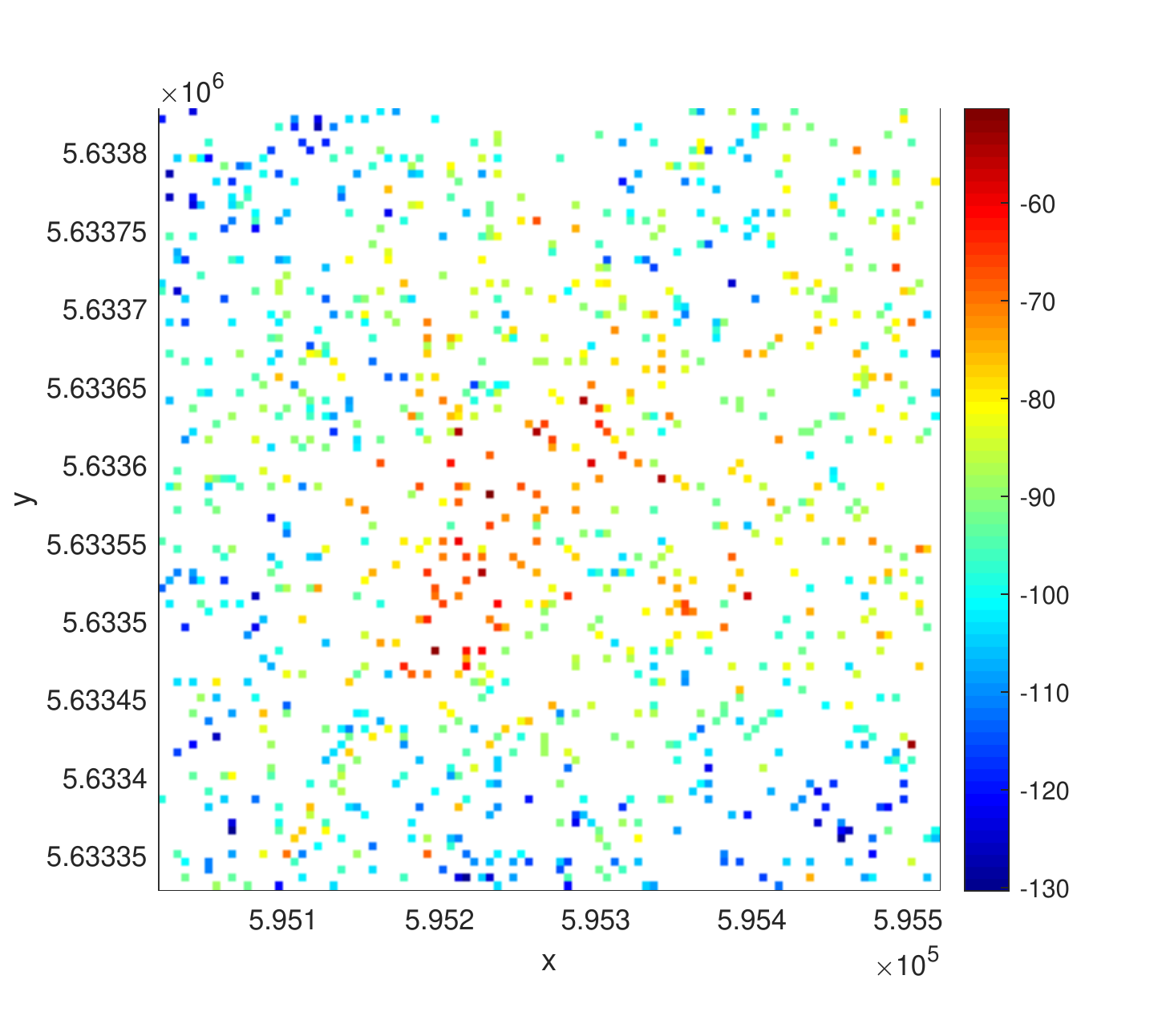}
		\caption{Scarce coverage map}
		\label{sparse}
	\end{subfigure}

	\centering
	\begin{subfigure}{0.23\textwidth}
		\includegraphics[width=\textwidth]{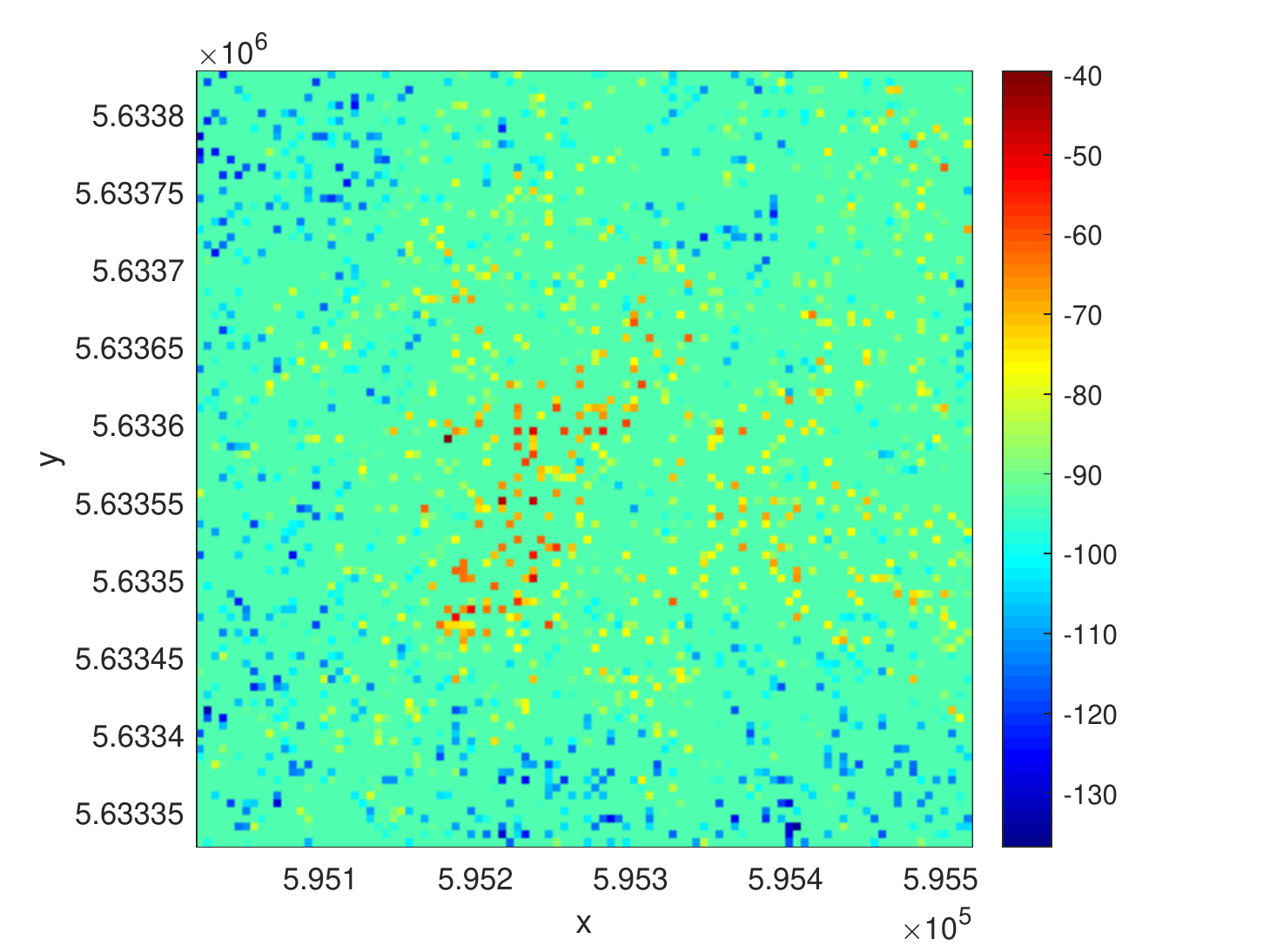}
		\caption{Moving average}
	\end{subfigure}
	\begin{subfigure}{0.23\textwidth}
		\includegraphics[width=\textwidth]{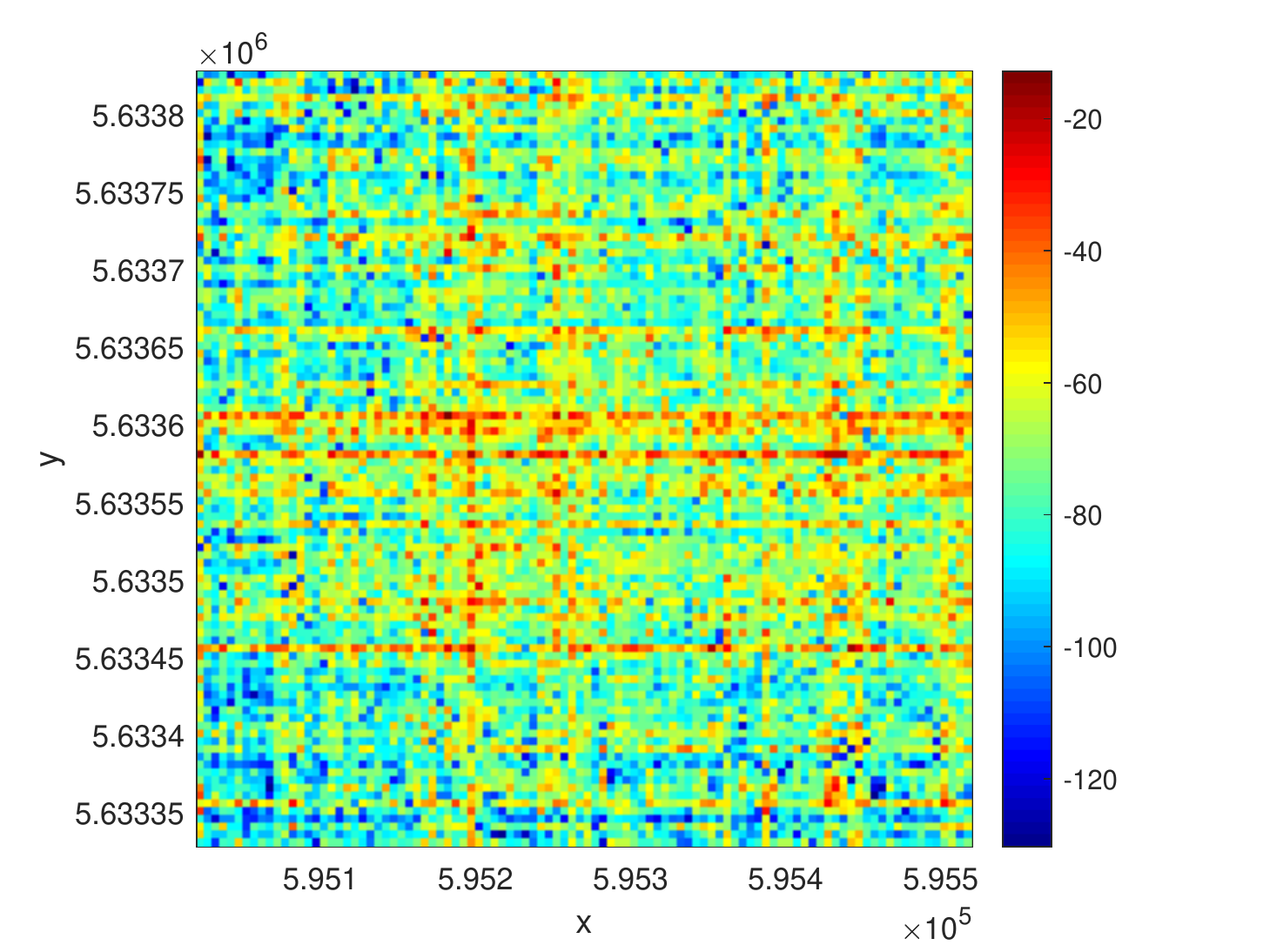}
		\caption{Matrix completion via SVT}
	\end{subfigure}
	\begin{subfigure}{0.23\textwidth}
		\includegraphics[width=\textwidth]{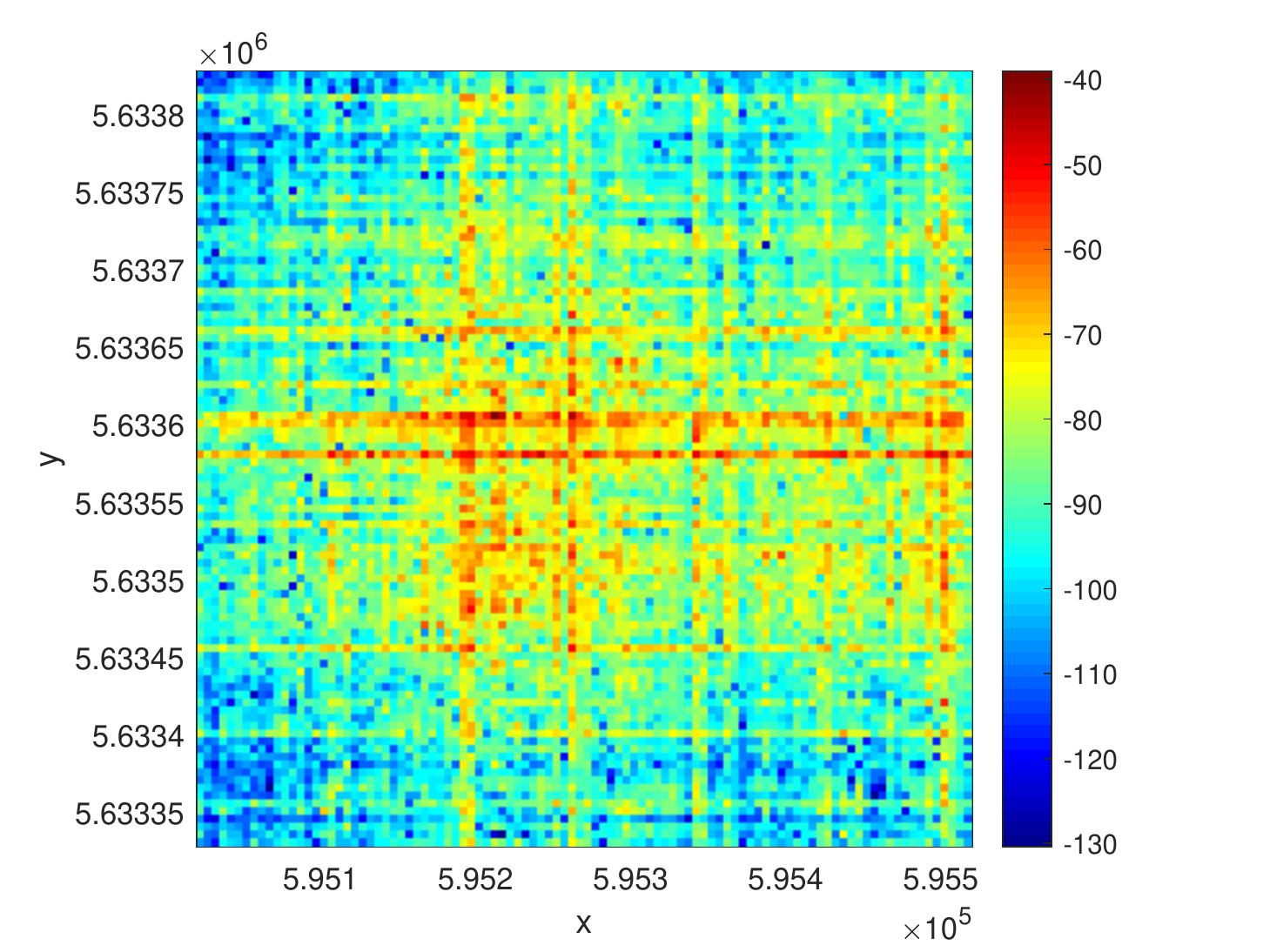}
		\caption{Matrix completion via FPC}
	\end{subfigure}
	\begin{subfigure}{0.23\textwidth}
		\includegraphics[width=\textwidth]{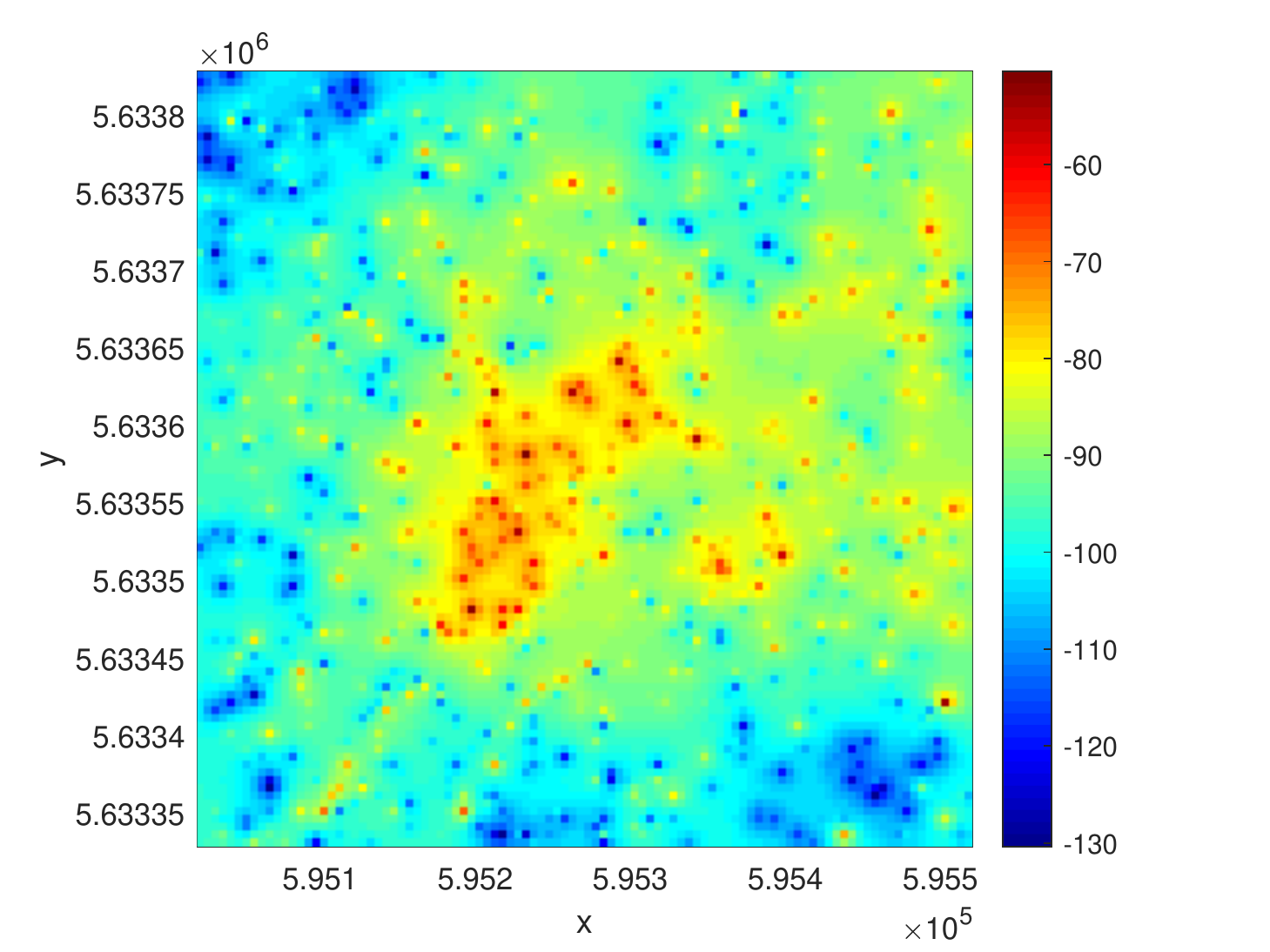}
		\caption{Inverse distance weighted}
	\end{subfigure}
	\begin{subfigure}{0.23\textwidth}
		\includegraphics[width=\textwidth]{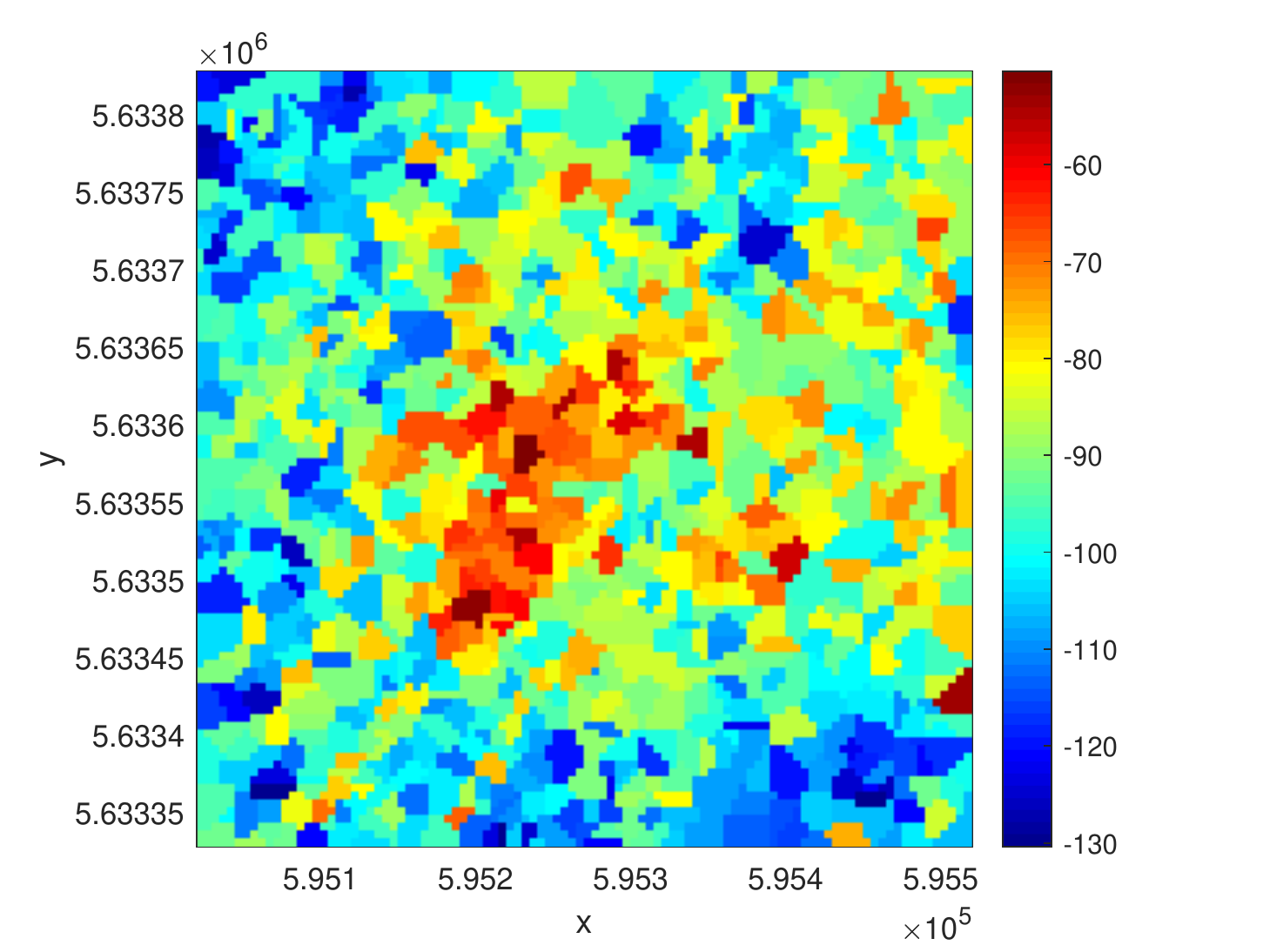}
		\caption{Nearest neighbor}
	\end{subfigure}
	\begin{subfigure}{0.23\textwidth}
		\includegraphics[width=\textwidth]{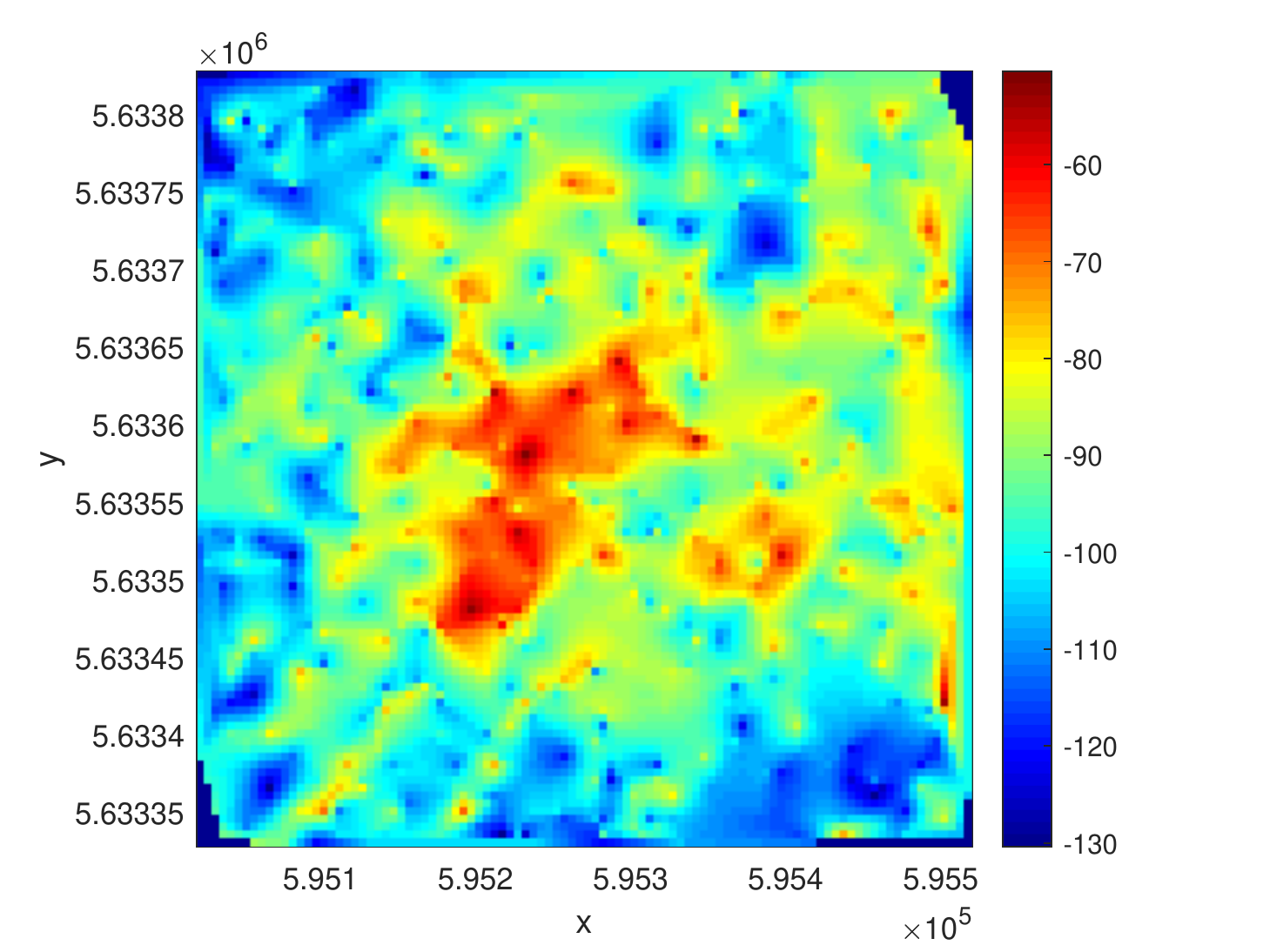}
		\caption{Natural neighbor}
	\end{subfigure}
	\begin{subfigure}{0.23\textwidth}
		\includegraphics[width=\textwidth]{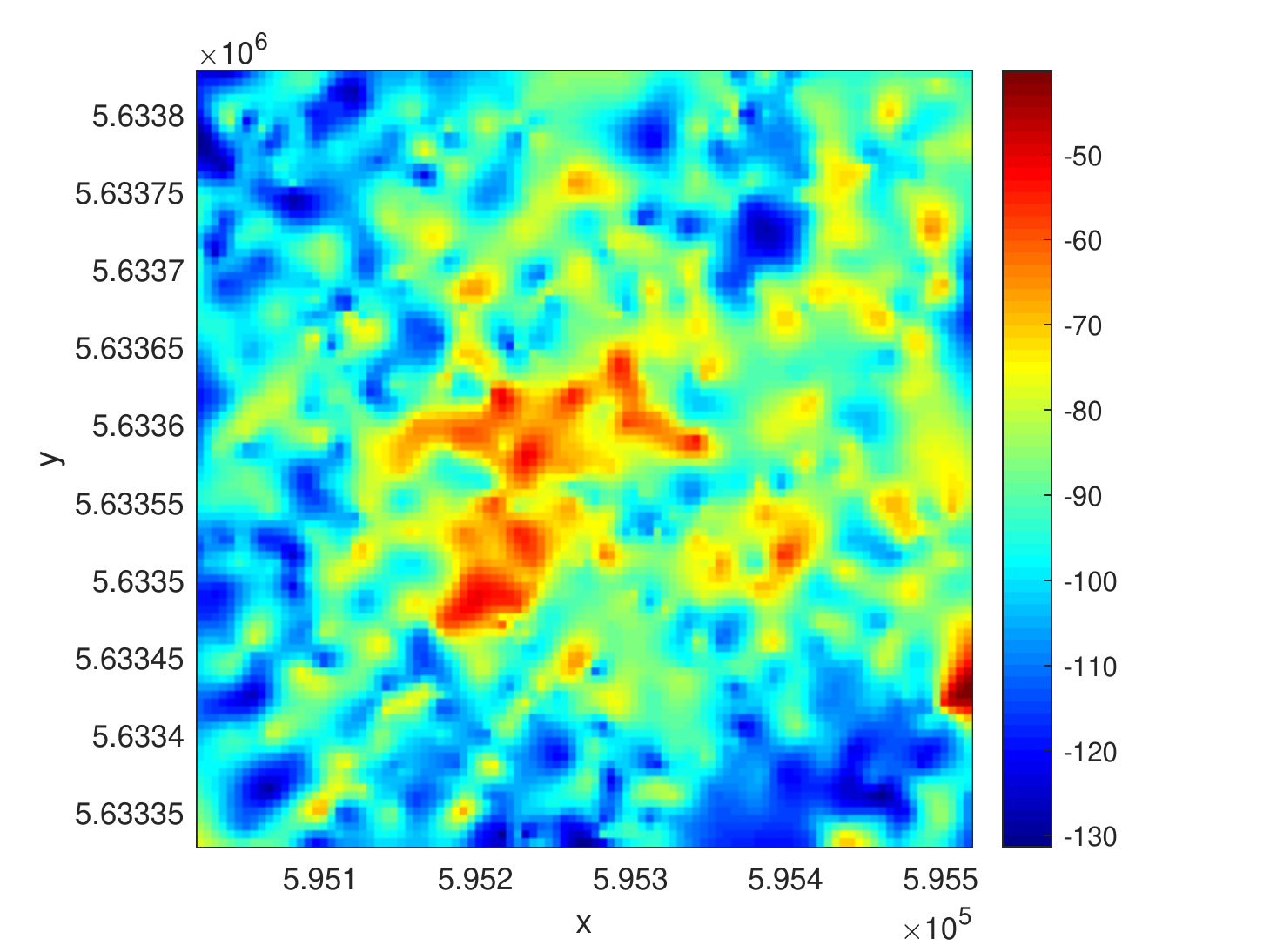}
		\caption{Spline}
	\end{subfigure}
	\begin{subfigure}{0.23\textwidth}
		\includegraphics[width=\textwidth]{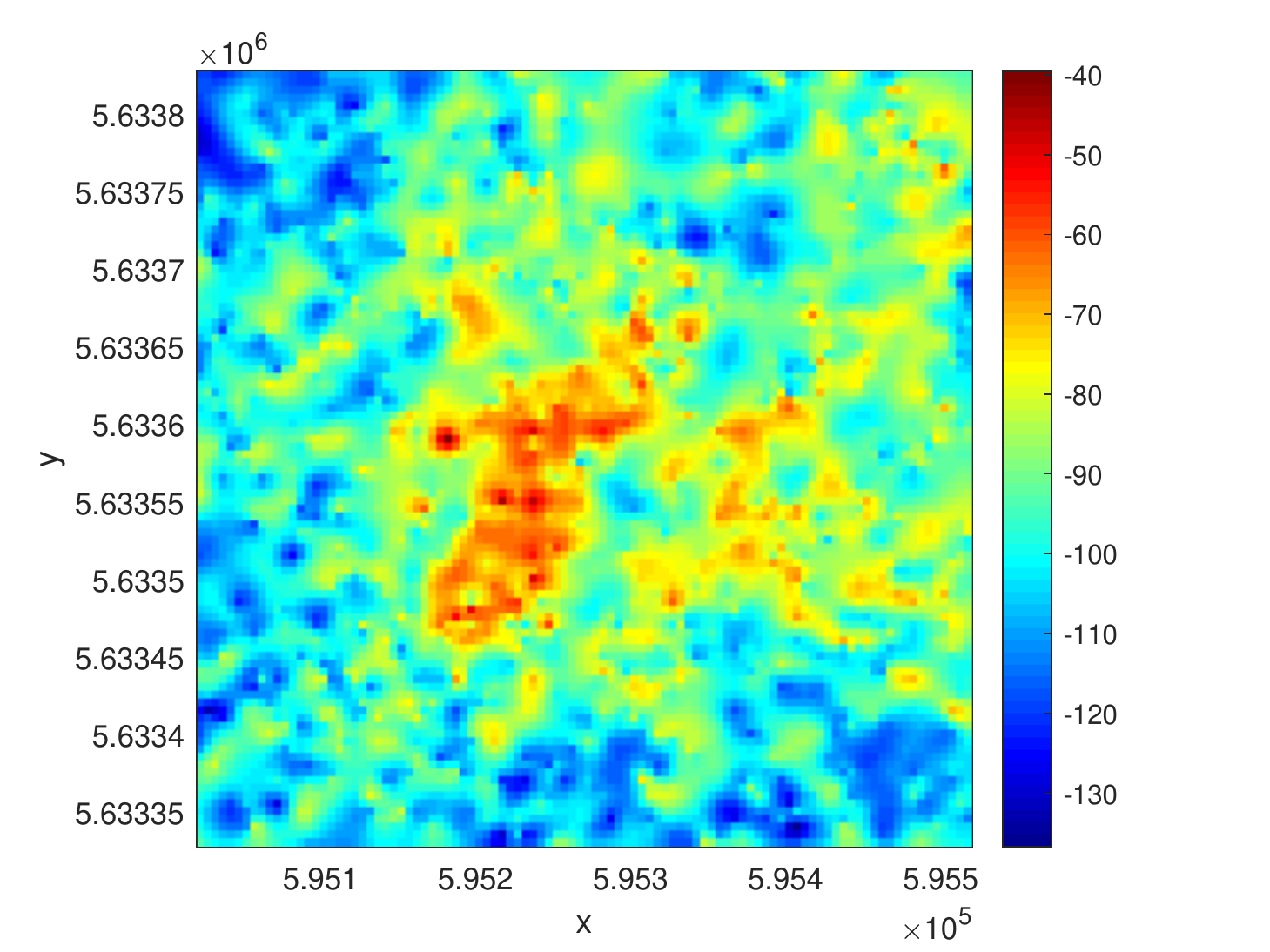}
		\caption{Kriging}
	\end{subfigure}
	\caption{Comparison of coverage map reconstruction techniques \cite{qureshi2020enhanced}.}
	\label{comparison}
		\vspace{-0.2in}
\end{figure*}

Another type of widely used data on which interpolation techniques are applied is the minimization of drive test (MDT) data \cite{qureshi2020enhanced}. 3GPP has standardized MDT that allows network performance estimation at a base station by leveraging measurement reports gathered at the user equipment (UE) without the need for drive tests \cite{release10}. The MDT reports contain network coverage related performance indicators (such as RSRP) measured at the UE. These reports are tagged with UEs' geographical location information and sent to their serving base stations \cite{my_letter}. MDT data can be scarce in areas of low user density, which will lead to inaccurate or sub-optimal coverage estimation models \cite{qureshi2020enhanced}.
 To address this problem, authors in \cite{qureshi2020enhanced} applied several interpolation algorithms, including the ones discussed in this section. Their results are illustrated in Fig. \ref{comparison} 
 and will be discussed further in the subsection pertaining to data enrichment technique used in each of the subfigures.
 
Different interpolation techniques can be applied in the cellular network context to address the data scarcity challenge. Each technique has its own set of advantages and disadvantages; we elaborate these techniques in this section.

\subsection{Matrix completion theory}
\label{sec:matrix_completion}
A recent work \cite{qureshi2020enhanced} applied matrix completion theory to cellular network data context. Assuming the coverage area is divided into bins, a coverage matrix $\bm C$ containing coverage indicator (such as RSRP measurements) is observed.
A scheme that jointly exploits matrix factorization theory and convex optimization is used to recover the missing data in $\bm C$ \cite{qureshi2020enhanced}. 

This leads to the following optimization problem in order to find the missing values in matrix $\bm C$:
\begin{equation}
\text{minimize} \quad  \text{rank} \{\bm P\} \nonumber\\
\text{subject to} \quad P_{ij}=C_{ij}\quad  (i,j) \in \Psi 
\label{min_rank}
\end{equation}
where  $\bm P$ is the decision variable in the optimization problem, the pair $(i,j)$ denotes the $i$-th row and $j$-th column of the matrices $\bm C$and $\bm P$ and  $\Psi$ is the set of locations corresponding to the observed entries $((i, j) \in \Psi$
 if $C_{ij}$ is observed).
However, the problem in \eqref{min_rank}  is known to be not only NP-hard, but also all known algorithms that provide exact solutions require time doubly exponential in the dimension $n$ in both theory and practice \cite{matrix_completion}.
However, the analysis presented in \cite{matrix_completion}  proves that the coverage values in vacant bins can be obtained with high accuracy by solving the following alternate convex optimization problem:
\begin{equation}
\text{minimize} \quad ||\bm P||_{*} \nonumber\\
\text{subject to} \quad P_{ij}=C_{ij}\quad  (i,j) \in \Psi 
\label{min_nn}
\end{equation}
\noindent where $||\bm P||_{*}$ is the nuclear norm and is given as:
\begin{equation}
||{\bm{P}}||_{*} = \sum_{k=1}^{n} \sigma_k( {\bm{P}})
\label{nuclear_norm}
\end{equation}
\noindent In \eqref{nuclear_norm}, $\sigma_k( {\bm{P}})$ denotes the $k$th largest singular value of ${\bm{P}}$. \eqref{min_nn} therefore aims to determine the matrix with minimum nuclear norm that fits the data.

The problem in \eqref{min_nn} can be solved with the singular  value-based threshold (SVT) algorithm presented in \cite{svt}. The SVT  algorithm solves the following problem:
\begin{equation}
\text{minimize} \quad \eta||{\bm P}||_{*} +\frac{1}{2} ||{\bm P}||_{F}^2\nonumber\\
\text{subject to} \quad \mathcal{O}_{\Psi}(\bm P) = \mathcal{O}_{\Psi}(\bm C)
\label{svt}
\end{equation}
\noindent where $\mathcal{O}_{\Psi}$ is the orthogonal projector onto the span of matrices vanishing outside of ${\Psi}$ so that the $(i,j)$th component of $\mathcal{O}_{\Psi}(\bm P)$ is equal to $P_{ij}$ if $(i,j) \in \Psi$ and zero otherwise. It is shown in \cite{svt} that the solution of the  problem of \eqref{svt} converges to that of \eqref{min_nn} as $\eta \rightarrow \infty$. 
The SVT algorithm is iterative and produces a sequence of matrices $\{\bm P,\bm Q\}$. At each step, a soft-thresholding operation is performed on the singular values of the matrix $\bm Q^t$.  Thus, by selecting a large value of the parameter, $\eta$ in \eqref{svt}, the sequence of iterates, $\{\bm P^t\}$ converges to a matrix which nearly minimizes \eqref{min_nn}. Starting with $\bm Q^0=\bm 0 \in \mathbb{R}^{(n \times n)}$, the algorithm inductively defines
\begin{equation}
\bm P^t=\text{shrink}(\bm Q^{t-1}, \eta) \label{shrink}\\
\bm Q^t=\bm Q^{t-1}+\Delta _i \mathcal{O}_\Psi (\bm C-\bm P^t)
\end{equation}
\noindent where $\{\Delta_i\}, i\geq 1$ is a sequence of scalar step sizes, until a stopping criteria is reached. The shrink function in \eqref{shrink} applies a soft-thresholding rule at level $\eta$ to the singular values of the input matrix. It is defined as
\begin{equation}
\text{shrink}(\bm Q^{t-1}, \eta)=\mathcal{S}_\eta(\bm Q^{t-1}) :=\bm U \mathcal{S}_\eta(\bm \Sigma) \bm V^* \\
\mathcal{S}_\eta(\bm \Sigma)= \text{diag} (  \{   (\sigma_k-\eta)_+    \}    ) \label{shrink_end}
\end{equation}
\noindent where $f_+=\text{max}(0,f)$. 
Equivalently, this operator is the positive part of $f$ and simply applies a soft-thresholding rule to the singular values of $\bm P$, shrinking them towards zero. 
$\bm U, \bm V$ are matrices with orthonormal columns and the singular values $\bm \Sigma$ are positive. $\bm U, \bm V$  and $\bm \Sigma$ are obtained from the singular value decomposition of matrix $\bm P$ of rank $r$:
\begin{equation}
\bm P=\bm U \bm \Sigma \bm V^*, \quad  \bm \Sigma=\text{diag}(\{\sigma_k\}), 1\leq k\leq r
\end{equation}
In case of the presence of random shadowing in the model, the stopping criteria of the algorithm can be modified as follows:
\begin{equation}
||  \mathcal{O}_{\Psi} (\bm P^t-\bm C)     ||_{F}^2 \leq (1+\zeta)m \phi^2
\label{stopping}
\end{equation}
\noindent where $\zeta$ is a fixed tolerance. 
The SVT algorithm is stopped when $\bm P^r$ is consistent with the data and obeys \eqref{stopping}. 
Therefore, the reconstruction matrix, $ \hat{\bm C}$ is the first $\bm P^t$ obeying \eqref{stopping}. 

Another similar rank minimization based algorithm used to recover the matrix $\bm C$ is the fixed point continuation (FPC) algorithm \cite{fpc}. While SVT is efficient for large matrix completion problems, it only works well for very low rank matrix completion problems. For problems where the matrices are not of very low rank, SVT is slow and not robust and therefore, often fails \cite{fpc}. To solve this problem, FPC-based algorithm is proposed in \cite{fpc}. FPC-based algorithm has some similarity with the SVT algorithm in that it makes use of matrix shrinkage as in \eqref{shrink}-\eqref{shrink_end}. However, it solves \eqref{svt} by leveraging operator splitting technique \cite{operator_splitting}.

\begin{algorithm}[h]
	\SetKwInOut{Input}{Input}
	\SetKwInOut{Output}{Output}

	\Input{sampled set $\Psi$ and sampled entries   $\mathcal{O}_\Psi (\bm C)$ , tolerance $\zeta$, parameter $\eta$, step size $\Delta$, increment $\alpha$, number of maximum iterations, $I_{M}$,  shadowing standard deviation $\phi$, and cardinality of $\Psi$ , $m$  }
	\Output{$\bm P^{opt}$}

Set $\bm Q^0=i_0 \Delta \mathcal{O}_\Psi (\bm C)$\\
Set $\tau_0=0$\\
{\bf for} $t=1$ to $I_{M}$\\
\hspace{5mm} Set   $h_t=\tau_{t-1}+1$\\
\hspace{5mm} {\bf repeat}\\
\hspace{10mm} Compute $[\bm U^{t-1}, \bm \Sigma^{r-1}, \bm V^{t-1}]_{h_t}$\\
\hspace{10mm} Set  $t_t=h_t+\alpha$\\
\hspace{5mm} {\bf until} $\sigma_{h_t -\alpha}^{t-1} \leq \eta$\\
\hspace{5mm} Set $\tau_r =\text{max} \{j:\sigma_{j}^{t-1}>\eta\}$\\
\hspace{5mm} Set $\bm P^t=\sum_{j=1}^{\tau_r}(\sigma_{j}^{t-1}-\tau) \bm u_{j}^{t-1} \bm v_{j}^{t-1}$\\
\hspace{5mm} {\bf if} $||  \mathcal{O}_{\Psi} (\bm P^t-\bm C)     ||_{F}^2 \leq (1+\zeta)m \phi^2$ {\bf then break}\\
\hspace{5mm} Set $Q_{ij}^t= \begin{cases} 
0   & \text{if} \quad (i,j) \not \in \Psi \\
Y_{ij}^{t-1} +\Delta(C_{ij}-P_{ij}^{t}) & \text{if} \quad (i,j) \in \Psi 
\end{cases} $ \\
{\bf end} $for \hspace{1mm} t$\\
Set $\bm P^{opt}=\bm P^t$
\caption{Singular value thresholding algorithm for finding missing coverage values}
\label{svt_algo}
\end{algorithm}

Authors in \cite{qureshi2020enhanced} use matrix completion for the task of interpolating missing RSRP values from MDT-based data. Fig. \ref{comparison} (e)-(f) is an illustrative example of their result. Authors in \cite{qureshi2020enhanced} conclude that this scheme is more likely to work well in small cells environments since matrix $\bm C$ will naturally be low ranked in such scenarios. This observation stems from the fact that propagation conditions are mostly dominated by line of sight in small cells and the standard deviation of shadowing is generally small. Moreover, the shadowing phenomenon that heavily determines coverage values, particularly in a small cell environment, remains correlated over small distances that separate users in the same small cell. However, the network scenario they consider consists of macro cell environment, therefore, the application of matrix completion to small cell environments needs further investigation.

\vspace{-0.5cm}

\subsection{Inverse distance weighted}
In this section, we first discuss the simplest form of inverse distance weighted (IDW) method, the simple IDW. Then we highlight several improvements in simple IDW interpolation and finally present an adaptive IDW method from literature.

\subsubsection{Simple IDW}
The simplest form of IDW method is also known as the Shepard’s method. It is based on the assumption that the distribution of signal samples is strongly correlated with distance.
To estimate the missing received signal strength value, ${\hat{c}}$ (at a particular bin location, $D$) in the matrix $\bm C$, weighted average of $N$ known signal strength values, $c_{k}$ from $N$ adjacent bins are used, where $k=1 \hdots N$. Each known received signal strength value is weighted with a weight that is equal to the inverse of distance, $d_k=d(D,D_k)$ between the location of the bin with missing RSRP value and location of the $k$-th bin and raised to the power $p$. 
Mathematically, the missing received signal strength value is calculated as:

\begin{equation}
\hat {c}=\begin{cases}\frac{\sum_{k=1}^{N} \frac{1}{d_k^p} c_{k}} {\sum_{k=1}^{N} \frac{1}{d_k^p}} \hspace{2mm} \text{if} \hspace{2mm} d_k \neq 0 \\
c_k \hspace{2mm} \text{if} \hspace{2mm} d_k = 0 
\end{cases}
\label{idw}
\end{equation}

The choice of $p$ is an important parameter in this method. For $p<1$, $\hat {c}$ remains no longer differentiable. Therefore, the exponent has to exceed 1 for the interpolation function to remain differentiable with respect to spatial coordinates (Cartesian coordinates $x$ and $y$ that are used in distance calculation)  \cite{idw_improvement}. It is shown by empirical testing that higher exponents tend to make the surface flat near all data points and the gradients over small intervals between data points are very steep. On the other hand, lower exponents tend to produce a relatively flat surface with short blips to achieve appropriate values at data points  \cite{idw_improvement}. When $p=0$ in \eqref{idw}, the missing coverage value is set equal to the weighted arithmetic average of the  neighboring coverage values and the recovery method is often termed as the `moving average method'.

Simple IDW method's disadvantages are that it leads to the production of the ``bull’s-eyes” effect, it is sensitive to  measurement outliers, it introduces significant errors in case of non-uniform distribution measurements or unevenly distributed data clusters, computational error becomes highly significant in the neighborhood of a data point, the calculation of missing value increases proportionally with the number of data points, leading to inefficiency of the method when the number of data points is large. Also, there is no way of pre-determining the optimal weighting power factor that will construct the most accurate RF-REM. The appropriate search radius also needs to be optimized. Another drawback is the lack of directionality, i.e., different configurations of co-linear points could yield the same results, attributing to the fact that only the distances from the missing location to the points with known locations are considered and not their direction \cite{survey_REM},\cite{idw_improvement}.

However, the advantages of simple IDW method include its efficiency and ease of comprehension since it is intuitive. This interpolation works best with evenly distributed points.

An illustrative example of IDW for REM interpolation using MDT-based RSRP measurements is shown in Fig. \ref{comparison} (f). It can be seen from the figure that although techniques like kriging in Fig. \ref{comparison} (j) outperform IDW in terms of accuracy of REM construction, IDW does outperform several techniques like moving average in Fig. \ref{comparison} (c) and is usually preferred for its reduced computational complexity. 
IDW has been widely used for REM construction in outdoor environments, such as in \cite{qureshi2020enhanced}, where authors use RSRP data to complete scarce REM using IDW.
Results in \cite{singh_sengar_construction_2018} also favor  the adoption of IDW for REM construction in a device-to-device network  crowd-sourcing scenario consisting of
Nakagami-m and Nakagami-lognormal channels.

\subsubsection{Improved IDW}
In order to address the drawbacks of simple IDW method in the preceding subsection, several improvements have been suggested in literature. 

The focus of the work in \cite{reliability} is on the reliable estimation of radio interference field with small number of measurements. For this purpose, different variants of IDW spatial interpolation method are employed which have proven robustness when dealing with limited number of observations \cite{reliability}.

Authors in \cite{idw_improvement2}, \cite{idw_improvement} and \cite{idw_improvement5} improve the weighting function by proposing a framework to intelligently select the nearby data points to be used in predicting the missing data point. This approach is developed keeping the overall density of the data points into consideration. 

Authors in \cite{idw_improvement} incorporate a direction factor, in addition to the distance factor in defining the weights. This direction factor is based on the cosine of angle of $D_i D D_j$, where $i \neq j$ and $i,j=1 \dots K$. If other data points $D_j$ are in approximately the same direction from $D$ as $D_i$, then the angles, $1-cos(D_i D D_j)$ are close to 0. On the other contrary, if other data points are in the opposite  $D$ from $D_i$, then the angles $1-cos(D_i D D_j)$ are close to 2. The direction factor in the improved weighting function in \cite{idw_improvement} leverages this fact. 

Other improvements to simple IDW  involve reduction of computational complexity and errors and  making  features of the interpolation function desirable, i.e., ensuring non-zero gradients at every location to achieve the desired partial derivatives for the function to remain differentiable \cite{adaptive_IDW}, \cite{idw_improvement}.

Since simple IDW assumes that the distance decay is uniform throughout the entire study area, it does not perform well in case of clustered data or data that depicts spatial variability. To address this problem, authors in \cite{idw_improvement3} suggested an improvement based on the weighted median of data in the neighborhood of missing data point. The weighting function in \cite{idw_improvement3} is a function of inverse-distance weights and the de-clustered weights that include the effects of distance and clustering among spatially correlated data in the estimator. 

In order to increase the accuracy of predictions through the IDW method, authors in \cite{idw_improvement4} proposed the use of piecewise least-square polynomial regression estimators to increase the accuracy, after evaluating fifteen different estimators using an extensive evaluation data set.

For  reducing the ``bull-eye" effect in simple IDW method, a  distribution-based distance weighting (DDW) technique is used \cite{idw_improvement5}. Weight calculations in DDW method are based on appropriate distributions according to available data, such as  Gaussian, Lorentzian and Laplacian distributions. Such a  distribution-based calculated ensures that if  data variations are very small, then the distribution will have a fairly sharp peak and will cause the weighting to be more sensitive to the distance. On the contrary, if data included in the interpolation are more spread out, a distribution with a larger variance would be a good choice and this would result in the distances having less impact on the weight calculations. 
 
 Authors in \cite{idw_improvement5} and \cite{idw_improvement6} propose another improvement to the IDW-based method, that incorporates temporal dimension in addition to spatial dimension. Although these approaches are  evaluated in the context of environmental data, such an approach can also be applied to wireless network data. In the approach in \cite{idw_improvement5}, time is treated independently from the spatial
 distance dimension and weights are calculated in two steps: using the inverse of 2D-spatial distance, followed by the inverse of the 1D-temporal distance \cite{idw_improvement5}.  Authors in  \cite{idw_improvement6} assume second‐order   
 non‐stationarity of both spatial and temporal distributions of the data, based on which they treat the space‐time
variables in their proposed method as a sum of independent spatial and temporal non‐stationarity components. Heterogeneous covariance functions are constructed to obtain the best linear unbiased estimates in spatial and temporal dimensions \cite{idw_improvement6}.

The applications of improved IDW techniques for  cellular network data are far less common than their application to the environmental modeling/geoscience domain \cite{idw_improvement4,idw_improvement5,idw_improvement6}. In wireless networks context, the study in \cite{reliability} used improved IDW accounting for the direction, the number and set of considered neighboring points and the slope of the interpolation function,  for radio interference field  estimation based on distributed spectrum use measurements. It  concluded that as compared to classical IDW, improved IDW experiences lower variance of mean absolute error but had more outliers \cite{reliability}.

\begin{table}
\centering
\caption{Improvements to IDW interpolation.}
\def\arraystretch{1.3}
 	\begin{tabular}{ |P{5cm}|P{2.8cm}|} 
 		\hline
 		 \bf	Improvement& \bf References\\ 
 		\hline 
 		Intelligent selection of data in neighborhood& \cite{idw_improvement2}, \cite{idw_improvement}, \cite{idw_improvement5} \\ 
 		\hline
 		Addition of directionality & \cite{idw_improvement}\\ 
 		\hline
 	Reduction of computational complexity & \cite{idw_improvement2},\cite{idw_improvement},\cite{idw_improvement5},  \cite{adaptive_IDW}, \cite{idw_improvement} \\
 		\hline
	Reduction of computational errors &  \cite{adaptive_IDW}, \cite{idw_improvement}, \cite{idw_improvement4} \\
  		\hline
  		 Addition of desirable features &  \cite{adaptive_IDW}, \cite{idw_improvement}\\ 
  		 \hline
  		  Extension to clustered/non-uniformly distributed data&\cite{idw_improvement3} \\ 
  		   \hline
  		    Addition of temporal dimension & \cite{idw_improvement5}, \cite{idw_improvement6}\\ 
  		    \hline
  		      Reduction of ``bulls-eyes" effect    & \cite{idw_improvement5} \cite{survey_REM}\\ 
  		      \hline
  	 	\end{tabular}
\end{table}

\subsubsection{Adaptive IDW}

The IDW method assumes that the distance-decay structure is uniform throughout the entire study area. However, recognizing the potential of varying distance-decay relationships over area, authors in \cite{adaptive_IDW} proposed a variation in the value of weighting parameter, $p$ according to the spatial pattern of sampled points in the neighborhood using information derived from empirical data.
Intuitively, when the unsampled location has highly clustered points around its neighborhood, a small $p$ is appropriate so that the nearest sampled values will not have an overwhelming influence on the estimated value. On the contrary, a large $p$ is desirable when data is spatially dispersed since the more reliable source for the estimate will likely be influenced from the closest location,  therefore, if a small $p$ value is used in this case, the contributions from local and more reliable sources will be small, resulting in less reliable estimates  \cite{adaptive_IDW}.

In order to adjust $p$  according to the spatial pattern of known data, authors in \cite{adaptive_IDW} first quantify the spatial pattern of sample locations in the form of nearest neighbor statistic:
 \begin{equation}
 R= r_o/r_e, \hspace{3mm} r_e= \frac{1}{2(M/A)^{0.5}}
 \end{equation}
 \noindent where $r_e$ and $r_o$ are the expected and observed average nearest neighbor distances respectively and $A$ is the area under consideration.
 
\begin{figure}[!t]
	\centering
	\includegraphics[width=7cm, height=3.7cm]{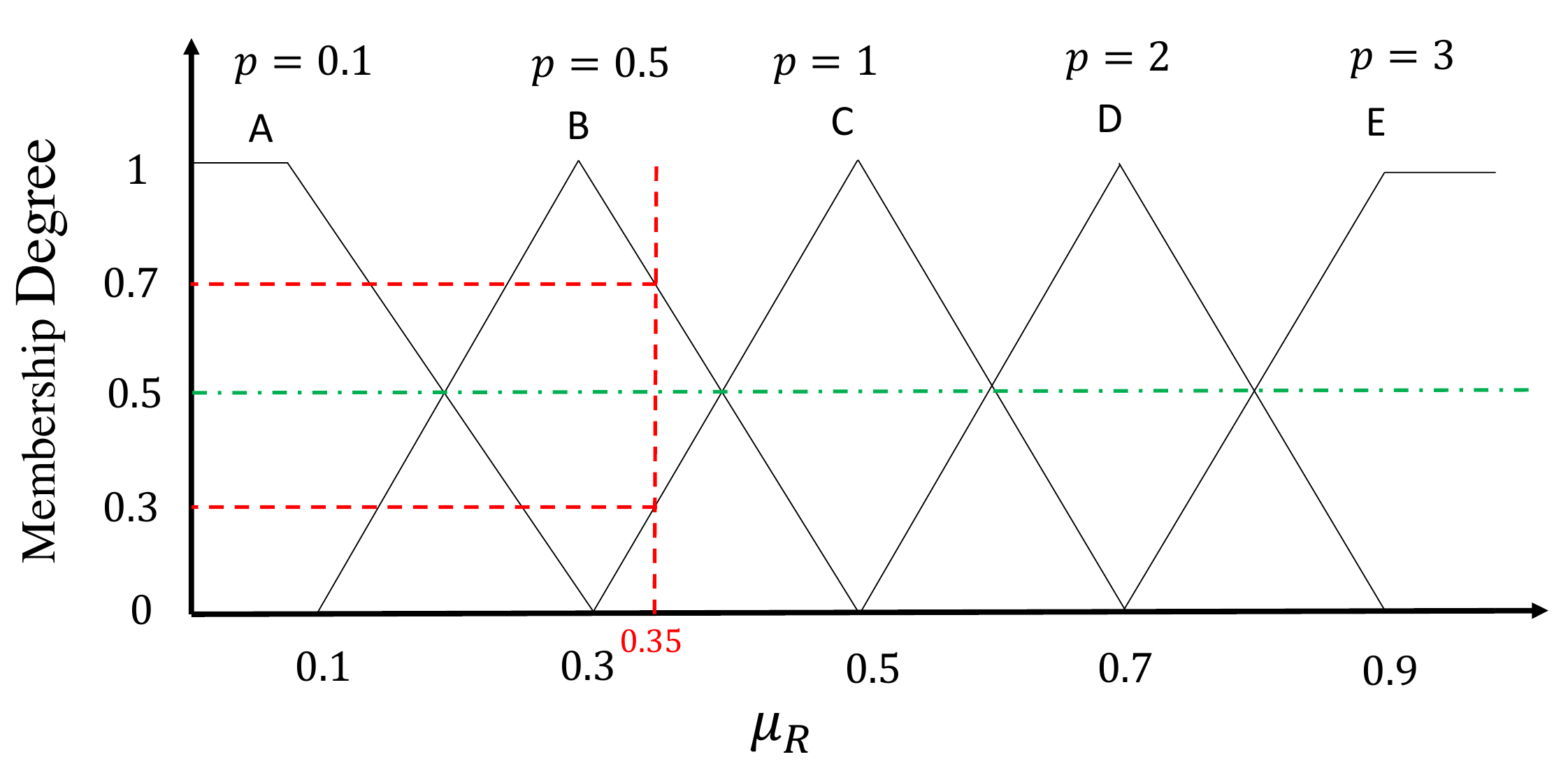}
	\caption{Triangular membership function for different adaptive distance-decay parameters (modified from 
		\cite{adaptive_IDW}).}
	\label{adaptive_IDW_membership}
 \vspace{-0.2in}
\end{figure}

 After normalizing R to get the normalized local nearest neighbor statistic, $\mu_R$, in the adaptive IDW method, this neighbor statistic carries a fuzzy membership that belongs to certain categories of $p$. This membership function is depicted in Fig. \ref{adaptive_IDW_membership}. As an example, $\mu_R$ corresponding to  $R$ of 0.8 will be 0.35, yielding two points in the membership degree (0.3 for category C and 0.7 for category B).
 The final $p$ would then be a weighted sum of these membership degrees and corresponding $p$ values (0.5 for category B and 1 for category C). Consequently, the final $p$ will be: $0.7 \times 0.5+0.3\times 1 = 0.65$. 
 
Adaptive IDW (AIDW) method can outperform IDW and work well in situations where local variability is relatively large or spatial correlation structure of the data is not strong or data is too limited to support data intensive methods, such as kriging. It is shown to outperform ordinary Kriging, when the spatial structure of data was such that it could not be modeled accurately by a variogram function \cite{adaptive_IDW}.

However, as compared to IDW, the AIDW method is computationally intensive as the distribution of $p$ has to be formulated to find the optimal set of parameter values, which require significant level of heuristics  \cite{adaptive_IDW}.

\subsection{Gradient plus inverse distance squared}

Gradient plus Inverse Distance Squared interpolation (GIDS) combines multiple linear regression and inverse distance based weighted coefficients for the interpolating missing data. By assuming that the  data of interest can be represented by a multivariate function, for the unsampled location, $D$, an ordinary least squared regression is done using $N$ neighboring locations. This yields the coefficients which represent the location gradients. If the measurements are taken at different heights, GIDS method can incorporate the elevation dimension in interpolation too. Assuming $D = (x,y,z)$ with corresponding coefficients $C_x, C_y, C_z$, representing the $x,y,z$ gradients respectively, the missing data point through GIDS can be estimated as \cite{comparative_REM}:

\begin{equation}
\hat{c} = \frac{\sum_{k=1}^{N}\left(c_k+C_x(x-x_k) +C_y(y-y_k)+C_z(z-z_k)\right)/d_k^2}{\sum_{k=1}^{N} 1/d_k^2}
\end{equation}

 The advantage of GIDS method is its ability to account for signal level gradients and elevation of the terrain at the interpolated location and at locations of the measurements. However, this method is very sensitive to the selection of neighborhood points as a small neighborhood selection would leave out important measurements and a large  neighborhood selection  may introduce noise \cite{survey_REM}.

GIDS has been used for REM construction in \cite{comparative_REM}, where authors conclude that when number available measurements are sufficient,
then Kriging outperforms GIDS in terms of
lower relative mean absolute error in most REM simulation scenarios. Note also that Kriging is highly sensitive to the performance metric used as it minimizes mean squared error (MSE), so performs best when MSE is used as evaluation metric.

\subsection{Modified Shepard's method}
The IDW based modified Shepard’s method (MSM) is a local interpolation that makes the estimation based on a real multivariate function, $f$, whose local approximation is  referred to as nodal functions. If $Q_k$ is the  output of the nodal function  of the data point $D_k$  (local approximation to $f$ at $x_k,y_k$), then the missing value using the MSM method can be written as a weighted average of the nodal functions within some radius influence (about the missing data point), $R_w$ in the following manner \cite{comparative_REM}, \cite{reliability}:
 \begin{equation}
 \hat{c} =
 \frac{\sum_{k=1}^{N} W_k Q_k}{\sum_{k=1}^{N}W_k}
 \end{equation}
First, the weights, $W_k$ are calculated by the following formula:

\begin{equation}
W_k=  \begin{cases}[R_w-d_k]/R_w d_k]^p \hspace{3mm} \text{if}  \hspace{2mm} d_k<R_w \\ 0 \hspace{3mm} \text{if}  \hspace{2mm} d_k \geq R_w   \end{cases}
\label{msm_weight}
\end{equation}

Then, another radius, $R_v$ around each known data point is considered and the weights are again calculated using \eqref{msm_weight}, this time, replacing $R_w$ with $R_v$. 

This technique can be extended to multivariate case but is dependent upon optimization of $R_w$, $R_q$ and $p$.  It is also shown to perform poorly if measurements lie in a low-dimensional subspace \cite{survey_REM}. However, this method can reduce the `bull's eye' effect as compared to classical IDW methods. 

An example of MSM application for the task of generating REM of total received signal power is illustrated in \cite{comparative_REM}. Authors in \cite{comparative_REM} use a wireless system simulator to simulate both indoor and outdoor scenarios with different levels of data scarcity.  Among the considered methods of Kriging, MSM and GIDS, MSM generally performs somewhere in between the other two. For example, when the measurement points increase from 38 to around 695, the relative mean absolute error (RMAE) reduces from 7.5\% to 1\% for Kriging, 8\% to 1.5\% for MSM, and 9\% to 2\% for GIDS. They thus conclude that although Kriging performs best in terms of interpolation error, but due its high computational complexity and weak performance when observation points are low, MSM may be preferred as it is more flexible and robust.

\subsection{Nearest neighbor}
The nearest neighbor (NeN) method is also known as proximal interpolation or point sampling. 
 Let $D_l$ be the nearest neighbor of the missing point, $D$ and $d(D,D_l)$ denote the distance between $D_l$ and $ D$, then $\min\{d(D,D_k) \} =d(D,D_l)$, $k=1 \dots N$.  In this case, the estimated value will be the same as the value in the nearest sampled location $l$. Mathematically, the weights, $\lambda_k$ can be represented as \cite{interference_map_estimation_cognitive}: 
 \begin{equation}
\lambda_k=\begin{cases} 1  \hspace{2mm} \text{if} \hspace{2mm} k=l \\
 	0 \hspace{2mm} \text{if} \hspace{2mm} k \neq l
 \end{cases}
  \end{equation}
  \noindent which leads to the missing point prediction as:
  \begin{equation}
 \hat {c} = \sum_{k=1}^{N} \lambda_k c_k = c_l
\end{equation}
Nearest neighbor method is known for its low complexity. Among the considered techniques in \cite{delaunay} for the task of interference map interpolation,  nearest neighbor interpolation is concluded to be the least complex method and  natural neighbor, linear, cubic and quadratic interpolation techniques have shown to exhibit comparable performances.

Although nearest neighbor approach is of low complexity, it results in sharp transitions between the individual signal level zones and increases noise, especially at the boundary of a given area, since it does not consider the influence of the sample data points apart from the nearest neighboring data point \cite{survey_REM}, \cite{modeling_carrier_aggregation}. 

Fig. \ref{comparison} (g) illustrates an example of using nearest neighbor interpolation to interpolate scarce RSRP measurements for constructing coverage maps. It can be seen from the figure that compared to methods like kriging in Fig.\ref{comparison} (j)  where the interpolated coverage map is smooth, nearest neighbor interpolation results in a representation that has more sharper transitions between adjacent values.

\subsection{Natural neighbor}
The natural neighbor (NaN) interpolation is based on Voronoi decomposition (tessellation) of a set of given points in the plane. The  received signal strength value at a particular location is found from a weighted average of $N$ from all available measurements which fall within its `natural neighborhood'. 
 
 The natural neighbors of any point are those associated with neighboring Voronoi polygons. If the 2-D point ${D}_k$ is a natural neighbor of the 2-D point $\bf D$, the portion of Voronoi region, $V_{D_k}$ stolen away by $\bf D$ is called the natural region of $\bf D$ with respect to ${\bf D}_k$. Initially, a Voronoi diagram is constructed of all the available coverage values. Then, a new Voronoi polygon is created around the interpolation point (missing coverage value). The proportion of overlap between this new polygon and the initial polygons is then used as weights. If we denote the Lebesgue measure of this natural region by $l_{{\bf D}_k}$, the natural coordinate associated to ${\bf D}_k$ is used as weights \cite{comparison_in_cognitive}:
 
\begin{equation}
 \lambda_{{\bf D}_k}({\bf D}) = \frac{l_{{\bf D}_k}({\bf D})}{\sum_{k}l_{ {\bf D}_k}({\bf D})}
 \end{equation}
 
The weights are thus  the ratio of the area of overlap to the total area of the new polygon. Once the weights are obtained, interpolation to find the missing coverage value can be carried out by a weighted sum of known coverage values.
 
The natural neighbor interpolation method performs well  with non-homogeneous distribution of measurements as well. However, its major drawback is that it can not find missing signal values that lie outside the convex hull of Voronoi polygons since it requires that the points to be interpolated be in the convex hull of the measurement locations as the Voronoi cells of outer data points are open-ended polygons with an infinite area \cite{survey_REM}.

Another scheme similar to natural neighbor using an area-wise multi-criteria triangulation-induced interpolation  algorithm which utilizes the linear interpolation to estimate the key performance indicators of the QoS inside a triangle with the known values of its three vertexes is proposed to reconstruct the coverage maps in \cite{liu_multi-criteria_2019}.

Fig. \ref{comparison}  (h) is an illustrative example of the result obtained by applying natural neighbor for the task of interpolating missing RSRP values from MDT-based data in \cite{qureshi2020enhanced}. An important observation is the interpolation at the corners of the coverage map in Fig. \ref{comparison} (h), that do not have any value due to the inability of natural neighbors to fill the missing values that lie outside the convex of Voronoi polygons as identified above. 

\subsection{Splines}
The spline method is also referred to as the radius basis  function and `rubber sheeting' \cite{survey_REM}.
It estimates the missing value by a mathematical function or piecewise defined polynomials called splines that minimizes the total surface curvature. This results in a smooth surface that passes exactly through the sampled points. This interpolation method is useful for estimating above maximum and below minimum points and for creating a smooth surface effect. 
However, because of this smoothing effect, the discontinuity in data might not be well estimated. Since it  uses slope calculations or change over distance to estimate the missing values, when the known data points are too close together or have extreme differences in values, this method does not work well.

There are different kinds of splines, such as linear, quadratic, cubic, biharmonic and thin-plate splines. For example, for thin-plate splines, the unknown value is estimated as \cite{comparison_in_cognitive}:

  \begin{equation}
\hat{c}= \sum_{k=1}^{N} w_k ||D-D_k||^2 \ln(||D-D_k||)
  \end{equation}

\noindent where $||.||$ is the Euclidean norm. $w_k$ can be obtained by solving $\bf Ow=i$, where $\bf i$ and $\bf w$ are the column vectors of input data points and weights respectively, while $\bf O$ is the matrix of output of the  basis function ($||D-D_k||^2 \ln(||D-D_k||)$ in this case) for all possible input values.

A visual example of splines in the case of REM construction of RSRP measurements is illustrated in Fig. \ref{comparison} (i). Authors in \cite{qureshi2020enhanced} conclude that Splines and Kriging have similar performance quantitatively in terms of relative recovery error (Frobenius norm of recovered interpolated matrix minus the ground truth matrix divided by Frobenius norm of ground truth matrix). 

\subsection{Kriging}
Kriging, unlike the other  methods discussed above, also takes into account the statistical relationships in additional to spatial relationships among the measured data points to estimate the missing values of data.

In Kriging, the weights are based not only on the distance between the measured points and the prediction location but also on the overall spatial arrangement of the measured points \cite{kriging, graph_signal_processing}. The weight coefficients are calculated by minimizing the variance of the estimation error, $\sigma_e^2$:
\begin{equation}
\sigma_e^2 = \mathbb{V} \hspace{1mm}[\hat{C_m} - C_m]
\label{kriging_opt}
\end{equation}

\noindent where $\mathbb{V}$ is the variance operator and $C_m$ is the missing coverage value located at the 2-D point, $\bf p$. 

The first step in kriging therefore involves creating a prediction surface map in order to uncover the dependency rules to make predictions. To achieve this, kriging first creates a semivariogram and covariance functions to estimate the statistical dependence values that depend on the model of autocorrelation. 
To solve the optimization problem in \eqref{kriging_opt}, semivariogram function, $\gamma$ is used to characterize the spatial correlation. 

The next step is to fit a model to the points forming the empirical semivariogram. A mathematical function is used to fit the empirical semivariogram as the theoretical semivariogram model to model spatial autocorrelation. 
There are many variants of kriging based on advanced and robust semivariogram models, such as simple kriging, 
block kriging, factorial kriging, kriging with a trend, dual kriging, universal cokriging, kriging with an external drift, indicator kriging, probability kriging, to name a few. A comparison of these variants is presented in \cite{liheap}, \cite{survey_environmental_sciences}.
Kriging weights then come from the semivariogram that was developed by analyzing the spatial nature of the data. These weights are a result of minimizing the variance in \eqref{kriging_opt}, which yield the following solution \cite{interference_map_estimation_cognitive}:
\begin{equation}
\begin{bmatrix}
{\bm \lambda} \\ 
\delta
\end{bmatrix} = 
\bf X^{-1} y
\label{extra_term}
\end{equation}
\noindent where $\bf X$ and $\bf y$ are defined as:
\begin{equation}
\bf X = \begin{bmatrix}
X_{1,1} & \cdots  & X_{1,N} & 1 \\
\vdots &  \ddots & \vdots & \vdots \\
X_{N,1} &  \cdots &  X_{N,N} &\vdots \\
1   & \cdots  &1 & 0 \\
\end{bmatrix},  \quad \bf y = 
\begin{bmatrix}
y_1 \\ \vdots \\ y_N \\ 1
\end{bmatrix}
\end{equation}
\noindent Each element of matrix, $\bf X$, $X_{i,j} = \gamma(|| {\bf p}_i - {\bf p}_j ||)$ and each element of the column vector $\bf y$, $y_i = \gamma (||{\bf p}-{\bf p}_i ||)$. The extra element in the weight vector solution in \eqref{extra_term}, $\delta$, is the result of fitting by assuming a mean trend component in the reconstructed coverage matrix.

Kriging is applied on RSRP measurements for REM construction in \cite{alam_clustering_2018, kriging_cs_rem}.
A more practical implementation of Kriging based approach using real data from the University of Colorado, Boulder campus has been demonstrated in \cite{practical_REM}.
In \cite{mao_constructing_2018}, the authors propose a REM construction  method by combining residual maximum likelihood-based radio propagation parameter estimation with Kriging-based transmission power prediction. They then benchmark the performance of their proposed algorithm 
with a path loss-based method and a Kriging-based method without prior fit of a path loss model, using the metric of  root mean square error (RMSE). Another  Kriging-based radio  environment map construction method based on mobile crowd sensing is proposed in \cite{han_radio_2019}. Authors in \cite{han_radio_2019} compare Kriging with the nearest neighbor and the inverse distance weighting interpolation algorithms and conclude that Kriging performs the best for their crowdsourced RSRP dataset.  Kriging is applied in the  context of a REM-enabled
spectrum sharing mechanism for performance analysis for mobile cellular networks in \cite{hosseini_tehrani_radio_2019}.
Authors in \cite{xia_radio_2020} propose an improved Kriging algorithm by 
combining the concept of affinity propagation clustering in ordinary Kriging algorithm for REM construction.
Another improvement over ordinary Kriging is the fixed-rank Kriging proposed in \cite{braham2014coverage}. However, it tends to neglect the small-scale
structured variations of the data, which may result in a loss of
accuracy \cite{alam_performance_2018}.
To overcome the limitations of ordinary and fixed-rank Kriging, authors in \cite{alam_performance_2018} propose covariance tapering based Kriging.
Neural network techniques are also applied to improve Kriging algorithm in \cite{sato_performance_2019}, \cite{appleby2020kriging}, \cite{mezhoud_hybrid_2020}.

In the domain of cognitive radio networks, authors in \cite{comparison_in_cognitive} compare three interpolation methods, namely, natural neighbor, kriging and spline for constructing interference cartographs from a scarce set of data. They conclude that both kriging and natural neighbor interpolations perform similarly when the channel  uncertainty is lower and that the average efficiency of all interpolation techniques improves with increased shadowing decorrelation \cite{comparison_in_cognitive}. Authors in \cite{carrier_aggregation} conclude that Kriging performs best among nearest neighbor and inverse distance weighted (IDW) methods.
Results in \cite{interference_map_estimation_cognitive} again demonstrate the superior performance of Kriging among nearest neighbors, IDW and triangular irregular network interpolation, but has demonstrated the robustness of IDW method overall.

Authors in \cite{comparative_REM} compare Kriging, Modified Shepard’s method (MSM) and Gradient plus inverse distance squared (GIDS) and IDW for creating radio environment maps. It is concluded that Kriging and IDW are most flexible among these methods and offer trade-off between the computational cost and accuracy.

Kriging has also been used in indoor environments, such as in \cite{indoor2}, where authors compare various interpolation techniques, including Kriging, splines, weighted moving average, Theissen polygons, trend surfaces, classification, in terms of accuracy, spatial distribution of measurements, measurement density and impact of a fixed location inaccuracy for the task of signal strength prediction in an indoor environment.
The results in \cite{indoor2} indicate that  Kriging is a fairly robust technique overall, across all considered scenarios.
Kriging has also shown to be
the method which is least sensitive 
to the deployment of the sensors as compared to nearest neighbor and inverse distance weighted in \cite{suchanski_radio_2019}, where the authors analyzed the impact of the
number of sensors on the REM quality in the context of military wireless networks. They used data from real field tests with 39 sensors 
in an area of 4 $km^2$.

Fig. \ref{comparison} (j) is an illustrative example of the result obtained by applying Kriging for the task of interpolating missing RSRP values from MDT-based data in \cite{qureshi2020enhanced}. Authors in \cite{qureshi2020enhanced} report that among the methods considered in Fig. \ref{comparison}, kriging method performs the best with the least quantitative relative recovery error (Frobenius norm of recovered interpolated matrix minus the ground truth matrix divided by Frobenius norm of ground truth matrix) of less than 0.15. This is because in contrast to other interpolation methods where the weights are only dependent on the distance, the weights in kriging are based on the overall spatial arrangement of the measured points too.

The major drawbacks of Kriging are that it requires a large number of measurement points in order to achieve high precision and it involves significant input from the user in order to select the best fit function for the semivariogram. Identifying the most appropriate theoretical variogram for the given data (especially if it exhibits large spatial heterogeneity) is critical in order for Kriging to perform well. 
Although Kriging has relatively high computational complexity, it is the most commonly applied technique in the literature \cite{kriging} \cite{Venezuela} due to its higher precision. As Kriging is geostatistical method, it also can estimate the variances of predicted values in the unsampled location.

\subsection{Lessons Learned}

Among the interpolation methods, Kriging has been most widely used in literature due to its high accuracy. However, it is computationally expensive. Simpler and less computationally demanding techniques, like IDW, are shown to work best for evenly distributed data points. Kriging, GIDS, MSM and Splines can be used in cases where extrapolation is required. However, when extrapolation is not required, IDW, natural neighbors and nearest neighbors are candidate choices. Among these, natural neighbors require all data points be inside the convex hull of location measurement. Another method, matrix completion, although has shown to be very promising in other domains, its applicability to small cell environments where it will most likely work best needs further investigation.

\section{Methods using contextual information}
\label{chap:contextual_info}
The preceding section discussed techniques that can be leveraged to address the data scarcity challenge when the only known information are the measured data and their locations. 
However, if some additional information other than the observed data is known,  we can employ 
other techniques leveraging that additional information, or use it to enhance the interpolation methods. 

This additional information can be knowledge of propagation model, such as path loss and other relevant parameters, transmitter parameters, such as transmit power or antenna patterns, transmitter location estimation, network geometry, or characteristics of the operating environment. It is then combined with observed scarce data to augment it. Based on the availability of known information, different indirect approaches can be employed. 
For example, authors in \cite{indirect_methods} estimate the transmitter power and location using received signal strength (RSS) measurements and empirical model to enrich REM. Similarly, authors in \cite{pesko_indirect_2015} calibrate propagation model using transmit power, antenna diagram, azimuth and tilt angles before generating more RSS data through it.

\subsection{Utilizing geometry of network}
\label{sec:network_geometry}

\subsubsection{Triangular method (interpolation using locations of data base stations)}
\label{sec:triangle}
\begin{figure}[!t]
	\centering
	\includegraphics[width=7cm, height=5.2cm]{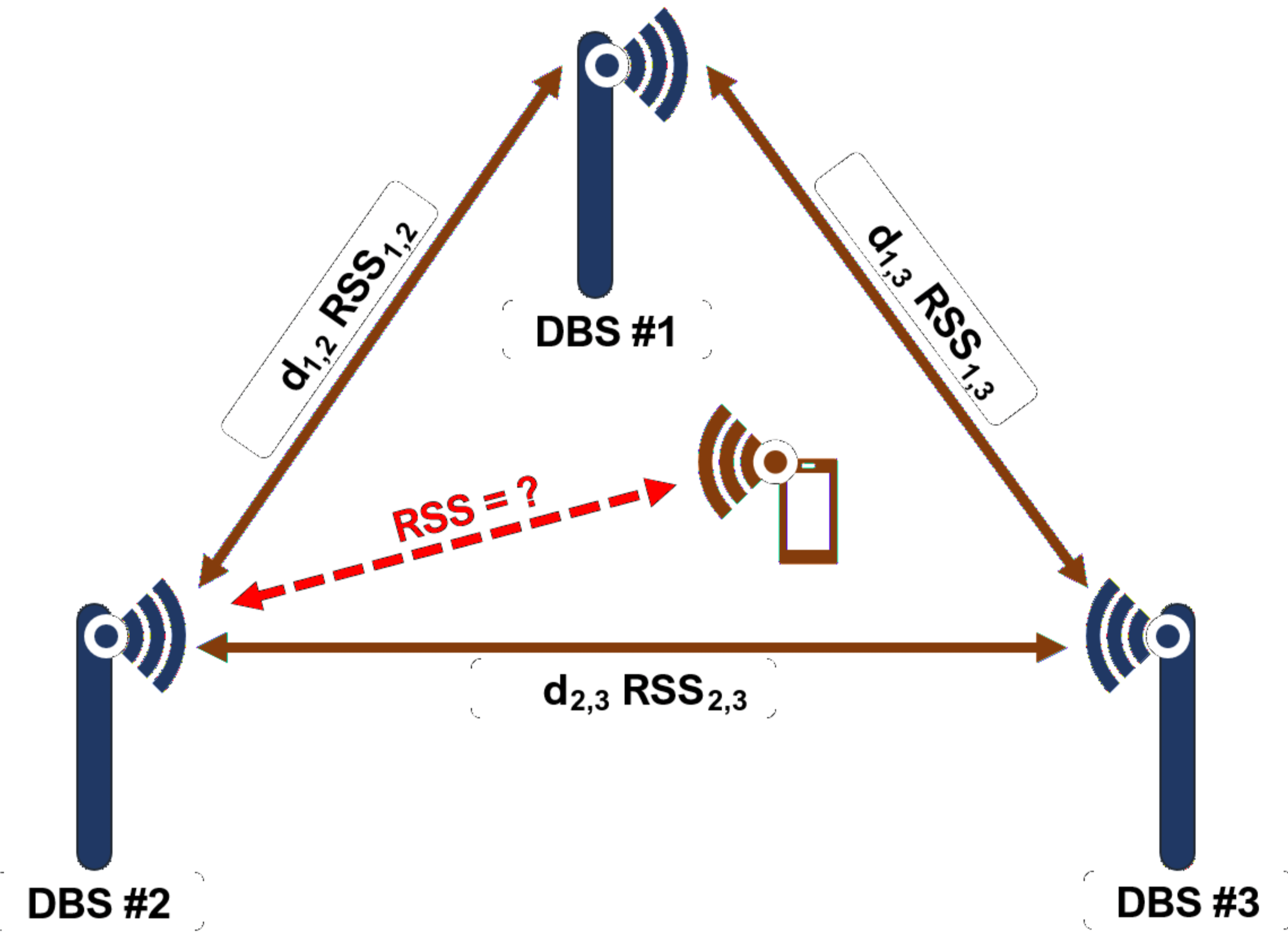}
	\caption{Leveraging dense base station deployment to enrich scarce data.}
	\label{spidernet}
 \vspace{-0.1in}
\end{figure}

One way to estimate measurements for bins with no user reports can be using the geometry of the base stations as shown in Fig. \ref{spidernet}. This is particularly suitable in ultra-dense deployment scenarios \cite{survey_cdsa}, where the data base stations (DBSs) are very densely deployed (by virtue of switching OFF DBSs to keep energy consumption and interference low).  These additional measurements, after appropriate transformation, can then be used to increase the accuracy of interpolation methods proposed above. However, this approach can complement only simple measurements such as received signal strength.

\subsubsection{Arc method (exploiting pattern among clusters in polar coordinates)}
\label{sec:cluster}
\begin{figure}[!t]
	\centering
	\includegraphics[width=7.1cm, height=4cm]{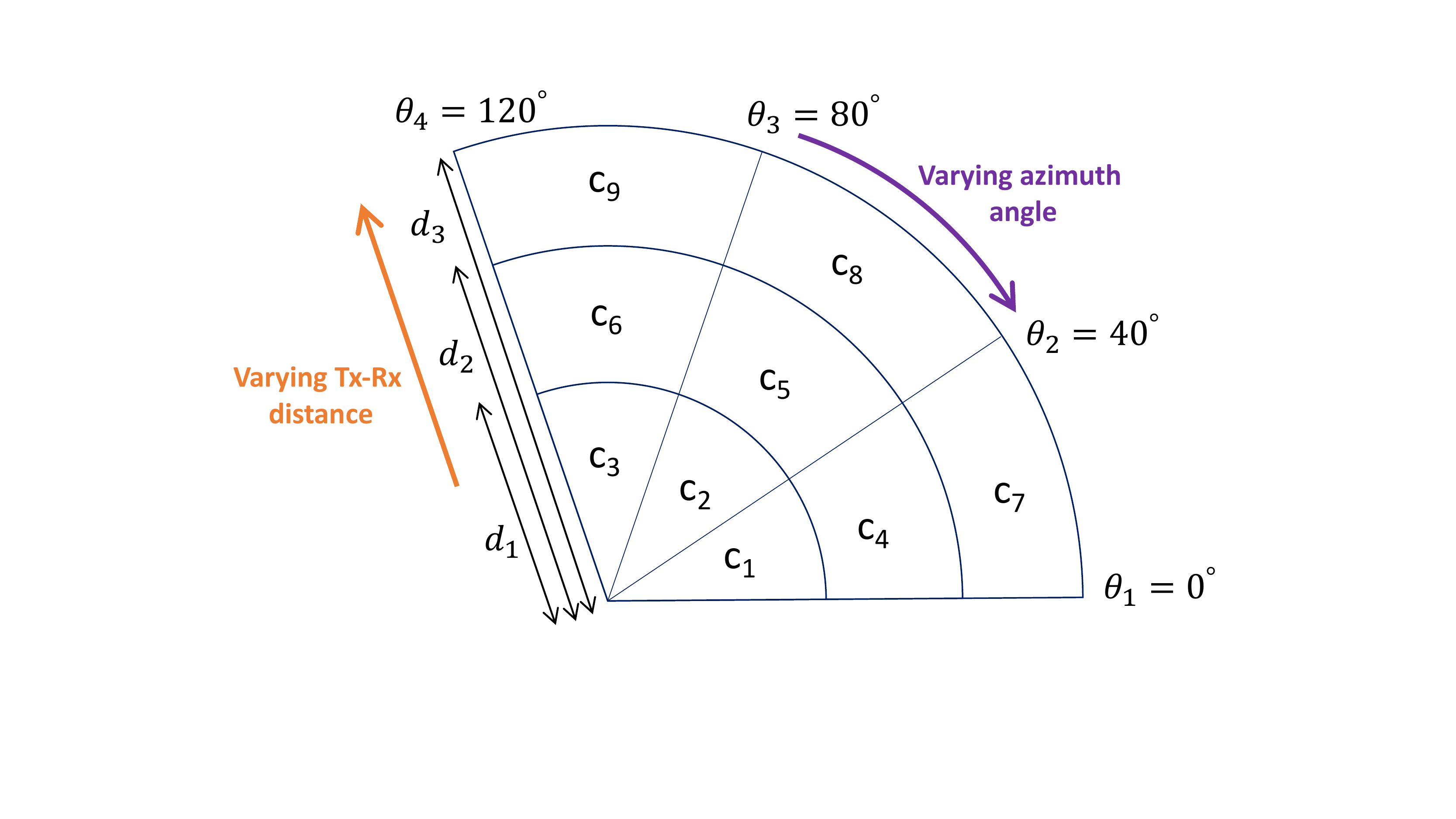}
	\caption{Leveraging cluster geometry to enrich scarce data.}
	\label{geometry_clusters}
  \vspace{-0.2in}
\end{figure}

Another way to enrich scarce data in a given network area can be by dividing the area into clusters into polar coordinates as shown in Fig. \ref{geometry_clusters}. 
Each cluster has a value that can show a given KPI, such as the average RSRP or SINR of the users in that cluster. To find the missing value in a particular cluster, geometric pattern among other clusters can be exploited, for example, if we travel along a particular circumference, we  observe that the Tx-Rx distance remains constant on that circumference and the only variation is in azimuth angle ($\theta_1$ to $\theta_4$ in Fig. \ref{geometry_clusters}). Conversely, if we traverse a path radially outwards, we can notice that the azimuth angle remains the same but there is variation in Tx-Rx distance ($d_1$ to $d_3$ in Fig. \ref{geometry_clusters} assuming base station is located at the center of the sector).
If we model the received signal strength as a function of azimuth angle and Tx-Rx distance, this pattern 
can be exploited to find the unknown signal strength values. 

Learning cluster values by exploiting this pattern using a supervised DNN has been proposed in \cite{jeff_online}. However, authors in \cite{jeff_online} has not used this approach to address the data scarcity challenge. In \cite{jeff_online}, correlations among their SINRs has been exploited to learn the locations of users at macrocells. However, we propose that such a model based on correlations among  SINRs of known clusters can also be used to find the missing SINR in another cluster.

\subsection{Through propagation modeling and transmitter parameter estimation}
\label{sec:context2}

\subsubsection{Received signal strength (RSS) based}
The RSS based method to recover scarce data is based on a combination of analytical models with statistical evaluation through measurements \cite{indirect_methods}. The RSS at a particular receiver, $i$ located at a distance, $d$ can be represented as:
\begin{equation}
P_i(d) = P_t-L-10 p \log_{10}(d)+ \phi
\end{equation}
\noindent where $P_t$ is the transmit power, $L$ is the free space path loss and $\phi$ represents a lognormal random variable for shadowing. $L$, $p$ and standard deviation of $\phi$ are environment dependent parameters.

After averaging out RSS measurements (in order to reduce random shadowing effect), and assuming the sample size of RSS measurements is large enough, the average RSS at a particular location can be estimated as:
\begin{equation}
P_i^{av}(d) \approx P_t-L-10 p \log_{10}(d), \hspace{1mm} \text{where} \hspace{1mm}P_i^{av}(d)  = \sum_{k=1}^{N} P_i^k(d)/N
\label{rss}
\end{equation}
After performing some algebraic manipulations, taking the anti-log of \eqref{rss} and representing $d$ is cartesian coordinates, \eqref{rss} can be transformed into a regression problem which can be expressed as a system of linear equation as follows \cite{live}: 
\begin{equation}
\begin{bmatrix}
10^{\frac{-L-P_1^{av}(d)}{5p}} & 2x_1 & 2y_1 & -1 \\
10^{\frac{-L-P_2^{av}(d)}{5p}} & 2x_2 & 2y_2& -1 \\
\vdots & \vdots & \vdots & \vdots\\
10^{\frac{-L-P_N^{av}(d)}{5p}} & 2x_N & 2y_N& -1 
\end{bmatrix}
\begin{bmatrix}
10^{\frac{P_t}{5p}} \\
x_t \\
y_t\\
x_t^2+y_t^2
\end{bmatrix} =
\begin{bmatrix}
x_1^2+y_1^2\\
x_2^2+y_2^2\\
\vdots\\
x_N^2+y_N^2\\
\end{bmatrix}
\label{rss2}
\end{equation}
\noindent where $x_t, y_t$ is the transmitter location and$(x_i,y_i)$ is the $i$-th receiver location. Therefore, by solving \eqref{rss2} using least-squares methods, we get estimates for transmit power, $P_t$ and the location of transmitter, $(x_t,y_t)$. These estimates can then be used to evaluate estimated received power at the missing location, by first calculating the Tx-Rx distance at the missing location and then using it to find RSS. A similar method combining transmitter localization estimation with Kriging is proposed in \cite{tsukamoto_highly_2018}.

Note that since path loss and shadowing parameters in the model are assumed to be known and are highly environment dependent, the quality of estimated is likely to be drastically affected if there is an error in estimation of propagation parameters, caused by, for example, high shadowing fading in the environment. However, this method is likely to improve if propagation conditions are not too drastic, for example, in rural areas and if the number of receivers with known measurements are large. It is also shown in \cite{live} that unlike IDW and Kriging, RSS-based method is not affected by the minimum distance between receiver and transmitter and therefore, is more robust  as compared to interpolation methods alone.

RSS algorithm was applied for the task of REM interference cartography generation in \cite{indirect_methods}. Results from \cite{indirect_methods}  show that the transmitter location estimation error decreases in an exponential manner as the number of sensor measurements increases.

\subsubsection{Received signal strength difference (RSSD) based}
The RSSD method is based on the received signal strength difference (RSSD) between two base stations or transmitters. It is assumed that transmit power is known, transmitter location, $(x_t,y_t)$ is estimated based on the idea that the ratio of the signal powers (or their differences expressed in dB) observed at two different receiver locations is related to the ratios of the transmitter to receiver distances. Specifically, the received power differences between any two receivers, located at $(x_a,y_a)$ and $(x_b,y_b)$ can be represented as \cite{indirect_methods}: 
\begin{equation}
P_{ab} = 5 p \log_{10} \left(\frac{\left(x_t-x_a\right)^2 +\left(y_t-y_a\right)^2}{\left(x_t-x_b\right)^2 + \left(y_t-y_b\right)^2}\right)
\label{tx_loc}
\end{equation}
The transmitter location in  \eqref{tx_loc} can then be estimated by solving a linear system of equations of the following form:

\begin{equation}
\begin{bmatrix}
1-\beta_{12} & -2(x_2-\beta_{12}x_1) & -2 (y_2-\beta_{12}y_1)\\
1-\beta_{13} & -2(x_3-\beta_{13}x_1) & -2 (y_3-\beta_{13}y_1)\\
\vdots & \vdots & \vdots \\
1-\beta_{1N} & -2(x_N-\beta_{1N}x_1) & -2 (y_N\beta_{1N}y_1) 
\end{bmatrix}
\begin{bmatrix}
x_t^2+y_t^2 \\
x_t\\
y_t
\end{bmatrix} = \nonumber \\
\hspace{3.4cm}\begin{bmatrix}
\beta_{12} (x_1^2+y_1^2) - (x_2^2 +y_2^2)  \\
\beta_{13} (x_1^2+y_1^2) - (x_3^2 +y_3^2) \\
\vdots\\
\beta_{1N} (x_1^2+y_1^2) - (x_N^2 +y_N^2)
\end{bmatrix}
\label{lin_sys}
\end{equation}

\noindent where $\beta_{ab} = \frac{\left(x_t-x_a\right)^2 +\left(y_t-y_a\right)^2}{\left(x_t-x_b\right)^2 + \left(y_t-y_b\right)^2}$. Solution to \eqref{lin_sys} by ordinary least squares using available receiver locations yields estimates for $x_t,y_t,x_t^2+y_t^2$. Once the transmitter  location has been estimated, the received signal level at any location can also be estimated by subtracting the path loss from transmitted signal power. As with RSS based method, this method is also dependent on selection of propagation parameters, such as  path-loss exponent and shadowing spread.

Performance comparison between RSS and RSSD based methods for REM construction was done in \cite{indirect_methods}.
 Results in \cite{indirect_methods} show that the transmitter location estimation error decreases in an exponential manner as the number of sensor measurements increases.  For example, as the number of measurements increase from 6 to 20,  the transmitter location error decreases from to 75 m to around 23 m for RSSD based approach and it decreases from around 24 m to approximately 12 m for the RSS based method.
 As can be seen quantitatively, RSSD algorithm outperforms RSS based method for all measurement densities.

\subsubsection{Angle of arrival (AOA) based}
Using prior knowledge of transmit power and using measurements from $N$ receivers with known locations, this method first estimates the angles of arrival at the locations of the measurements and combines them with the received signal powers to estimate the location of the transmitter. Once the location of the transmitter and its transmit power is available, any appropriate propagation model can be applied to estimate unknown data at different locations.

The signal model for received signal  at $i$-th receiver is modeled as \cite{sun_simple_2010}:
\begin{equation}
{\bf R}_i=  \sqrt{\alpha}(d_i) {\bf h}(\theta_i) s + {\bf {n}}_i
\end{equation}

\noindent where $s$ is the complex baseband transmitted signal with known transmit power, $d_i$ is the unknown distance between the unknown transmitter and receiver, $\theta_i$ is the unknown angle by which the signal reached the $i$-th receiver and  ${\bf {n}}_i$ is additive white Gaussian noise vector. The $(\theta_i, d_i)$ pair represents a unique position. The directional and attenuation characteristics of the channel $\bf h$ can be modeled by:

\begin{equation}
 {\bf h}(\theta_i) = \begin{bmatrix}
 1 \\ \exp(j \frac{\pi}{2} \sin(\theta_i))
 \end{bmatrix}, \hspace{2mm} \alpha(d_i) = \phi \left(\frac{c}{4 \pi f}\right) d_i^{-p}
\end{equation}
 
For the recovery of missing measurements, first, the angle of arrival based on the received signal strength is estimated at each receiver and then a fusion of these estimates is performed. For angle of arrival estimation, authors in \cite{sun_simple_2010} apply the multiple signal classification (MUSIC) algorithm and obtain estimated of the pair $(\theta_i, d_i)$, that translate into a location estimate for the $i$-th receiver:

\begin{equation}
\begin{bmatrix}
\hat{x}_t^i \\ \hat{y}_t^i
\end{bmatrix} =
\begin{bmatrix}
x_i \\ y_i 
\end{bmatrix} +
\begin{bmatrix}
\hat{d}_i  \cos(\hat{\theta}_i) \\
\hat{d}_i  \sin(\hat{\theta}_i)
\end{bmatrix}
\end{equation}

Next, these estimated locations are transferred to a central network that combines these estimates. One way to combine these estimates can be through simple averaging. Another fusion method proposed in \cite{fusion} obtains the following over-conditioned system from the estimates:

\begin{equation}
\begin{bmatrix}
-x_1 \sin(\hat{\theta}_1) + y_1 \cos(\hat{\theta}_1) \\
\vdots\\
-x_N\sin(\hat{\theta}_N) + y_N \cos(\hat{\theta}_N) \\
\end{bmatrix} \approx 
\begin{bmatrix}
-\sin(\hat{\theta}_1)  & \cos(\hat{\theta}_1) \\
\vdots &\vdots\\
-\sin(\hat{\theta}_N)  & \cos(\hat{\theta}_N) 
\end{bmatrix}
\begin{bmatrix}
\hat{x}_t \\ \hat{y}_t
\end{bmatrix}
\end{equation}

Solving this system of equations through least squares solutions yields the transmitter location, which can then be combined with known transmit power and a suitable propagation model to estimate signal strengths at unknown locations.

Authors in \cite{sun_simple_2010} use AOA based method for interference source localization to interpolate REMs. Authors in \cite{sun_simple_2010} compare the AOA based method with simple averaging method (where averaging of sensor estimates by all sensors is done) and SNR based method in Section \ref{snr-based-sec}, where sensor results are weighted by each sensor's SNRs. The AOA method outperforms the other two methods at low SINRs.

\subsubsection{Signal to noise ratio (SNR) based method}
\label{snr-based-sec}
The initial steps of this method are similar to AOA based method in which the estimation step  
at each receiver enables the estimation of the angle of arrival and the received signal power. However, in the later step, combination of the location estimates is done through SNR-aided fusion.
The basic idea of this approach is the observation that receivers far away from the  transmitter yield worse location estimates. Hence the receiver results are weighted with their respective receiver's SNR, $\Gamma_i$ as follows \cite{sun_simple_2010,kakalou_radio_2019}: 

\begin{equation}
\begin{bmatrix}
\hat{x}_t \\ \hat{y}_t
\end{bmatrix}=\sum_{i=0}^{N} \frac{\Gamma_i}{\sum_{k=1}^{N}\Gamma_k} \begin{bmatrix}
\hat{x}_t^i\\ \hat{y}_t^i
\end{bmatrix}
\end{equation}

\noindent where the received SNR at the $i$-th receiver is:

\begin{equation}
\Gamma_i(d) = E \left[\frac{\alpha(d_i)P_t}{N_oB}\right]
\end{equation}
\noindent with $N_o$ being the  noise power density and $B$ being the bandwidth of the receiver.

The SNR based method has been used for interference source localization  for cognitive radio scenarios  to interpolate REMs in \cite{sun_simple_2010}. Authors in \cite{sun_simple_2010} conclude that AOA based method using tens of sensor nodes with two antennas in an area of 2500 m $\times$  2500 m can meet the location error requirement of FCC, which is $\pm$ 50 m and outperforms AOA based method at moderate to high SINR.

\subsubsection{Self-tuning method}
\label{sec:stm}
Another method utilizing propagation parameters but also taking the antenna pattern into account is the self-tuning method (STM) is proposed in \cite{pesko_indirect_2015}. In addition to  leveraging characteristics of the operating environment, it performs estimation of the transmitter location, antenna parameters, transmit power and  parameters of the propagation model such that the error between available measurements and predicted data is  minimized.

Using the scarce data collected, the STM first estimates transmitter parameters and calibrates the propagation model. This is then used to predict missing data, such as signal levels. 
Among these transmission parameters, the location of transmitter is calculated using localization algorithms based on parameters such as angle of arrival or timing advance,  time of arrival or time difference of arrival. Then, based on the transmitter location, distance from transmitter to receiver is calculated. This distance is then used in an appropriate propagation model. As an example, if the Okumura-Hata model is used, the received power at a particular location can be represented as:
\begin{equation}
\hspace{-16mm}P_r = P_t - A_o-A_1\log_{10}(d)- A_2 \log_{10}(H_{e}) -\nonumber \\
\hspace{3mm}A_3\log_{10}(d) \log_{10}(H) +3.2 (\log_{10}(11.75H_m))^2 -\nonumber\\
\hspace{7mm} 44.49 \log_{10}(f)+ 4.78 (\log_{10}(f))^2 - L_{d} - L_{c}+G
\end{equation}

\noindent where $P_t$ is the transmit power, $d$ is the transmitter-receiver distance, $f$ is the operating frequency, $L_d$ represents the diffraction loss, $L_c$ is  the loss through terrain clutter,  $H$ is the height of transmitter and $A_o, A_1,A_2,A_3$ are the constant coefficients. $G$ represents the antenna gain and can be represented as \cite{pesko_indirect_2015}:
\begin{equation}
G= G_{\max} - F_\theta + F_\theta \abs{\cos^{p_1} \left(\frac{\theta_{azi}-\theta_u}{2}\right)}- F_\phi + F_\phi \abs{\cos^{p_2} \left(\frac{\phi_{tilt}-\phi_{u}}{2}\right)}
\end{equation}

\noindent where $\phi_{tilt}$ is the tilt angle of the antenna, $\phi_{u}$ is the vertical angle from the reference axis (for tilt) to the user. $\theta_{azi}$ is the angle of  orientation of the antenna with respect to horizontal reference axis i.e., positive x-axis,  $\theta_{u}$ is the angular distance of the user from the horizontal reference axis. $G_{max}$ represents the maximum antenna gain and $F_\theta$ and $F_\phi$ are the front to back ratios in both directions, whereas the antenna form is approximated with the cosine functions to the power of $p_1$ and $p_2$

We suggest that another option for a more practical directional antenna  model defined by 3GPP and utilized in \cite{analy3} can be as follows:

	\begin{equation}
	G=\lambda_{\phi}\left(G_{\max}-\min\left(12\left({{\phi_{u}-\phi_{tilt}}\over{B_{\phi}}}\right)^{2},A_{\max}\right)\right)+\nonumber\\ \hspace{10mm}\lambda_{\theta}\left(G_{\max}-\min\left(12{\left({{\theta_{u}-\theta_{azi}}\over{B_{\theta}}}\right)^{2},A_{\max}}\right)\right)
	\label{gain_db}
	\end{equation}

\noindent The additional antenna parameters in this model are the half power vertical and horizontal beamwidths, $B_\phi$ and $B_\theta$ respectively and the side and back lobe attenuation, $A_{\max}$.

Having defined a suitable propagation and antenna model, the optimal antenna, transmitter and propagation environment parameters can then be obtained by minimizing  the mean squared error between the measured and estimated signal strengths.  Authors in \cite{pesko_indirect_2015} solved this optimization problem in a non-least squared sense, using prior knowledge of the  bounds for the parameters to be optimized. 

 After solving the optimization problem by a suitable algorithm, the optimized parameters are applied in the calculation of signal levels at unknown location to augment the existing data. 
 
Note that $L_d$ and $L_c$ require knowledge of the propagation environment, such as access to clutter database of a mobile operator or knowledge of the digital elevation model \cite{pesko_indirect_2015}. 
 Also, antenna parameters knowledge through antenna datasheets or antenna diagrams is required in this method.

STM has been applied for constructing the radio frequency layer of REM in \cite{pesko_indirect_2015}. When 1000 measurements are used, STM method obtains the lowest RMSE of 5, followed by Kriging with RMSE of 17.5, while IDW attains the highest RMSE of 22.5 \cite{pesko_indirect_2015}.

\subsection{Lessons Learned}
The methods discussed in this section can be used in cases where some additional contextual information is known. Based on the network geometry, triangular method can be used in the case when transmitter locations are known, and arc method can be used in cases where transmitter locations are not known. When the propagation environment parameters are known, along with the transmit power and receivers’ SNR, the SNR-based method can be used. However, if SNR is not known, but antenna characteristics are known, the STM method can be a potential candidate solution. There are also methods such as AOA based method, RSS,  RSSD based method that do not require antenna or SNR information, but instead make use of mathematical equations/models after estimating or using prior knowledge of the transmit power and location. However, since these methods are mostly based on analytical models, they inherit some assumptions.  
\section{Machine learning methods}
\label{chap:ml}
{
Several machine learning techniques such as generative adversarial networks (GANs), autoencoders, transfer learning and few-shot learning techniques can be leveraged to address the training data scarcity challenge in radio access networks. In certain RAN use-cases involving higher dimensional datasets, these neural network based techniques can be trained with much less training data (or with higher performance for the same amount of data) due to their efficient learning ability for higher-dimensional datasets as compared to previously mentioned interpolation and contextual information based methods \cite{chahrour2022comparing}.
Examples of scarce data and use cases in RAN where ML techniques have shown superior performance than other techniques, include CDR data for traffic map prediction \cite{hughes_generative_2019,zhang_zipnet-gan_2017}, MDT data for outage detection \cite{zhang_generative_2020}, cell trace data for performance analysis \cite{aoki_few_shot_2020}, RSS data for pathloss prediction \cite{wang2021indoor, masood2022interpretable}, RF data for radio map generation \cite{han_radio_2020, han_two-phase_2020} and configuration data for performance prediction \cite{parera_transfer_2019, zappone_wireless_2019, chuai_collaborative_2019, li_tact_2014, parera_transfer_2020, parera2021anticipating, moradi2019performance, larsson2021source}. 
}

\subsection{Generative adversarial networks}
\label{sec:gans}
Generative adversarial networks (GANs) success in image processing has been well established \cite{gan_images1}-\nocite{gan_images2}\nocite{gan_images3}\nocite{gan_images4}\cite{gan_images5}. Although this concept has widely been used in image processing, it can also be used in wireless communications.  
In wireless communications context, the works that utilize GANs are limited to \cite{hughes_generative_2019, zhang_generative_2020,han_radio_2020, han_two-phase_2020, erpek_deep_2018, shi_generative_2019, nabati_using_2020, zhang_zipnet-gan_2017}.
While GANs have been widely used for image data, its application to tabular data remains relatively limited. The works that use GANs on tabular data in a non-cellular network data context include \cite{engelmann_conditional_2020, tanaka_data_2019, gao_generative_2020,  camino_improving_2019,xu_modeling_2019,xu_synthesizing_2018}. However, similar concepts can be applied to wireless data domain too.

\begin{figure}[!t]
	\centering
	\includegraphics[width=0.99\columnwidth]{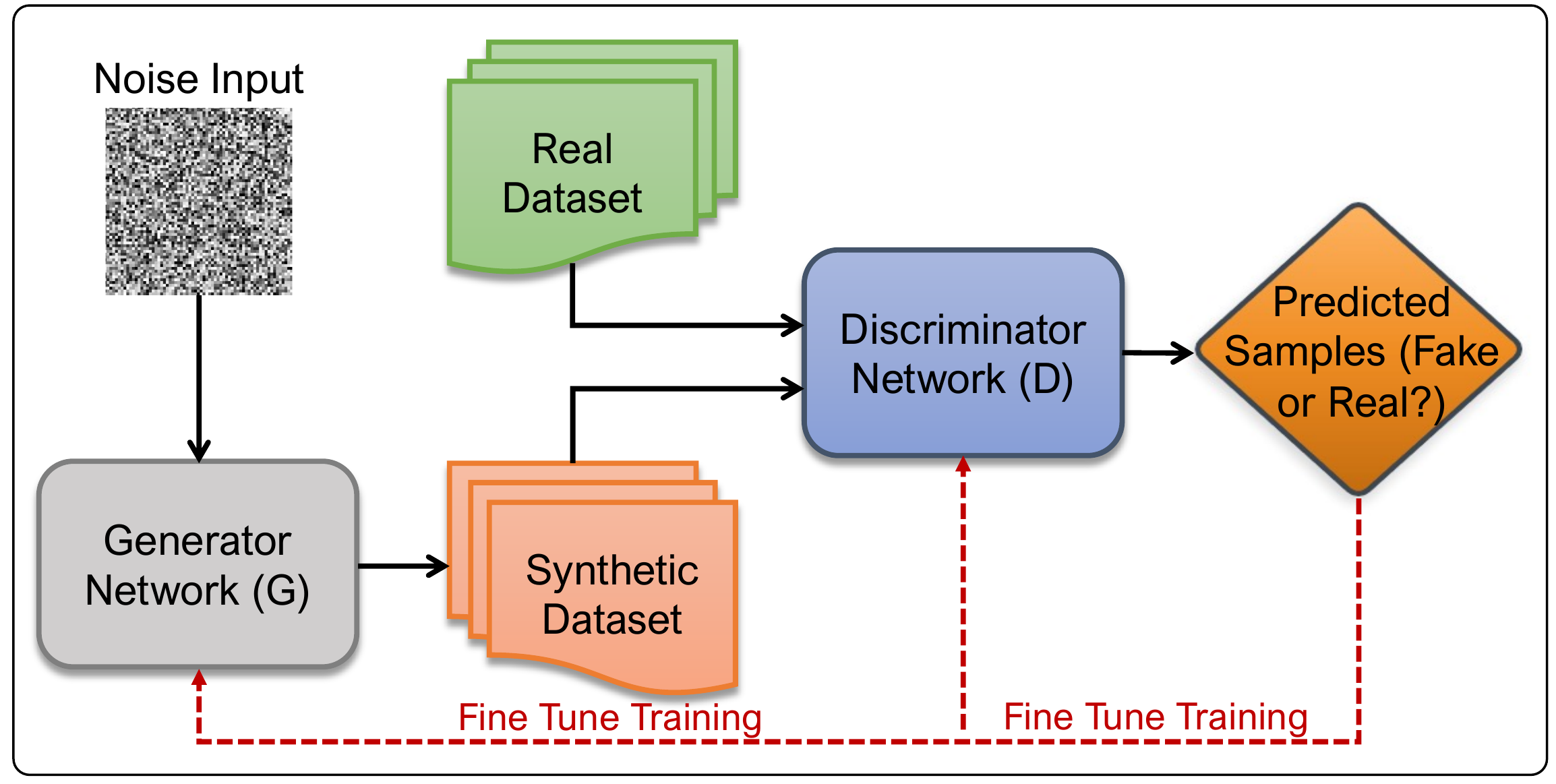}
	\caption{Conventional GAN architecture.
	}
	\label{GAN_basic_idea}
 \vspace{-0.1in}
\end{figure}

The basic idea of GAN illustrated in Fig. \ref{GAN_basic_idea} is to generate large amount of synthetic data building on small amounts of real data which will not be distinguishable from real data.
The intuition behind GANs is to exploit the potential of deep neural networks (DNNs) to both model nonlinear complex relationships (the generator) as well as classify complex signals (the discriminator). 
In GAN, a two-player minimax game is set between the discriminator DNN and generator DNN as follows:
\begin{equation}
\hspace{-1cm}\min_{G} \max_{D} V (D,G) =
\mathbb{E}_{x \sim{p_{\rm data}(x)}} \left[\log D(x)\right] +  \nonumber\\
\hspace{3.5cm}\mathbb{E}_{z \sim{p_{z}(z)}} \left[\log (1- D(G(z)))\right]
\end{equation}

where $V (D,G)$ is the value function over which training happens, the latent variable $z$ is randomly drawn from prior distribution $p_{z}(z)$, $x$ is sampled from 
$p_{\rm data}(x)$,
generator $G$ is a mapping from the latent variable $z$ to data space and the discriminator is a scalar function of data space that outputs probability that input was genuine. 
Other types of loss functions for the discriminator and generator for different types of GANs are described in \cite{lucic2017gans}.
In each training epoch, the generator iterates its weights to produce synthetic data trying to fool the discriminator DNN. The discriminator DNN on the other hand, tries to discriminate between real data and generated data. In theory, when Nash equilibrium is reached between the generator DNN and discriminator DNN, the pair of DNNs will provide us a generator that can exactly duplicate or reproduce the distribution of the real data so that the discriminator would be unable to identify whether a sample is synthetic i.e., whether it is generated by the generator DNN or it is from the real data. At this point, the synthetic data generated by the generator DNN are indistinguishable from the real data, and are thus as realistic as possible.

\begin{figure}
	\centering
	\captionsetup[subfigure]{justification=centering}
	\begin{subfigure}{0.24\textwidth}
		\includegraphics[width=\textwidth]{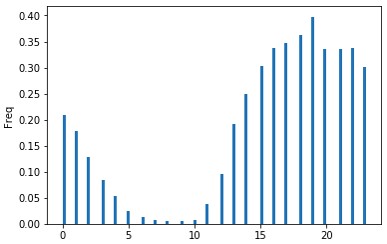}
		\caption{\scriptsize{{Real Calls Start Hours }}}
		\label{Real_SH}
	\end{subfigure}
	\begin{subfigure}{0.24\textwidth}
		\includegraphics[width=\textwidth]{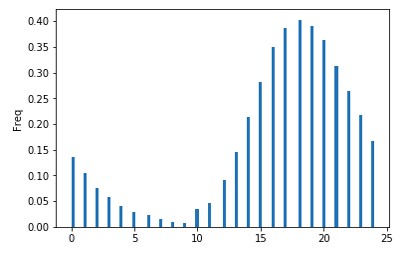}
		\caption{\scriptsize{Synthetic Calls Start Hours}}
		\label{Synthetic_SH}
	\end{subfigure}
	\centering
	\begin{subfigure}{0.24\textwidth}
		\includegraphics[width=\textwidth]{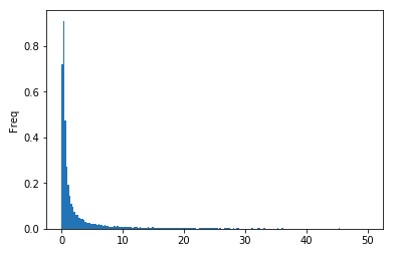}
		\caption{\scriptsize{Real Calls Durations (min)}}
		\label{Real_CD}
	\end{subfigure}
	\begin{subfigure}{0.24\textwidth}
		\includegraphics[width=\textwidth]{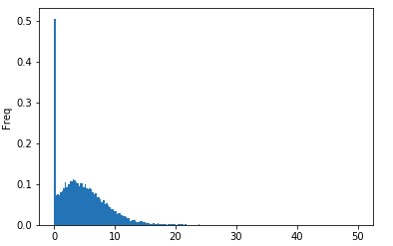}
		\caption{\scriptsize{Synthetic Calls Durations (min)}}
		\label{Synthetic_CD}
	\end{subfigure}
	\caption{ Leveraging GAN for enriching the  scarce training data \cite{hughes_generative_2019}.}
	\label{GAN}
  \vspace{-0.2in}
\end{figure}

To assess the efficacy of GAN-based approach outlined above, as a preliminary study recently published in  \cite{hughes_generative_2019}, GAN was leveraged to generate  synthetic call data records (CDRs) data and thus increased training dataset size by enriching the real scarce CDR from \cite{milan} with realistic synthetic data. 
CDRs data are selected as preliminary  case study because  CDR data can be used  by a large number of SON solutions such as in \cite{leveraging_intelligence_CDR}, \cite{spatiotemporal_hasan}. Real network traces  with  call durations  and call start time stamps, provided by one of the leading mobile operators in USA, were used in this study to train the GAN. The discriminator was trained beginning with 20,000 data points (from a record of several hundred thousand). Once  the discriminator could reliably differentiate between the real data taken from the record and randomly generated CDR data with  two features i.e., call duration and start time, the generator was trained. After the generator was generating data that the discriminator perceived  to be real, we used the trained generator to produce another 20,000 CDR data samples. Figs. \ref{Real_SH} and \ref{Real_CD}  and  represent the distribution of the real data used to train the discriminator. Figs. \ref{Synthetic_SH} and \ref{Synthetic_CD}  show the  distribution of the 20,000 synthetic data points produced by the trained generator. These preliminary results show the high similarity between real and synthetic data produced by the proposed GAN based approach.

Other GAN-based approaches in cellular networks context include the
use of GANs to address the imbalance data issue in cell outage detection \cite{zhang_generative_2020}
Authors in \cite{zhang_generative_2020} use an LTE simulator to get RSRP and RSRQ data and combine GAN with AdaBoost to improve classification performance of imbalanced data for cell outage detection in self-organizing cellular networks.

A radio environment maps estimation algorithm leveraging a GAN-based pixel regression framework (PRF) for underlay cognitive radio networks using incomplete training data is proposed in \cite{han_radio_2020, han_two-phase_2020}.
In these works, the authors first transform the radio environment maps estimation task into a pixel regression through color mapping. Then they
extract helpful information from the incomplete training data, design a feature enhancing module for the PRF algorithm, which intelligently learns and emphasizes the important features from the training images.  Finally, they train the PRF to reconstruct the radio environment maps in the target area. 
Three indicators are used to test the proposed algorithm: the visual display of the radio environment maps, the estimated power spectrum of primary users, and the average REMs estimating error against different numbers of secondary users.
Results are bench-marked with IDW and Kriging with the exponential semi-variogram estimation.

Moreover, authors in \cite{zhang_zipnet-gan_2017}, while drawing inspiration from image processing
design a deep-learning architecture tailored to mobile networking, which combines Zipper Network (ZipNet) and GAN models.
Using the open-source Telecom Italia's  dataset \cite{milan}, they infer fine-grained mobile traffic patterns to monitor city-wide mobile traffic via the GAN.

However, GANs suffer from many challenges, such as vanishing gradients, oscillations, modal collapse and the design of suitable evaluation metrics to evaluate their performance.

\subsection{Autoencoders}
\label{sec:autoencoders}

\begin{figure}[!t]
	\centering
	\includegraphics[width=0.99\columnwidth]{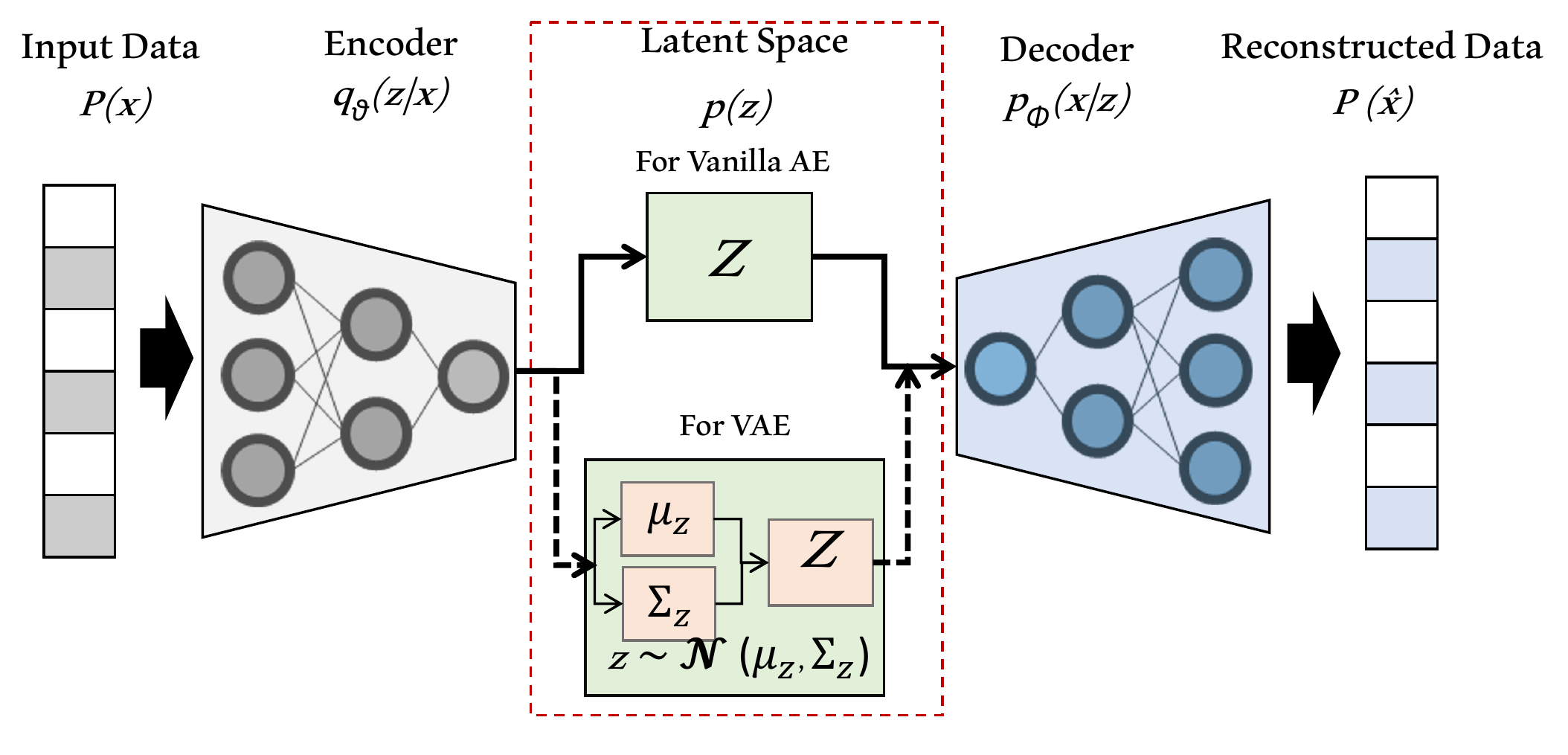}
	\caption{A conventional vanilla autoencoder and variational autoencoder (whose internal representation is described by a probability distribution). 
	}
	\label{fig:ae}
  \vspace{-0.2in}
\end{figure}

{
Unlike GANs, which come in the class of implicit density methods (where the prior distribution of latent features is not known), some generative methods fall under explicit density method, meaning that the distribution of latent features is explicitly defined. One such method is a type of autoencoder, namely variational autoencoder (VAE). Autoencoders are basically neural networks consisting of an encoder and decoder, that encodes the input to a point in latent space, by performing non-linear dimensionality reduction (Fig. \ref{fig:ae}). The parameters of the encoder and decoder are optimized during training to minimize the reconstruction loss, as the autoencoder learns to reproduce its input. On the other hand, as illustrated in Fig. \ref{fig:ae},  variational autoencoders encode the input into a multi-variate distribution (e.g., normal distribution) in latent space, described by the mean and variance vector where the length of the vector is equal to the number of dimensions in latent space.  This probabilistic representation ensures that the latent space has good properties, such as variability of the latent space, thus making the model more robust and achieve better performance as compared to traditional autoencoders.}

VAEs are used in literature \cite{yuan2020anomaly, rajendran_saife_2018} to handle labeled training data scarcity problem for anomaly detection use-cases in RAN. In these use-cases labeled training data is severely imbalanced and traditional machine learning techniques are not able to distinguish the anomalies from the majority data. As a case study, authors in \cite{yuan2020anomaly} used VAEs for anomaly detection and root cause analysis (RCA) in radio access networks. The data used in the analysis includes key performance indicators (KPIs) that indicate network quality of service (QoS), as well as key quality indicators (KQIs) that indicate user quality of experience (QoE). The anomaly detection module focuses on detecting the performance degradation in RAN, whereas the RCA module tries to find the root cause of detected anomalies. The proposed anomaly detection module takes time series of KPIs/KQIs from a cell as an input to the VAE model and outputs their respective anomaly score based on the  
error from the VAE model when it tries to reproduce its input.
The RCA module is trained by auto-labelling the anomaly labels in a semi-supervised fashion using KQI rules, e.g., high PRB usage, over coverage, weak coverage, etc. The proposed AI-based approach is then tested in a live O-RAN compliant network for closed loop automation, resulting in 25\% increase in downlink rate and 8\% increase in RRC connection establishment with zero human cost in the entire process.

Similarly, adversarial autoencoders are a type of variational autoencoders which combines the architecture of autoencoders with GANs adversarial loss for regularization. Authors in \cite{rajendran_saife_2018} demonstrated the effectiveness of adversarial autoencoders for detecting anomalous behavior in wireless spectrum using power spectral density data. Manual spectrum management, especially in emerging dense and heterogeneous networks is inefficient and can only detect limited anomalies. Therefore, automated spectrum monitoring solutions are becoming more crucial than ever before. Along with anomaly detection, the proposed model in \cite{rajendran_saife_2018} shows a semi-supervised wireless band classification accuracy close to 100\% on datasets using only 20\% of the labeled samples.

\subsection{Transfer learning}
\label{sec:transfer_learning}

For data streams where latent features are too little to allow the use of GANs, matrix completion or other interpolation techniques identified above, the transfer-learning paradigm  \cite{pan_survey_2010, zappone_wireless_2019} can be leveraged. 

\begin{figure}[!t]
	\centering
	\includegraphics[width=0.99\columnwidth]{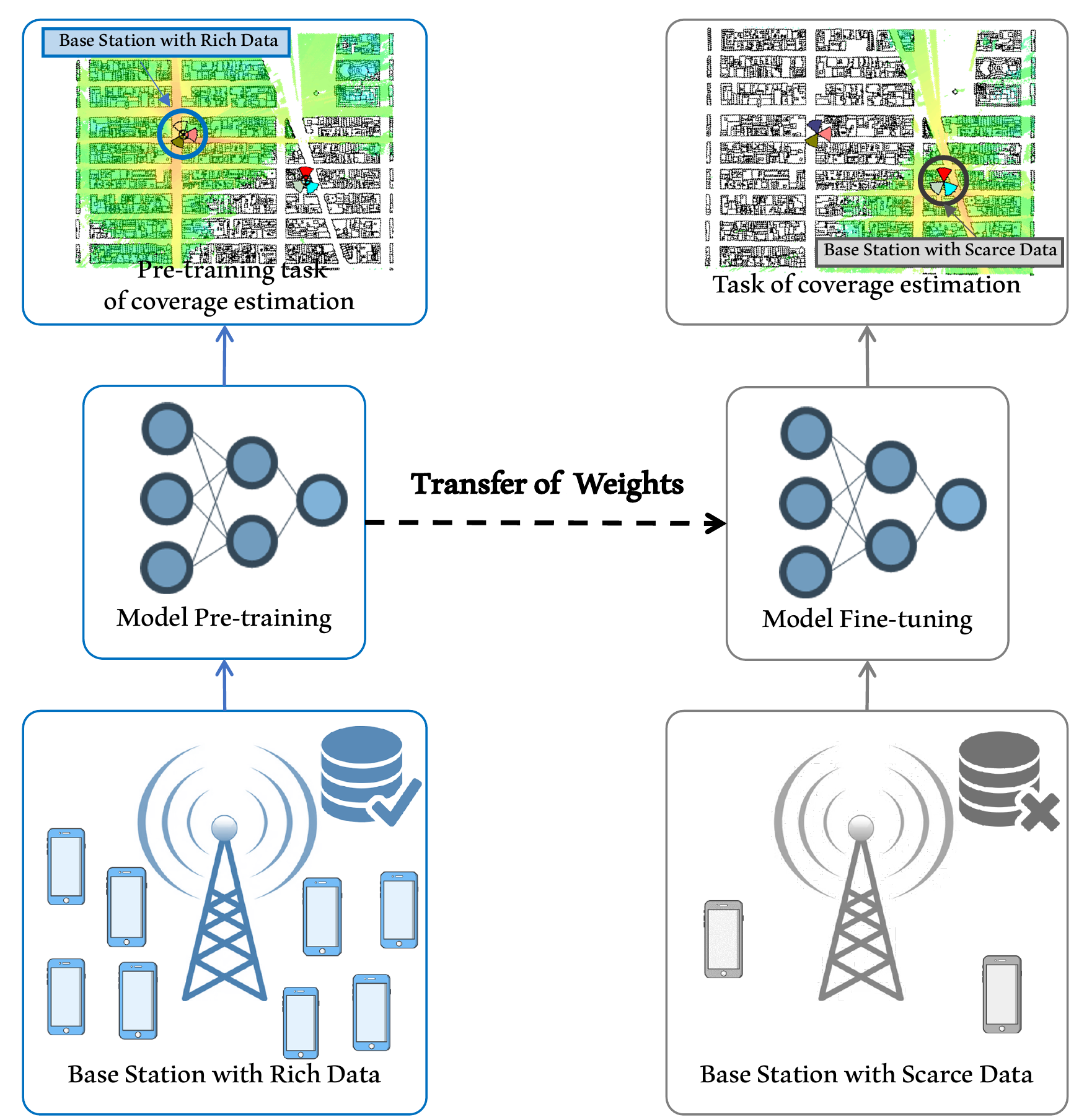}
	\caption{An example of transfer learning in deep neural networks for coverage estimation. The feature network (source model) is pre-trained on a large dataset (from BS with rich data). The target model is  created by transferring the knowledge learned from the source model, e.g., weights of the model. This model is then trained/fine-tuned using the scarce dataset (from BS with scarce data). 
	}
	\label{fig:tl}
 \vspace{-0.2in}
\end{figure}

{ Transfer learning aims to help improve the learning of the target environment (target model) by transferring the knowledge learned from another similar environment (source model). One way of achieving that is by model fine-tuning, where a larger source dataset is used to pre-train a neural-network based model (source model) and fine-tuned using the target scarce dataset (as illustrated in Fig. \ref{fig:tl}).} 

In cellular network context, similarities among cells can be leveraged for determining when to use transfer learning. To quantify similarities among the cells, one approach is to use Wasserstein distance measure \cite{8260947}. Given two random variables $f_i$ and $f_j$ with marginal distributions $P(f_i)$ and $P(f_j)$ respectively, let \(\psi\) denote the set of all possible joint distributions that has marginals of $P(f_i)$ and $P(f_j)$. Then Wasserstein distance between them is defined as:

\begin{equation}
W(f_{i},f_{j})=\underset{P_{f_if_j}\in\psi}{\inf} \int{|f_i - f_j|P_{f_if_j}(f_i,f_j)d_{f_i}d_{f_j}}
\label{eq:wasserstein}
\end{equation}

\noindent The inf in \cref{eq:wasserstein} gives joint distribution with $f_i$ and $f_j$ having smallest distance while maintaining the marginals.

{
Several works have been carried out in the literature using transfer learning to address data scarcity problem for network performance prediction \cite{chuai_collaborative_2019, parera_transfer_2020, parera2021anticipating, zappone_wireless_2019, li_tact_2014, moradi2019performance, larsson2021source}.
As a case study, authors in \cite{chuai_collaborative_2019} proposed to use transfer learning for parameter configuration in cellular networks. In this work, contextual bandit algorithm is leveraged along with transfer learning to optimize parameter configurations for uplink power control and user scheduling using cell KPI/counter data. Cell state measurements e.g., the number of total users within the cell, the number of active users, the average channel quality indicator (CQI) of the cell, etc. are collected for each cell at each hour, and the goal is to minimize the ratio of users with experienced throughput less than 5Mbps for each cell. Live field tests in a real cellular network consisting of 1700+ cells show a significant performance improvement of 20\% by optimizing five parameters for two weeks, thereby demonstrating  the effectiveness of the proposed scheme. 

A transfer actor-critic learning framework for energy saving in cellular radio access networks is proposed in \cite{li_tact_2014}. This work utilizes the transferred learning expertise in historical periods or neighboring regions for predicting traffic load variations for BS ON/OFF switching.
The problem of predicting the signal strength in the downlink of a real LTE network, where the antennas can be tuned to operate with different antenna tilt configurations is addressed using transfer learning in \cite{parera_transfer_2020}. The authors show that augmenting the data from the source domain by adding data available from other tilts configurations of the same antenna improves the performance of the proposed transfer learning approaches.
Transfer learning for channel quality and active UEs prediction is proposed in \cite{parera2021anticipating}, using KPI/counter data from a commercial LTE network. The results show how transfer learning can be carried out across pairs of cells working at different frequencies, or at the same frequency in different locations and how to pick suitable candidate cells across the city for the transfer learning task. 
Transfer learning is also particularly helpful in tasks that require frequent model retraining, due to changes in the operational environment during execution, such as learning performance model for a cloud service \cite{moradi2019performance}. Authors in \cite{moradi2019performance} show that the number of new measurements required to compute a new model are reduced by an order of magnitude in most cases using transfer learning, as compared to training the new model from scratch, when evaluated on traces collected from a testbed running video-on-demand service, under various load conditions. 
However, finding suitable transfer candidates, or where to transfer is another challenging research question that remains unfocused in most of the works discussed earlier. Authors in \cite{larsson2021source} argue that the choice of source domain can either yield `transfer gain', or further decrease the performance of the baseline model, commonly known as `negative transfer', and proposed two source selection approaches to mitigate this issue. A key result from their study is that source selection should encourage diversity of the data in source domain rather than similarity between source and target cell, especially in scenarios with few samples in target domain as the similarity between the underlying distributions of both domains cannot be reliably measured.
}

 \begin{table*}
 \renewcommand*{\arraystretch}{1.3}
    \begin{adjustbox}{width=\textwidth,center}
    \newcommand*\rot[1]{\hbox to1em{\hss\rotatebox[origin=br]{-25}{#1}}}
        \begin{threeparttable}
        \tablefootnote{Free license means free for academia use and in some cases under a signed contract by the lab head.}
        \centering
        \caption{Comparison of different simulators for solving data scarcity
problem.}
        \begin{tabular}{p{4.5cm}p{0.7cm}p{0.7cm}p{0.7cm}p{0.7cm}p{0.7cm}p{0.7cm}p{0.7cm}p{0.7cm}p{0.7cm}p{0.7cm}p{0.7cm}p{0.7cm}p{0.7cm}p{0.7cm}} 
        \hline
        {\bf Feature} &  \multicolumn{14}{c}{\bf Simulator}\\
                \hline
        &\rot{GTEC \cite{GTEC}}& \rot{OpenAirInterface \cite{sim8}}&\rot{5G-K \cite{sim9}} & \rot{X.Wang et al. \cite{XWang}} & \rot{V.V.Diaz et al. \cite{Diaz}} &\rot{ns-3 \cite{sim5}}&\rot{OMNeT$++$ \cite{sim6} }&\rot{NYUSIM \cite{NYUSIM}}&\rot{MATLAB/SIMULINK \cite{sim4}} & \rot{C-RAN \cite{C-RAN}} & \rot{OPNET \cite{sim7}} & \rot{Vienna 5G \cite{sim10}} & \rot{Atoll \cite{forsk}} & \rot{SyntheticNET \cite{simx}} \\
        \hline
         Scheduling support & \ding{52} & \ding{52}& \ding{52} &\ding{52}& & \ding{52} & \ding{52}& \ding{52} &  & \ding{52}& \ding{52} &\ding{52} & \ding{52} & \ding{52} \\
         \hline
         mm-Wave support    &  & &  & &\ding{52} &\ding{52}  &\ding{52} &\ding{52} &  & \ding{52}&  &\ding{52} & \ding{52} & \ding{52} \\
         \hline
         Adaptive numerology    &  & &  && &  & & & \ding{52} & &  &\ding{52} & \ding{52} & \ding{52} \\
          \hline
         QCI support &  & &  && &  & & & \ding{52} & &  & & \ding{52} & \ding{52} \\
         \hline
          Parallelized offline traces and time-independent KPIs pre-generation for reduced online computational cost &  & &  & & &  & & &  & &  & & \ding{52} & \ding{52} \\
          \hline
          Realistic antenna patterns modeling &  & &  & & &  & & &  & & \ding{52} & & \ding{52} & \ding{52} \\
           \hline
          Signaling overhead modelling &  & &  & & &  & & &  & &  & &  & \ding{52}\\
            \hline
        Realistic mobility modeling &  & &  & & &  & & &  & &  & &  & \ding{52}\\
            \hline
        AI based pathloss modeling &  & &  & & &  & & &  & &  & &  & \ding{52}\\
            \hline
    500+ COPs modeling &  & &  & & &  & & &  & & & &  & \ding{52}\\
         \hline
     Realistic HO management&  & &  & & &  & & &  & &  & &  & \ding{52}\\
     \hline
     Realistic mobility pattern&  & &  & & &  & & &  & &  & &  & \ding{52}\\
     \hline
     Python based to enable data processing and easy incorporation of ML libraries &  & &  & & &  & & &  & &  & & & \ding{52}\\
     \hline
     Free license* & \ding{52} & \ding{52} &\ding{52} &\ding{52} &\ding{52} &\ding{52} &\ding{52} &\ding{52} & &\ding{52} &\ding{52}&\ding{52}& &\ding{52}\\
     \hline
        \end{tabular}
          \label{simulators_table}
  \end{threeparttable}
  \end{adjustbox}
\end{table*}
\subsection{Few-shot learning}
\label{sec:few shot}
Few-shot learning (FSL) is another branch of machine learning that addresses the performance degradation problem of deep learning algorithms when the training dataset size is small. Using prior knowledge, FSL can master new tasks from a limited number of examples \cite{wang_generalizing_2020}. This type of learning is primarily motivated from the ability of humans to learn from only a few examples. Therefore, FSL can eliminate expensive data collection efforts and help in building suitable models for rare cases of limited supervised data \cite{wang_generalizing_2020}.

FSL can be used for classification, regression and even reinforcement learning tasks using only few labeled, input-output and state-action examples respectively. However, the most common application scenario for FSL is \textit{``N-way-K-shot classification"}, where a classifier is built for distinguishing between N classes, each having only K examples per class. When only one example with supervision is available, it is referred to as One-Shot Learning and when no example is available, it is called Zero-Shot Learning.

\begin{figure}[!t]
	\centering
	\includegraphics[width=0.99\columnwidth]{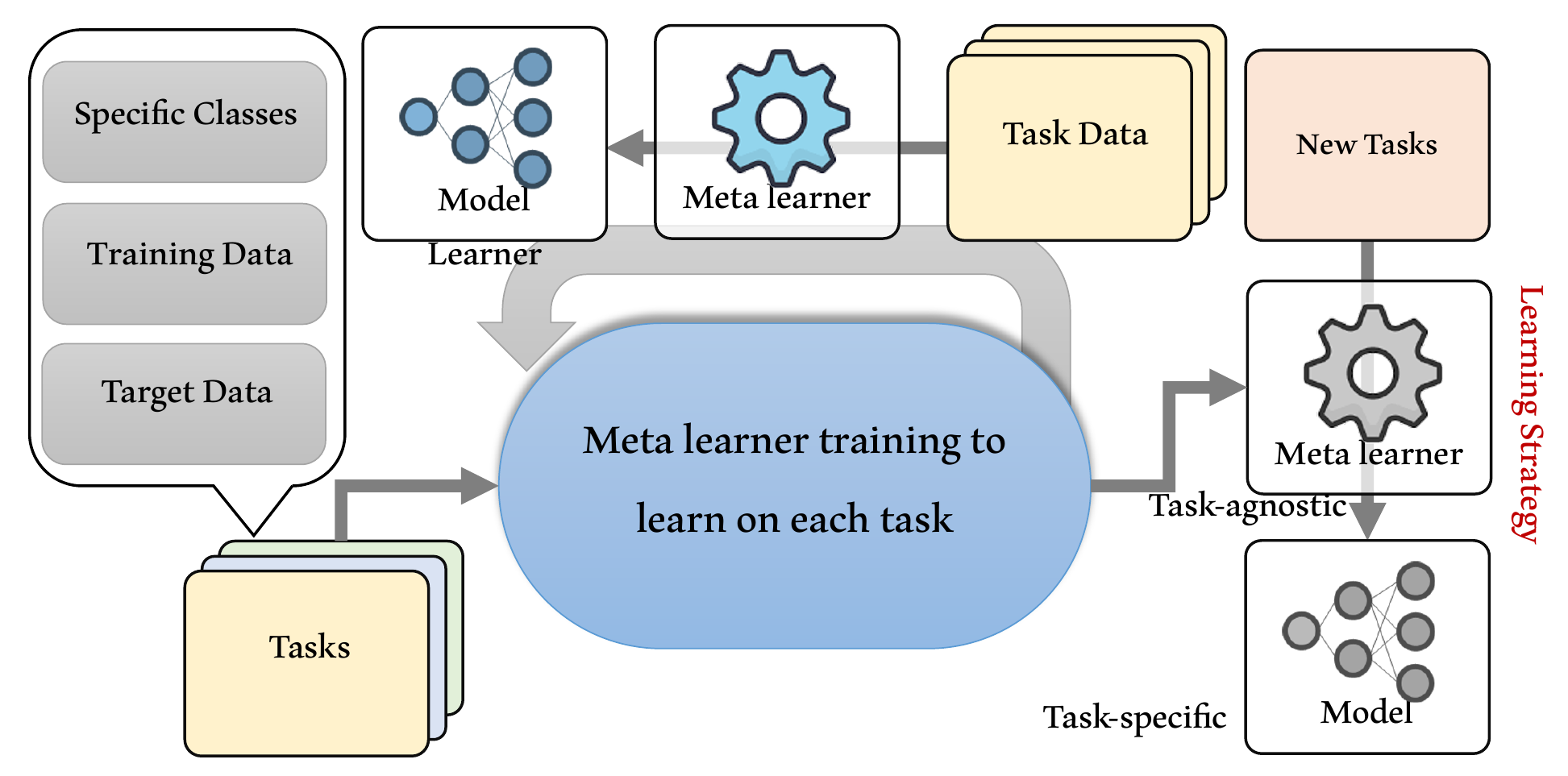}
	\caption{Meta learning-based methods can learn a learning strategy from a family of tasks by developing a task-agnostic learner. The learning strategy (or task-agnostic knowledge) can then be used to improve the learning of a new few-shot learning task from that task family \cite{shtok_2019_fsl}.}
	\label{fig:meta_learning}
  \vspace{-0.2in}
\end{figure}

FSL is a very active area of research these days and the methods being proposed in the literature for solving the few-shot problem can be broadly classified in two different branches: 1) Meta learning, and 2) Metric learning. { The key idea in Meta learning-based methods (as shown in Fig. \ref{fig:meta_learning}) is to distill the experience of multiple learning episodes from a distribution of related tasks. This learning to learn strategy can improve the future learning performance on new few-shot learning tasks, thus developing a task-agnostic learner with improved data and compute efficiency \cite{shtok_2019_fsl, 9428530}.} Examples of methods include Model Agnostic Meta Learning \cite{finn2017model}, Task-Agnostic Meta Learning \cite{jamal2019task} and Meta-transfer Learning \cite{sun2019meta}. These methods are good at out-of-distribution tasks and can handle varying and large shots well, but their model and architecture are intertwined and their optimization process is challenging \cite{finn2019meta}. On the other hand, Metric learning-based methods learn to compare query set (test set) with support set (few-shot training set) by learning transferable representations in semantic embedding space using a distance loss function (learn to compare). Examples include Siamese Neural Networks \cite{koch2015siamese}, Matching Networks \cite{vinyals2016matching}, Prototypical Networks \cite{snell2017prototypical}, Relation Networks \cite{sung2018learning} and Graph Neural Networks \cite{garcia2017few}. As compared to meta learning-based methods, these are relatively simple, entirely feedforward, computationally fast and easy to optimize, but harder to generalize to varying shots and to scale to very large shots \cite{finn2019meta}.

{
A few works have been carried out using few-shot learning to address training data scarcity issue in cellular networks. Authors in \cite{aoki_few_shot_2020} use prototypical networks, a few-shot learning-based algorithm for performance metrics analysis in LTE networks. They used eNodeB trace data from live network and classified individual eNodeBs into different performance classes based on their KPIs. Their results show an improved performance as compared to baseline DNN, 1-D CNN and 2-D CNN.

Authors in \cite{wang2021indoor} show that meta learning can be used in mmWave smart factory environment to frame the indoor pathloss prediction task as a meta-task comprising of multiple tasks. Authors show that meta-learning based CNN-based model trained on a meta-task of multiple beams can outperform conventional training methods. Specifically, the prediction RMSE of the proposed meta-learning based CNN model show a gain of 70\% in terms of prediction accuracy as compared to floating-intercept (FI) model, and a gain of 55\% as compared to conventional CNN based model.

Authors in \cite{shen_lorm_2020} use self-imitation via transfer learning to achieve few-shot learning for the resource management (network power minimization) problem in Cloud Radio Access Networks (C-RAN). Their simulation results show that few-shot learning is able to achieve similar performance even with scarce and unlabeled training data, as compared to a model that is trained without few-shot learning even with labeled data. These results show the power of few-shot learning in scenarios where labeled training data is not available or is very scarcely available.
}

\subsection{Lessons Learned}
{ 
Based on the covered literature, we can see that all the above-mentioned ML/DL techniques work well for modeling high-dimensional datasets, however, they differ in terms of their applicability. For instance, both GANs and autoencoders can only generate quality synthetic data if their training data contains some latent information about their environment. In situations where the scarce dataset is not representative of the environment from which it is collected,  few-shot learning and transfer learning techniques can be used. Both, however, rely on the availability of auxiliary datasets to help them learn the target environment from unrepresentative training data. Transfer learning requires data from a similar domain or task to gain insights and then transfer that knowledge to the task at hand. few-shot learning requires data from a lot of different (but not necessarily similar) task/domain to learn the unfamiliar environment. These takeaways are also illustrated in Fig. \ref{fig:flowchart} for the benefit of the reader.
}

\begin{figure}[!t]
\vspace{-0.1in}
	\centering
	\captionsetup[subfigure]{justification=centering}
	\begin{subfigure}{0.24\textwidth}
		\includegraphics[width=\textwidth,height=2.6cm]{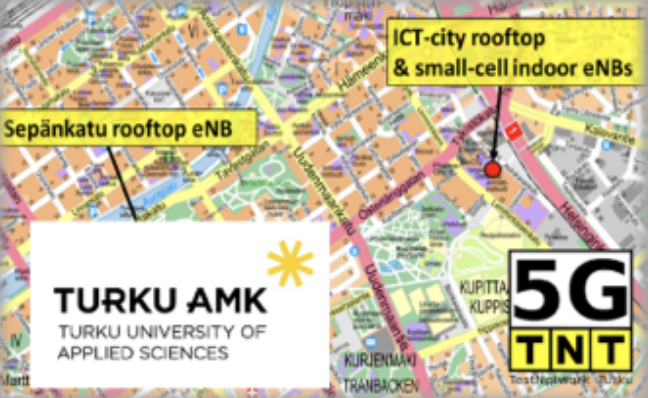}
		\caption{\scriptsize{{Turku Testbed}}}
		\label{Turku}
	\end{subfigure}
	\begin{subfigure}{0.24\textwidth}
		\includegraphics[width=\textwidth, height=2.6cm]{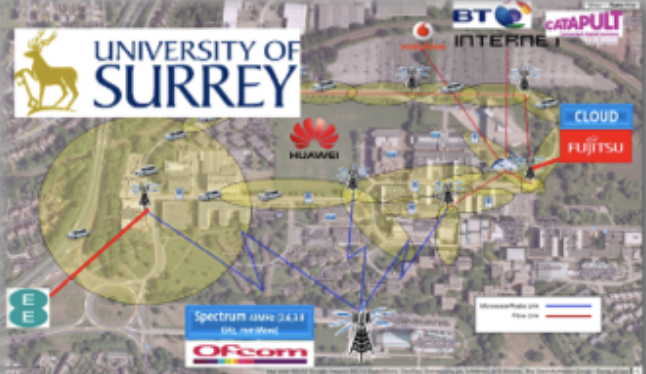}
		\caption{\scriptsize{Surrey Testbed}}
		\label{Surrey}
	\end{subfigure}
	\begin{subfigure}{0.24\textwidth}
		\includegraphics[width=\textwidth,height=2.6cm]{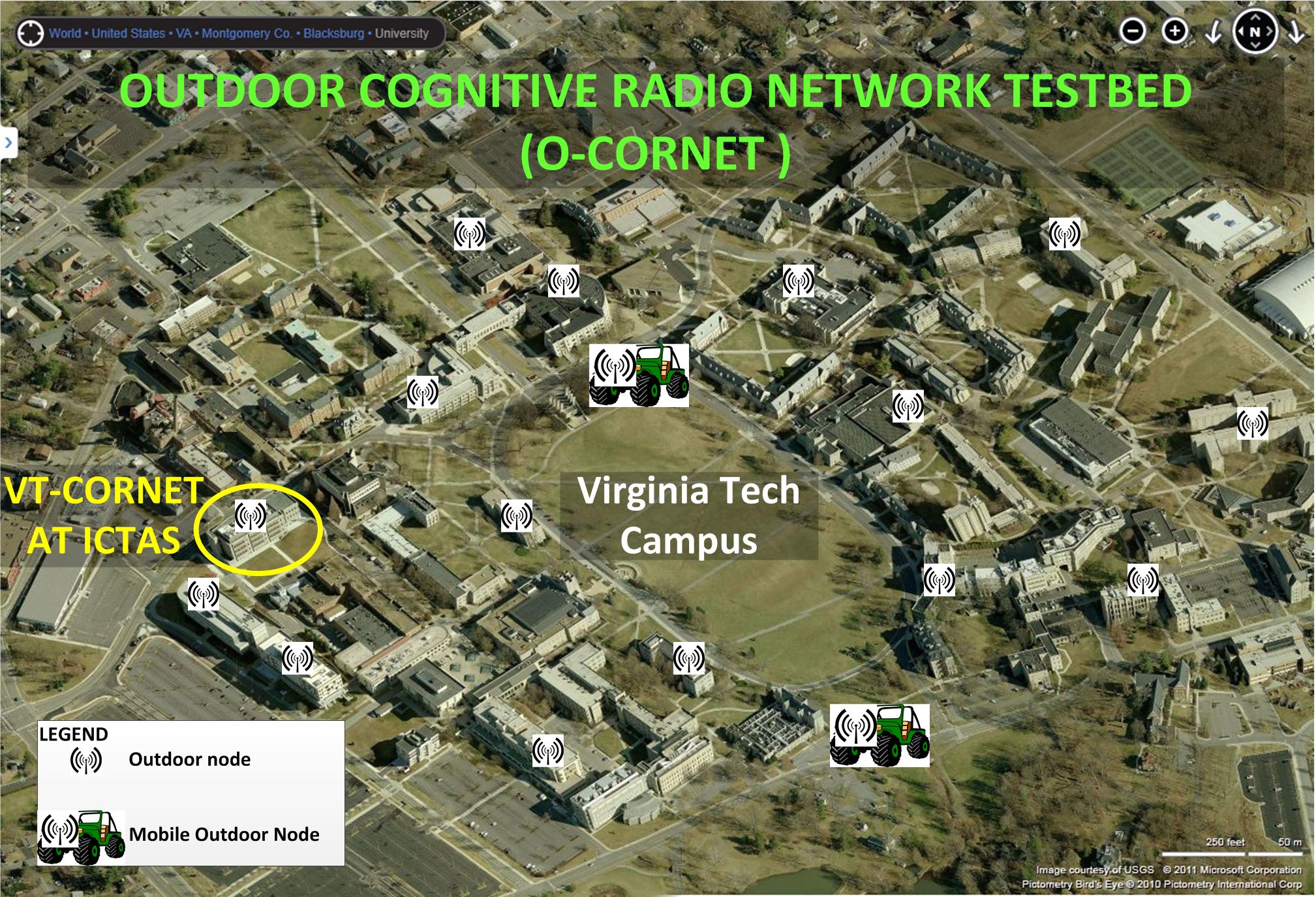}
		\caption{\scriptsize{CORNET Testbed}}
		\label{virginiatech}
	\end{subfigure}
	\begin{subfigure}{0.24\textwidth}
		\includegraphics[width=\textwidth,height=2.6cm]{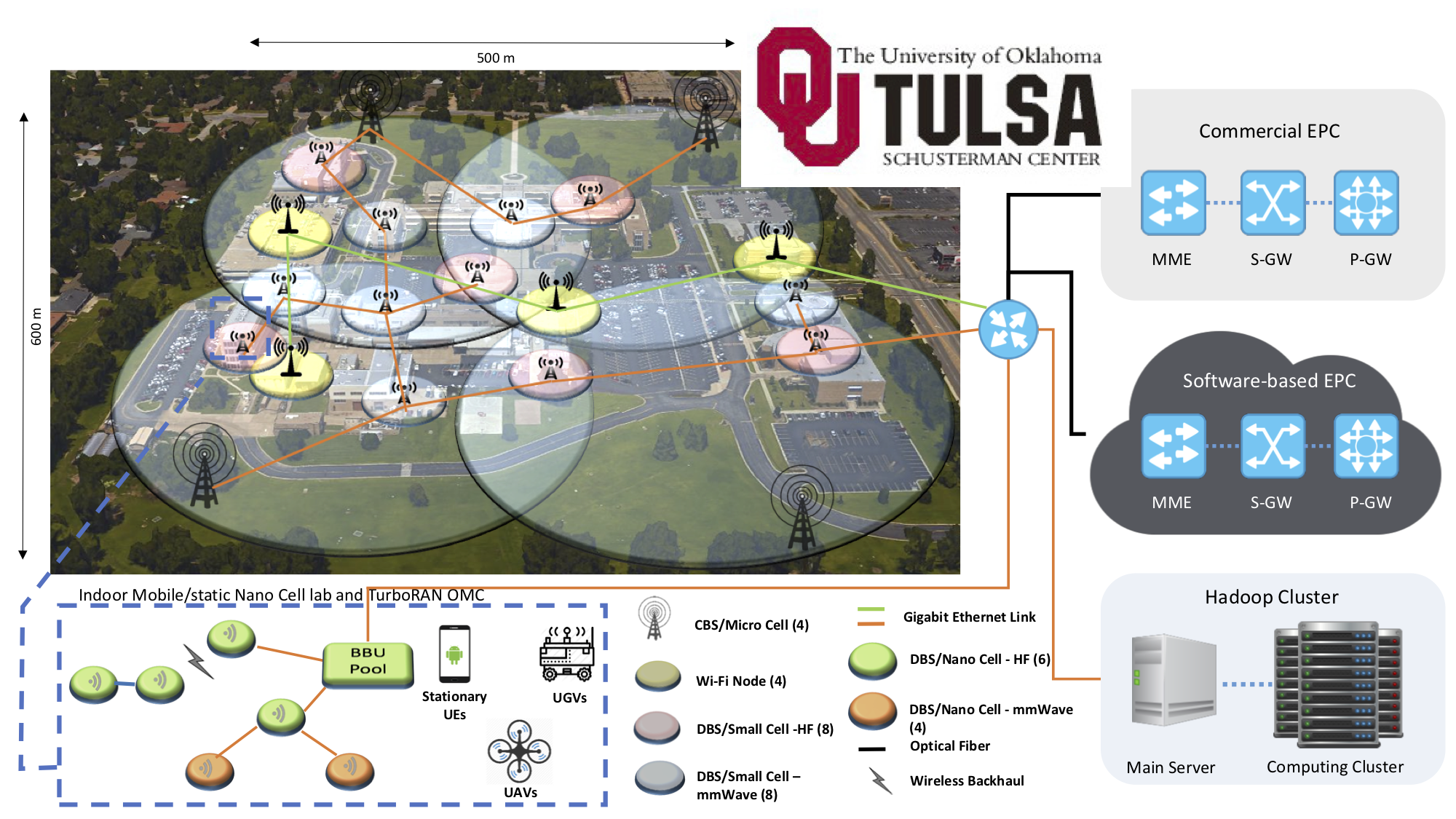}
		\caption{\scriptsize{OU-Tulsa Testbed}}
		\label{OU-Tulsa}
	\end{subfigure}
	\centering
	\begin{subfigure}{0.24\textwidth}
		\includegraphics[width=\textwidth,height=2.6cm]{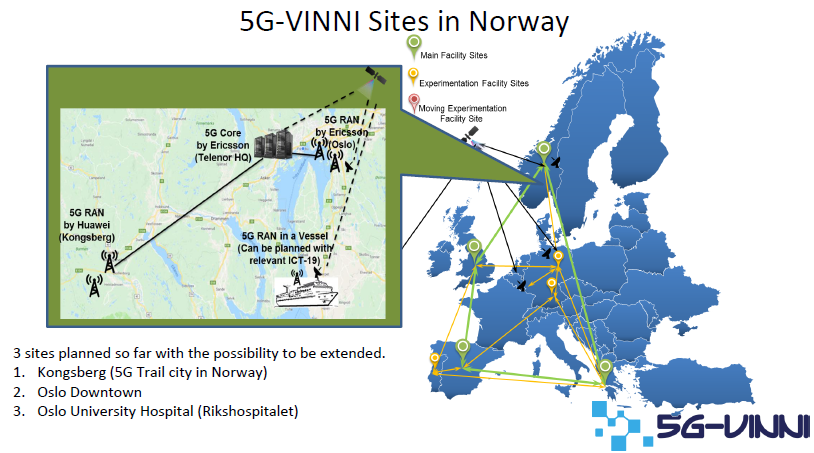}
		\caption{\scriptsize{5G-VINNI Norway}}
		\label{vinni_norway}
	\end{subfigure}
	\begin{subfigure}{0.24\textwidth}
		\includegraphics[width=\textwidth,height=2.6cm]{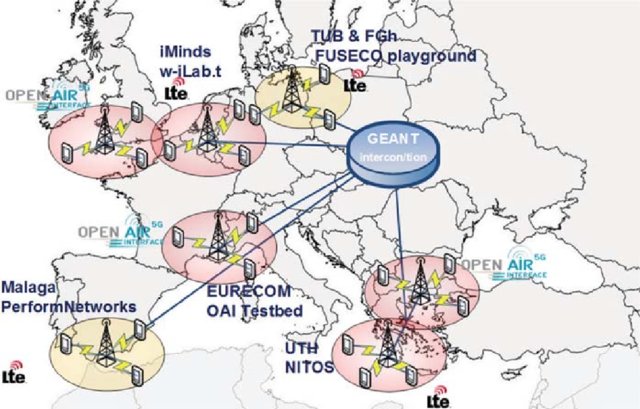}
		\caption{\scriptsize{FLEX Testbed}}
		\label{flex}
	\end{subfigure}
	\begin{subfigure}{0.24\textwidth}
		\includegraphics[width=\textwidth,height=2.6cm]{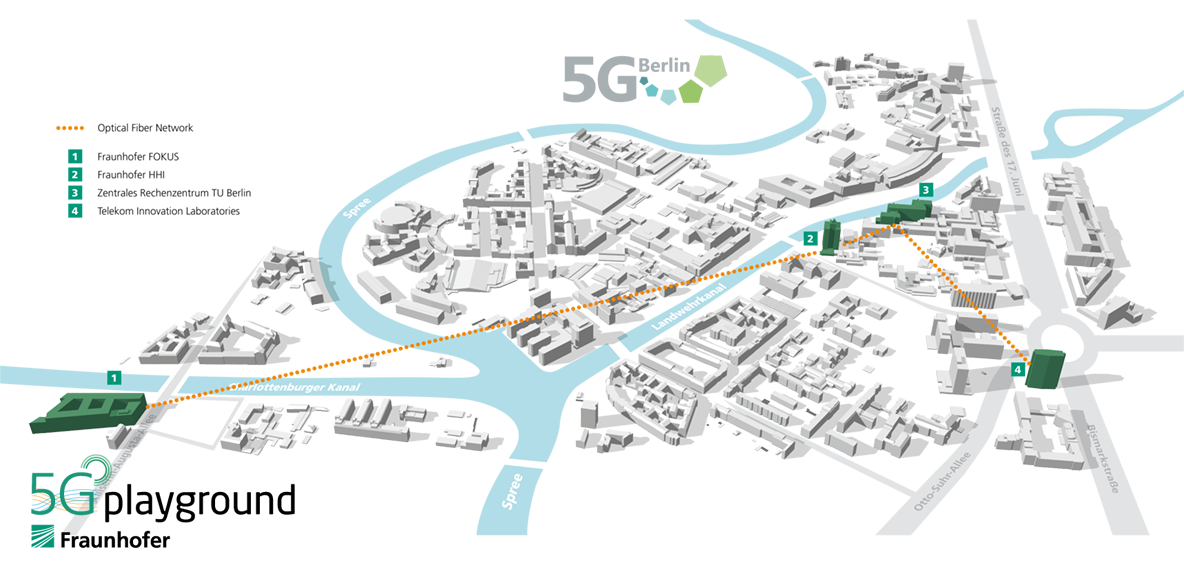}
		\caption{\scriptsize{5G Playground Testbed}}
		\label{playground}
	\end{subfigure}
	\begin{subfigure}{0.24\textwidth}
		\includegraphics[width=\textwidth,height=2.6cm]{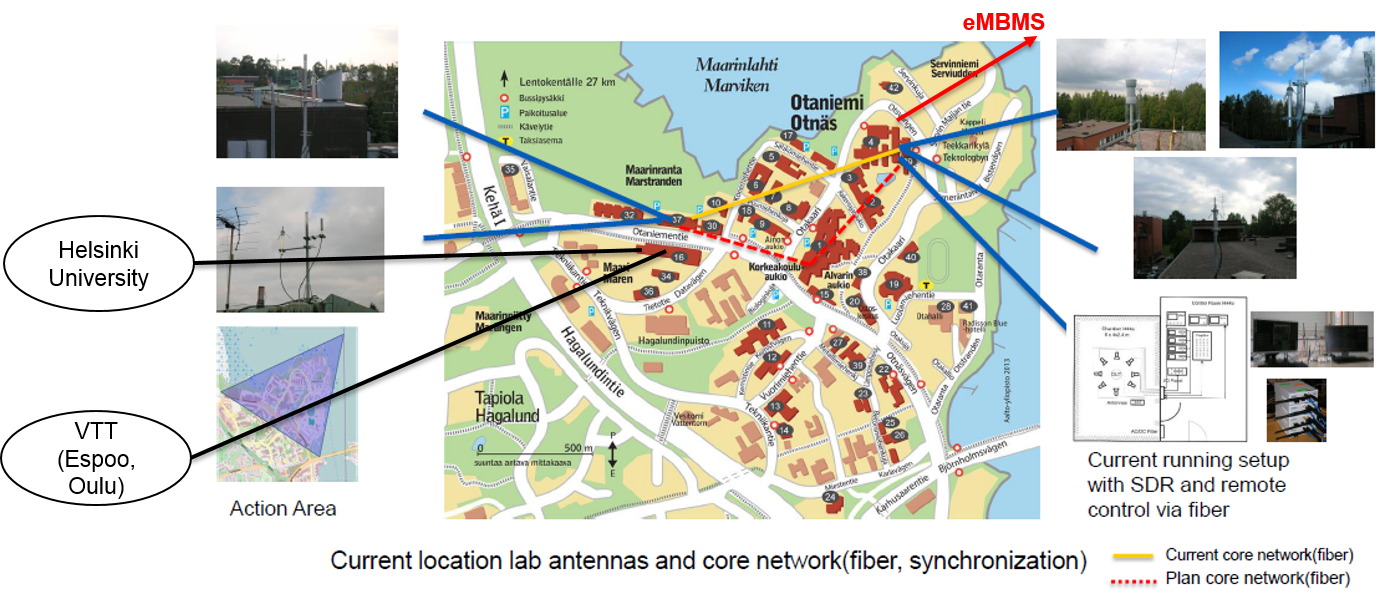}
		\caption{\scriptsize{5G Test Network, Espoo}}
		\label{finland}
	\end{subfigure}
	\begin{subfigure}{0.24\textwidth}
		\includegraphics[width=\textwidth,height=2.6cm]{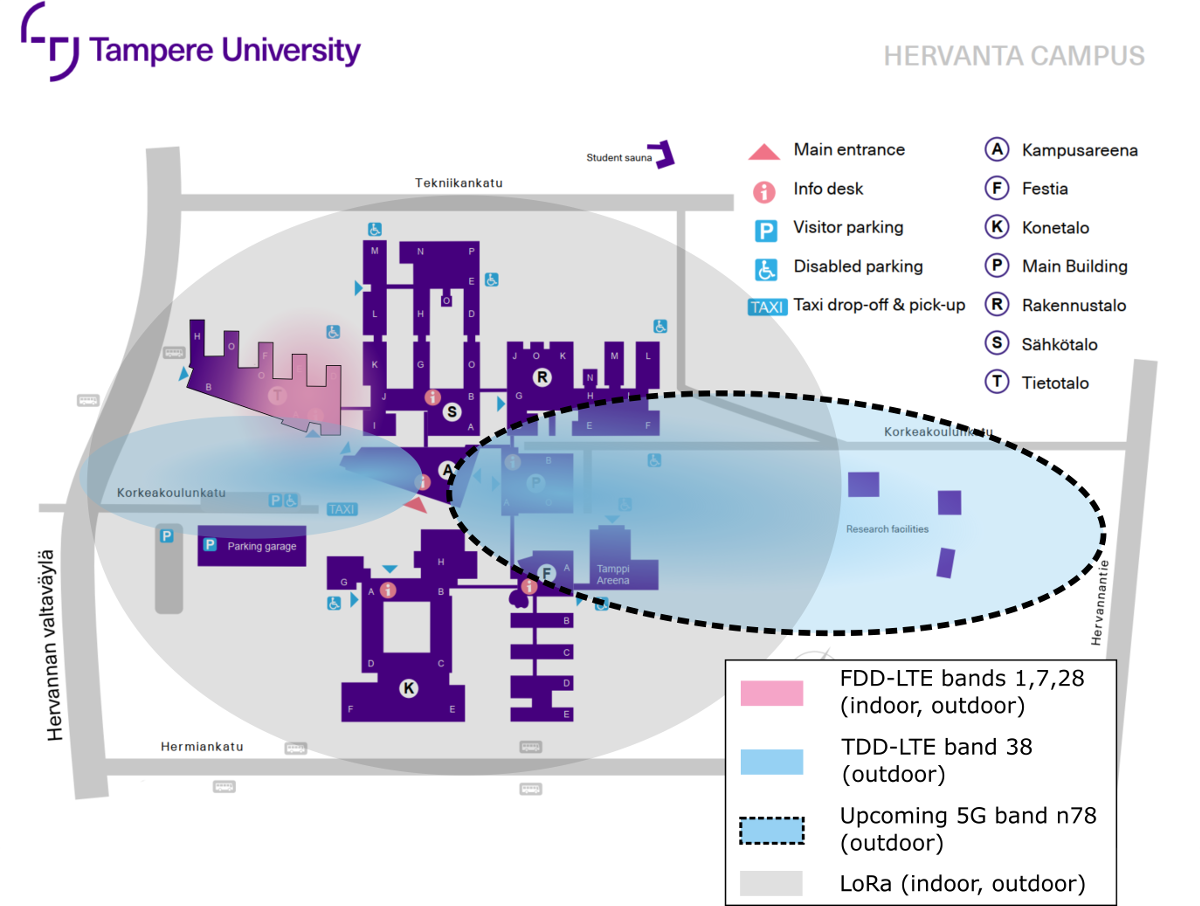}
		\caption{\scriptsize{5G Test Network, Tampere}}
		\label{tampere_finland}
	\end{subfigure}
	\begin{subfigure}{0.24\textwidth}
		\includegraphics[width=\textwidth,height=2.6cm]{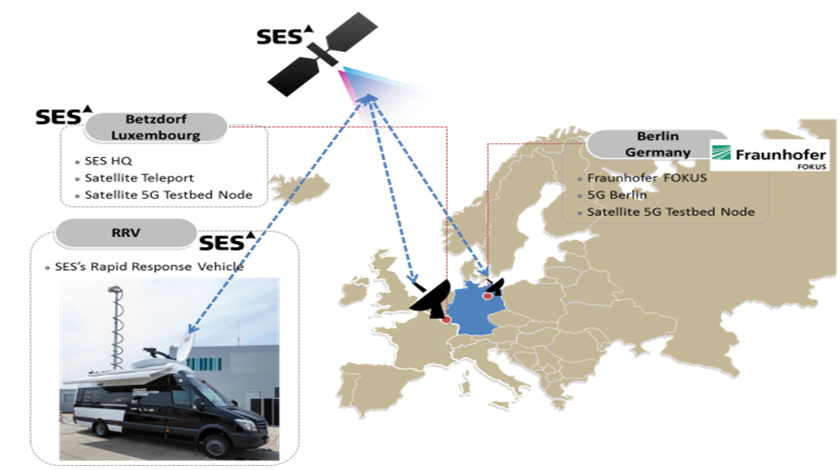}
		\caption{\scriptsize{5G-VINNI Berlin}}
		\label{move}
	\end{subfigure}
		\begin{subfigure}{0.24\textwidth}
		\includegraphics[width=\textwidth,height=2.6cm]{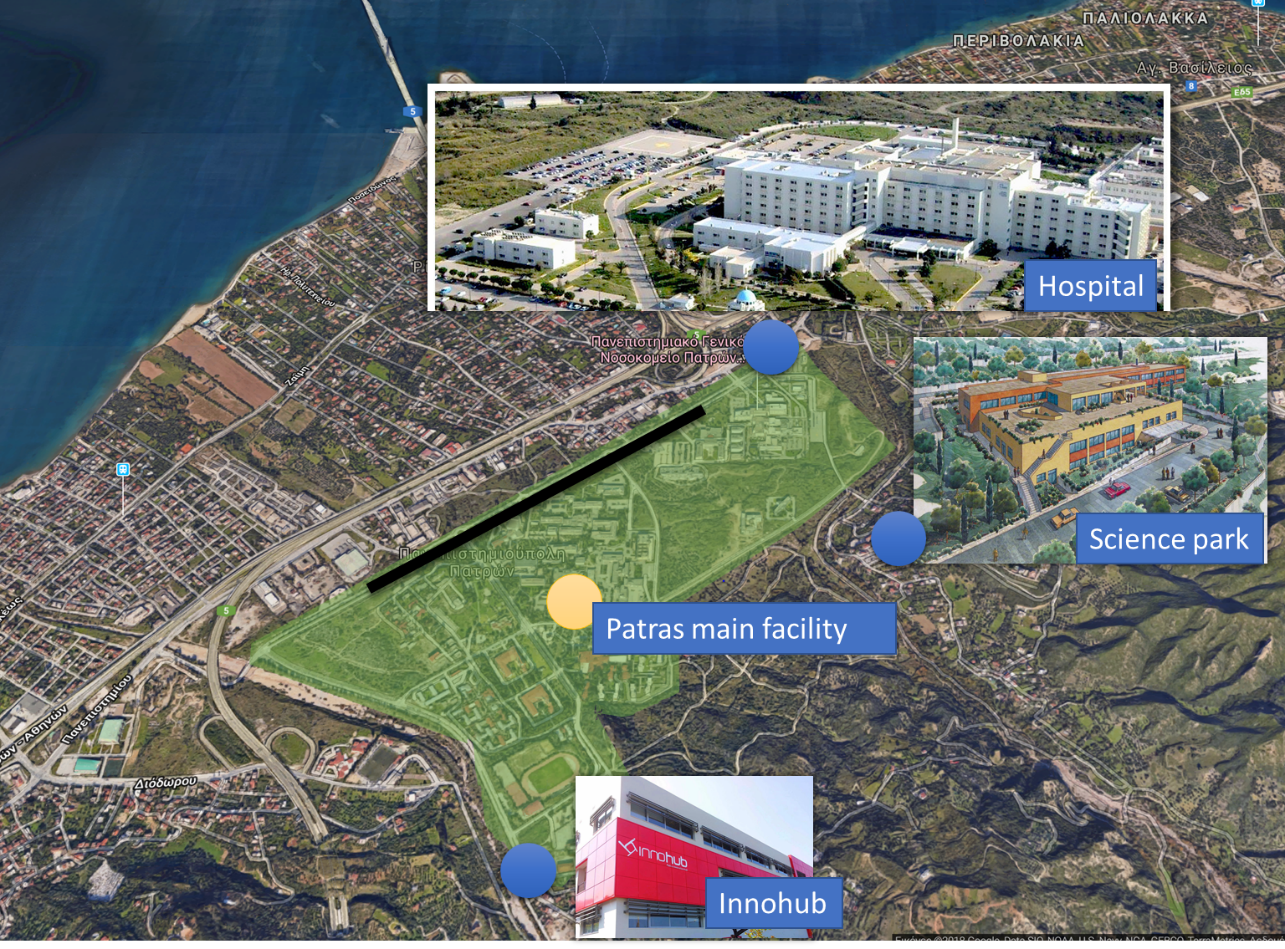}
		\caption{\scriptsize{5G-VINNI Greece }}
		\label{greece}
	\end{subfigure}
	\caption{Some current and emerging 5G testbeds.}
 \vspace{-0.1in}
	\label{testbeds_figures}
\end{figure}

\begin{figure*}[t]
\vspace{-0.1in}
	\centering
	{\includegraphics[height=13cm, width=19cm]{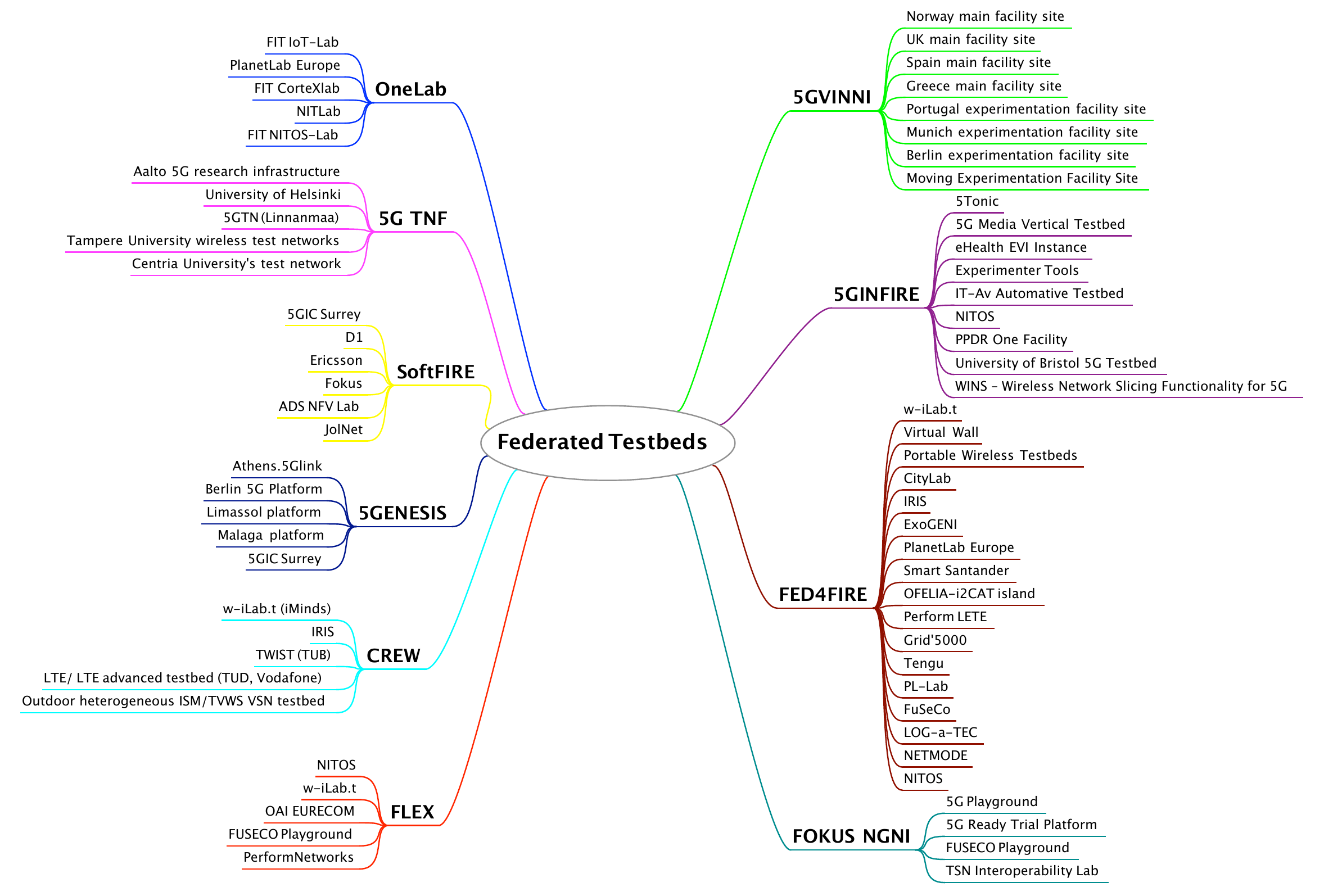}}
	\caption{Federated Testbeds.} \label{federated}
\end{figure*}

\section{Synthetic data generation}
\label{chap:synthetic}
The  techniques mentioned in previous sections are likely to work well when the scarce available data is somewhat representative of the whole data or exhibits some degree of correlation. In situations where the available data is scarce and non-representative, the methods presented in preceding sections are likely to perform poorly. Likewise, in other scenarios, the available data can be big, but still not representative. In these cases, the solution lies in either resorting to get real data or generate synthetic data. In this section, we will present ways to generate synthetic data  through simulators.

\subsection{Simulators}
\label{sec:simulators}
System level simulators are widely used in both industry and academia due to limitations of analytical models and field experiments. 
Apart from the limitation of mounting Base Stations (BSs) on predefined locations, the support of antenna height, tilt, transmission power etc. for individual BSs is absent in the analytical model. Furthermore, stochastic geometry-based models are unable to capture the network dynamics which include mobility management and transmission latency. On the other hand, field trials exhibit the most realistic modeling of network performance, evaluation and tuning. However, this approach is impractical owing to the cost and time effort required to conduct field trials on a large scale, and with the high probability of significant network performance impairment of live mobile network during the trial phase.

A list of existing simulators along with a comparison of their features is presented in Table \ref{simulators_table}. For more details on these simulators, the reader is referred to two existing surveys on simulators; \cite{simulator_survey1} that compares 4G and 5G simulators, and \cite{simulator_survey2} that gives  the summary of the most significant 5G simulators.

As observed from Table \ref{simulators_table}, none of the simulators is based on comprehensive 5G standard incorporating all aspects outlined in the standard. To tackle this problem, SyntheticNET simulator built on Python platform was developed by the AI4Networks Research Center at the University of Oklahoma \cite{simx}. The SyntheticNET simulator is modular, flexible, microscopic and versatile, built-in compliance with the 3GPP Release 15. This simulator supports features like adaptive numerology, actual hand over (HO) criteria and futuristic database-aided edge computing to name a few. Instead of an objected-oriented programming (OOP) based structure like existing simulators, SyntheticNET simulator supports commonly used database files (like SQL, Microsoft Access, Microsoft Excel). Site info, user info, configuration parameters, antenna pattern etc. can be directly imported to the simulator. As a result, the simulation environment is more realistic and closer to actual deployment scenarios.  For further details of this simulator, the reader is referred to \cite{simx}.

Python based platform and the flexibility of different input and output data formats in SyntheticNET simulator can assist in solving the data scarcity challenge by generating ample amounts of synthetic data to enrich the available scarce real data, which can then be used to implement different Self Organizing Networks (SON) related features or  AI based network solutions \cite{bson}. Mobile operators can use it for planning, evaluating or even optimization of beyond 5G networks. Research community can also benefit from it by implementing the new ideas on data generated from this 3GPP-based realistic 5G network simulator.

\begin{figure}[!t]
	\centering
	{\includegraphics[width=
\columnwidth]{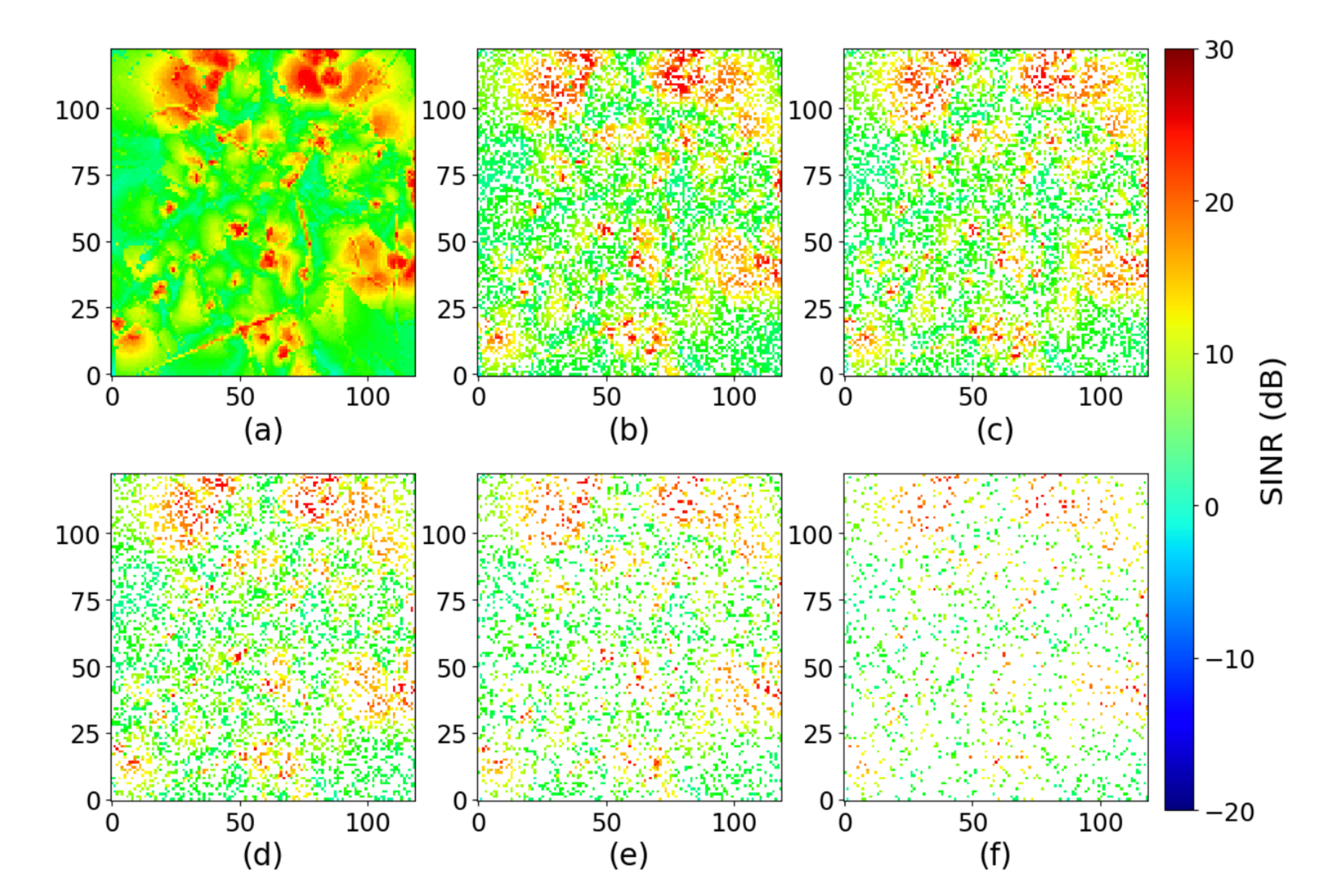}}
	\caption{ Network coverage maps with various user densities (a) Full coverage
map (203 UEs/cell) (b) 100 UEs/cell 
(c) 80 UEs/cell (d) 60 UEs/cell 
(e) 40 UEs/cell 
(f) 20 UEs/cell 
\cite{riaz2022deep}.} \label{fault_1}
\end{figure}

\begin{figure}[!t]
	\centering
	{\includegraphics[width=
\columnwidth]{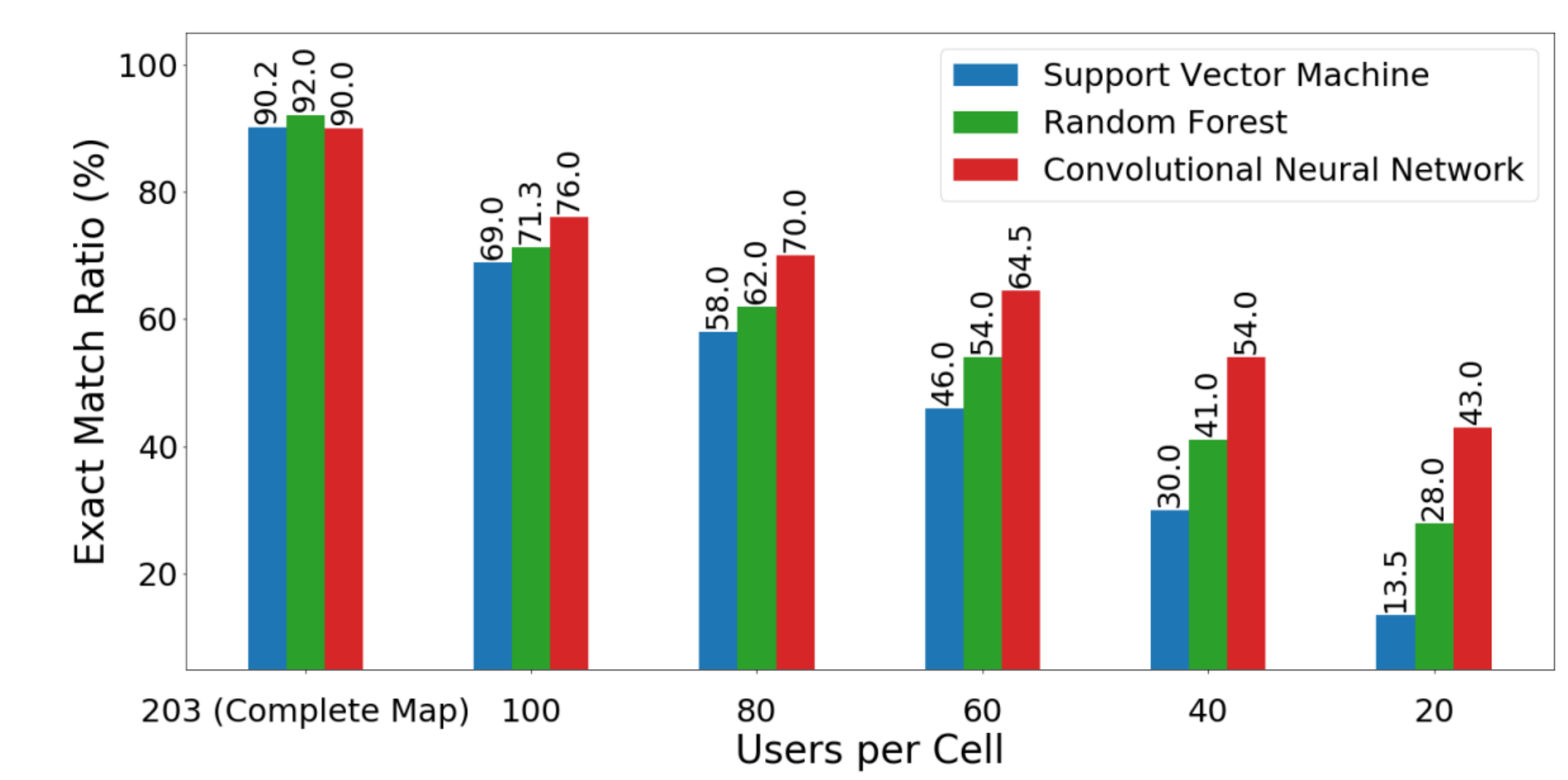}}
	\caption{ Performance comparison of ML models on scarce and complete coverage maps data 
\cite{riaz2022deep}.} \label{fault_2}
\end{figure}

Fault diagnosis using synthetic data from Atoll simulator is used in \cite{riaz2022deep}.  Authors in \cite{riaz2022deep} consider 4 types of faults characterized by cell outage, low transmit power, excessive antenna uptilt, and excessive antenna downtilt. The SINR maps obtained in these scenarios are scarce as shown in Fig. \ref{fault_1}. Authors in \cite{riaz2022deep} then analyse the performance of several ML-based algorithms for fault diagnosis in Fig. \ref{fault_2}, where the UE density on x-axis corresponds to the network depiction in Fig. \ref{fault_1}. As compared to complete coverage maps, a drastic drop in diagnosis accuracy is observed for the ML models on scarce data, where the exact match ratio (EMR) drops from 90.2\% to 69\% and from 92\% to 71.3\% respectively, as the density of users drops from 203 to 100 users/cell. Performance continues to deteriorate as the number of users decreases per cell.

\begin{figure}[!t]
	\centering
	{\includegraphics[width=
\columnwidth]{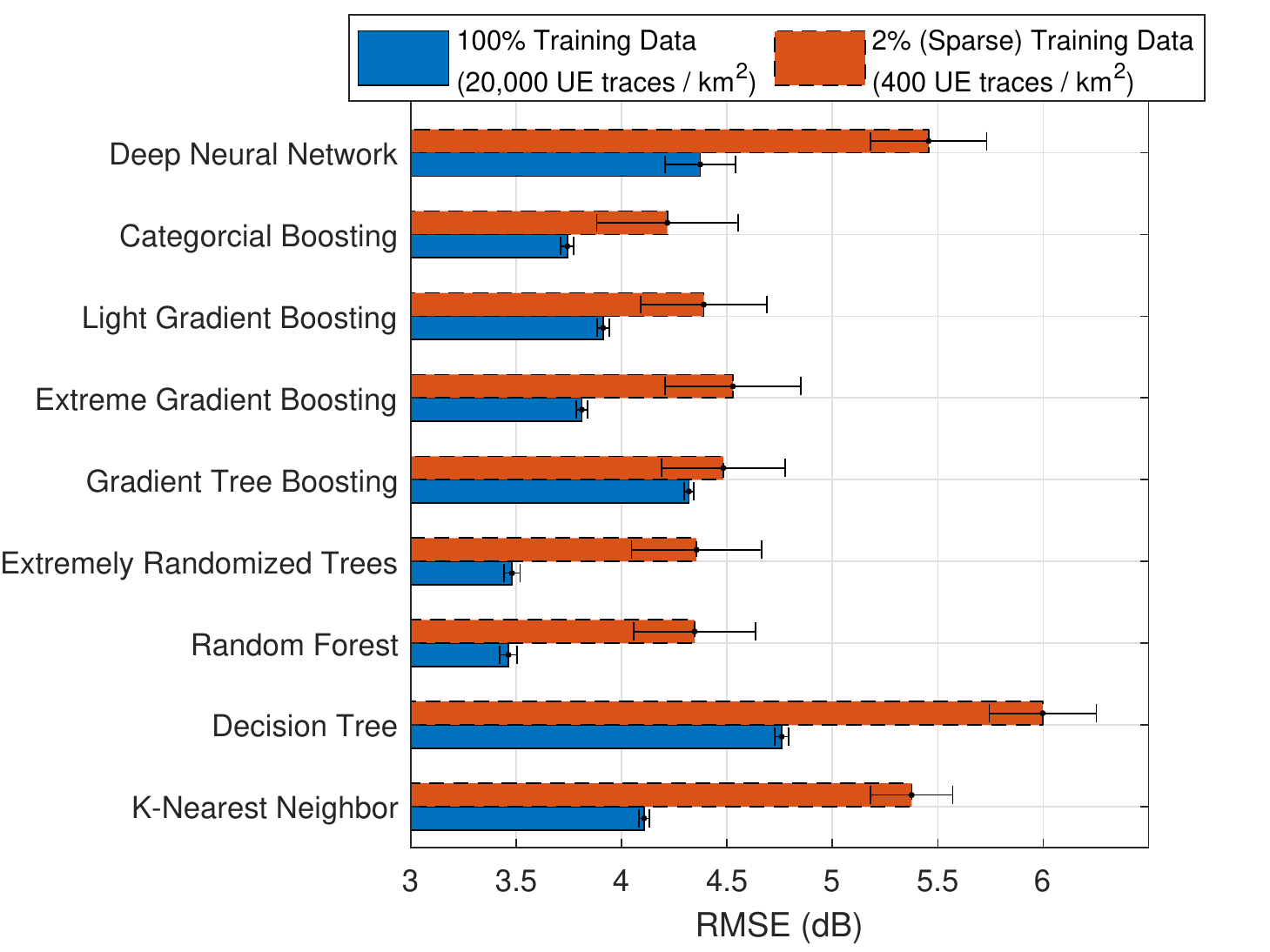}}
	\caption{Comparison of RSS prediction error when the ML based prediction models are trained using scarce and enriched synthetic data. Height of bars represent the mean value and error bar represent the standard deviation using 5-fold Repeated Cross Validation. Enriched synthetic data leads to a reduction in RSS prediction error (RMSE) \cite{masood2022interpretable}}. \label{fig:prop_model_result}
  \vspace{-0.3in}
\end{figure}

Another example of data generated through simulators include system features data (such as BS horizontal/vertical separation, transmit power, operating frequency, antenna beamwidth and gain) and environment features (such as propagation distance, clutter types, BS height, diffraction points, number of building penetrations in each clutter type) to create a machine learning based prediction model for 3D pathloss and received signal strength (RSS) \cite{masood2022interpretable} to overcome the challenges of conventional and ray tracing based path loss modeling. This work investigated the model performance under varying data scarcity levels (UE density). Fig. \ref{fig:prop_model_result} is a key numerical result from this study, which shows how the augmentation of scarce training data (from 400 UE traces/$\rm km^2$ to 20,000 UE traces/$\rm km^2$) leads to significant reduction in RMSE (RSS prediction error) for most ML algorithms used for path loss and ultimately RSS prediction.

 Another simulator generated data in \cite{farooq2022data} includes the dataset of RSRP, SINR, and handover success rate (HOSR) against the rarely explored mobility configuration and optimization parameters, namely A5 time to trigger, A5 threshold 1 and 2. The A5 parameters are usually fixed to a gold standard value or adjusted through hit and trial due to  
the valid reluctance of network operators to test all parameter combinations in the live network. To overcome this issue, synthetic data from a 3GPP-compliant simulator was generated.
This type of data was  then used to develop a closed loop solution for optimizing seldom explored A5 parameters by jointly maximizing RSRP, SINR and HOSR \cite{farooq2022data}.

\vspace{-0.1in}
\subsection{Lessons learned}

Synthetic data using simulators can be used to augment data in situations where the available data is non-representative. Simulators are also a good candidate to generate training data for transfer learning or meta-learning techniques. Although most simulators are link level, system level simulators are also there. The choice of simulators depends on what features (e.g., scheduling support, mmWave, adaptive numerology, mobility and pathloss modeling, COPs, etc.) are supported and Table \ref{simulators_table} can assist the reader for this purpose. Based on the available literature, SyntheticNET  has the most features supported.

\section{Real data generation}
\label{chap:real}
The preceding techniques, with the exception of using simulators, are likely to work well when the scarce available data is somewhat representative of the whole data or exhibits some degree of correlation.
 In situations where the available data is scarce or big but non-representative, the solution lies in obtaining  real data. 

One way of getting access to real data can be utilizing historic logs of data gathered by other researchers. However, these logs might become outdated quickly with the emergence of new technologies, heterogeneous deployments or change in traffic patterns, number of users, construction of buildings and other terrain changes. Another way of generating real data can be through the use of mobile phone applications. However, what if researchers require data for scenarios which are not yet deployed in a real network? The techniques presented in previous sections (except simulators), all require some starting real data but with the advent of AI based next generation networks, there exists the  potential of new or anticipated scenarios which do not exist in a real network. In such cases, testbeds to generate real data are going to be the best option for wireless communications community.
\subsection{Phone applications and parametric subscriber/third-party data}
Many smartphone  applications offer the ability to log parameters such as RSRP, RSRQ, SNR,  events occurring (handover, cell re-selection), serving time, speed, height, cell ID, along with timestamp and location (latitude, longitude) information). 
As an example, one of the studies  \cite{spatiotemporal_hasan}, used a 
novel methodology of utilizing smartphone application, based on the idea of participatory sensing, to collect real LTE network data for building, training and evaluating the performance of mobility prediction schemes in live network \cite{spatiotemporal_hasan}. 
The data in this case was the handover information of the user.
An android application, ``LTE Discovery'' was installed on the smartphone to log the timestamp and new cell IDs around the OU-Tulsa campus. This information was then used to build a semi-markov model for mobility prediction. 

The quality of data gathered through smartphone applications, however, depends on a number of factors, including measurement capabilities of different smartphones and GPS error inaccuracy for measuring heights and positions.  Smartphones equipped with barometers are likely to give a better estimate of heights in scenarios with varying terrains. In addition, transmitter parameters, such as type of antennas and their characteristics remain unknown, unless the network operator is involved. When the network operator is involved, it is possible for the subscriber to obtain parametric data from them. However, that type of data may be limited to a certain number of possible configurations. For this reason and for potential new scenarios, the solution may lie in resorting to testbeds. 

\subsection{Testbeds}
\label{sec:testbeds}
Field trials using testbeds generate real training data and provide the most realistic picture of the network.  An aerial view of some of these testbeds is presented in Fig. \ref{testbeds_figures}. 
We have summarized the existing and emerging testbeds in Table \ref{testbeds_table} to make readers aware of current and emerging platforms to access real data.
Most of these testbeds are open, i.e., available to external experiments. This will foster collaboration among different academic  institutions as well as with industry, which will in turn enable the utilization of these existing facilities to the fullest and accelerate quality research in the field.

Apart from individual testbeds, several federations or consortiums of testbeds have been formed around the world. Some key federated testbeds comprising of the testbeds in Table \ref{testbeds_table} are presented in Fig. \ref{federated}.

Examples of data collected from testbeds include data for scenarios that are not fully and widely deployed yet, e.g., mmWave channel measurement data consisting of direction of user movement with respect to BS-UE link, distance
resolution,  the number of user locations and whether blockage is present or not \cite{jain2020mmobile}. This type of data can be used for building beam tracking algorithms. Other examples of data include received signal strength indicator, electric vector magnitude, packet and bit error rate data from CORNET testbed \cite{sharakhov2014visualizing} and massive MIMO data from LuMaMi testbed such as signal to noise ratio (SNR) and bit error rate for different antenna configurations and modulation schemes \cite{malkowsky2017world}. These types of data can provide flexibility to researchers for design and testing network scenarios 
using a much wider range of parameters, which is difficult to obtain from network operators otherwise, due  to the  high probability of network impairment when varying parameters too much in live networks.

\begin{figure*}[!t]
	\centering
	{\includegraphics[height=23cm, width=16cm]{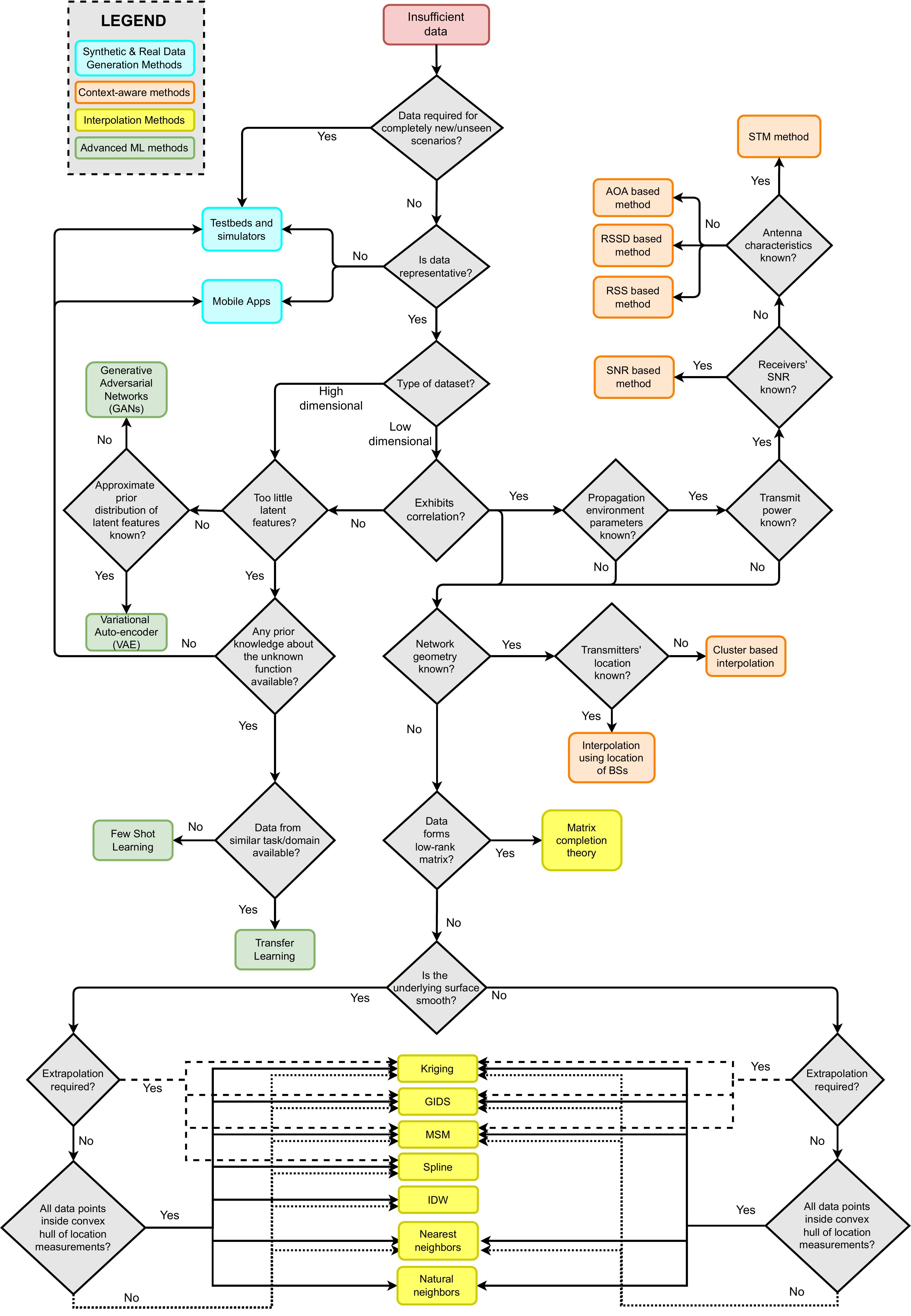}}
	\caption{Decision flowchart for the selection of data augmentation technique for handling scarce datasets in mobile networks.} \label{fig:flowchart}
\end{figure*}

\subsection{Lessons learned}

One way of getting access to real data to augment scarce data can be utilizing historic logs of data gathered by other researchers. However, these logs can become outdated. Lack of diversity in the COP-KPI data is another problem when data is  obtained through logs.
Testbeds is another way to generate real data and is particularly useful to test new or anticipated scenarios which do not exist in a real network. Key features of several federations and individual testbeds around the world have been presented in \ref{testbeds_table}  that can assist the readers in the choice of testbed for their works. 

\section{Conclusion and discussion}
\label{chap:conclusion}

In this paper, we have presented an overview of key techniques in literature to address the  data scarcity challenge and presented some emerging new techniques that can be applied to radio access networks in the wireless communication domain to solve this problem. 

Table \ref{tab:sparsity} summarizes the data augmentation techniques for handling scarce datasets in mobile networks. The typical use cases targeted in existing literature include mobile traffic maps generation using scarce CDR data, spectrum sensing, MDT-based outage detection, CSI/RSS for localization, BS trace data for performance analysis, 
network power minimization, optimizing BS Tx power using UE SINR data, network parameter configuration optimization for power control and user scheduling, resource allocation, traffic load based energy saving, CQI and RSS prediction, radio environment map reconstruction, channel estimation in Massive MIMO systems and discovering user patterns using user trajectory data. The tools in existing literature to address these use cases include GANs and its variants, transfer learning, autoencoders, interpolation techniques, simulators and testbeds. While these techniques have proved to be beneficial for particular use cases, the generalization ability of a particular technique to different scenarios remains a challenge. Another notable challenge is the applicability of these techniques to highly dynamic or mobile environments. Efforts are  also being made to reduce the training time of machine learning based models and modifying them for more robustness.

It should be noted however, that the success of any technique for solving the data scarcity challenge depends on a number of factors, including type of data under consideration, number of transmitter and receivers, distributions of users and base stations in a given area, distribution of measurement data,  level of accuracy required, measurement capability of receivers, dynamics of propagation environment, propagation modeling accuracy, time and computational resources available. Also, highly dynamic spatio-temporal environment would greatly hamper the outputs of techniques covered in this paper. In that case, using data through simulations and testbeds may provide the best option.
Further options on addressing the data scarcity challenge for highly dynamic environments is out of the scope of this work and can be  considered as part of a future study.
Therefore, while a certain technique might work well in a particular scenario, it is likely to perform poorly in other scenarios. It should also be noted that the selection of a performance metric to assess the accuracy of a particular method is important too. As an example, if the metric of mean residual error is used to access Kriging accuracy, it would always yield zero, since this type of  interpolant satisfies the unbiased-ness condition, and so some other performance metric, like the average relative error would be more appropriate in this case.
 
Finally, based on the analysis from literature and domain knowledge, in order to assess the applicability of a particular method, the tree diagram in Fig. \ref{fig:flowchart} is aimed to assist researchers and network operators in choosing the appropriate techniques based on available information.
We start the figure by the red box, ‘Insufficient data’. The first question in the decision figure is whether the data required is for completely new or unseen scenarios (e.g., 6G drones to terrestrial networks that are not yet deployed) or whether the data required is for scenarios already present in today's networks. In the former case, the only options are utilizing testbeds and simulators to depict new use cases. In the latter case, if the data is non-representative (i.e., very few data points are available that might not represent the scenario very well), the options are again to generate more synthetic data through simulators or real data through testbeds and mobile applications.

However, if the data is representative, low dimensional in nature (e.g., spatial only), and exhibits some correlation (e.g., RSRP values that are correlated with distance), the choice of methods depends on whether the propagation environment parameters (e.g., frequency, path loss exponent) are known or not. If these parameters are known, along with knowledge of receivers’ SNR and transmit power (through e.g., operator), then SNR based method in Section \ref{snr-based-sec} can be used. If transmit power is known, but receivers’ SNR is not known, but antenna characteristics (e.g., antenna tilt, patterns) are known, then the STM method in Section \ref{sec:stm} can be used. If SNR is not known, and antenna information is also not available, then based on the propagation environment and transmit power information only, three methods described in Section \ref{sec:context2}, AOA, RSSD and RSS can be used.

If the low dimensional data is correlated, but we do not have information about propagation environment or transmit power, choice of interpolation method can be done on the based on other contextual information, such as network geometry, which if known, leads to cluster-based interpolation in Section \ref{sec:cluster}. If, along with network geometry, transmitter locations are also known, then the triangle method in Section \ref{sec:triangle} can be a possible choice. If, however, the network geometry is also not known, but the data forms a low-rank matrix (e.g, ultra-dense high frequency scenario), then matrix completion in Section \ref{sec:matrix_completion} can be a choice. Otherwise, decision is made by assessing whether the underlying data surface is mathematically smooth or not.  By smooth, we mean differentiable and continuous surface. In case of smooth surface that requires extrapolation of data, kriging, GIDS, MSM, and Splines can be used and where extrapolation is not required, all interpolation methods in Section \ref{chap:interpolation} can be used with the exception of natural neighbors, which can be used only if all data points are inside the convex hull of location measurements. In the case of non-smooth surface that requires extrapolation, kriging, GIDS, MSM can be used, and if the non-smooth data surface requires interpolation only, then kriging, GIDS, MSM, Nearest neighbors, natural neighbors are the choices, since splines and IDW can be used on smooth data surfaces only. The exception here is again natural neighbors, which can be used only if all data points are inside the convex hull of location measurements.

If the low dimensional data does not exhibit any correlation, we arrive at the decision block that coincides with the case of high dimensional data (e.g., spatio-temporal tabular data with multiple features) nature of data. In these cases, if the data has many latent features, then VAEs in Section \ref{sec:autoencoders} can be used given the prior distribution of latent features is known or can be approximated, otherwise GANs discussed in Section \ref{sec:gans} can be the choice since they do not require the knowledge of prior distribution of latent features. On the contrary, if the low dimensional data does not exhibit any correlation and also does not have enough latent features, then the decision is made based on the availability of any prior knowledge about the distribution of data, which if unknown, leads to the augmentation of data through testbeds, simulators and mobile application, and if known, leads to the possible solution of transfer learning (if data from a similar domain is available), otherwise,  few-shot learning can be the choice.

\section{Future  directions}

Since the advanced machine learning methods, such as GANs, transfer learning and few short learning are much less explored for different telco use-cases, as compared to techniques such as interpolation methods, more investigation of these techniques in telco domain in needed. Particularly the potential of transfer learning remains unexploited. Future work focused on questions on what to transfer, where to transfer and how transfer while taking into account domain knowledge of RAN may help avail the full potential of transfer learning for wireless networks.

Similarly, in GANs, research questions such as how much minimum data is needed to train a generator for given type of RAN data and problem is an important direction to exploit the full potential of GANs and their limits on synthesizing RAN data. A recent work explores this question \cite{naveed2022assessing} indicating significance of this research direction.

Moreover, solutions that have the scalability to generate high dimensional data, robustness to highly dynamic real environments and the capability to take conditional context of the required network conditions into account can also be another future direction. 

Another research direction worth exploring to address the data sparsity challenge in wireless communication domain is by leveraging active learning \cite{active_learning}, which harnesses the power of machine learning together with the experience from domain expert.

Most current machine learning based approaches to enrich training data are predominately used as black-box models, allowing little interpretability. Therefore, another future direction can be to design gray-box (or hybrid) machine learning models (e.g., GANs) by combining domain knowledge and analytical modeling with machine learning. This can bring model interpretability and therefore improved ability to extrapolate beyond the exposed training data distributions. 

Validating the recent and new developed methods and solutions on real data from operators and testbeds can also be a focus of future work. 

There is also a need for datasets in this domain to be publicly accessible to enable the research community to devise practical solutions that can be benchmarked. One such initiative in this direction was taken in the form of CRAWDAD repository \cite{crawdad}.

Recent advancements in Open RAN might also help the data scarcity challenge as Open RAN  introduces a set of open standardized interfaces to interact, control and collect data from every node of the network \cite{d2022orchestran}. However, the issue stemming from sparsity of data (resulting from operators trying a limited range of COPs that leads to a sparse data distribution) will still remain as Open RAN will not allow experimentation on a live network.
Consequently, the exploration and advancements of the techniques discussed in this survey will be required.

\onecolumn
\renewcommand*{\arraystretch}{1.2}
\begin{small}
\begin{longtable} 
		{ |P{2.3cm}|P{4.0cm}|P{10.7cm}|} 

			       		\caption{Worldwide existing and emerging testbeds for solving data scarcity problem. }		             \label{testbeds_table} \\
    	\hline 
		\bf \scriptsize Testbed& \bf  \scriptsize Location & \bf \scriptsize Key Features \\
	    \hline 
	\begin{tabular}{@{}c@{}@{}}  NITOS \\ \cite{nitos1} \cite{nitos2} \end{tabular}& \begin{tabular}{@{}c@{}@{}}  NITlab, University of \\Thessaly (UTH), Volos, \\Greece  \end{tabular} &  \begin{tabular}{@{}c@{}} - Open (facilities available to external experimenters) \\ - Over 100 wireless indoor and outdoor nodes \\- 45 nodes equipped with a mixture of Wi-Fi and GNU-radios \\ - One Cloud installation with 200-cores \\  - Multiple wireless sensor network deployments\\- Cameras, temperature and humidity sensors  \\ -  Software defined radio testbed with 10 USRP devices\\- Two programmable robots provide mobility \\ - WiMAX/3G/LTE technologies  \\ - 5G virtual infrastructure provisioning by 5GINFIRE \cite{nitos_5g}
	 \end{tabular} \\
	\hline
\begin{tabular}{@{}c@{}@{}} 	6GIC \\ \cite{5gic1} -\nocite{5gic2}\nocite{5gic3} \cite{5gic4} \end{tabular}&ICS, University of Surrey, Guildford, UK &\begin{tabular}{@{}c@{}} - 4G LTE, 5G NR, 6G (ongoing) \\ - 4km$^2$ comprising indoor and outdoor environments \\ 	- Outdoor:  4G ultra-dense C-RAN comprising 3 macro cells, 39 LTE-A TDD \\ small-cell sites, operating at 2.6 GHz, 1x 4G FDD site operating at\\ 700 MHz,  8x 5G NR TDD sites, operating at 3.5 GHz \\
	-  Indoor:  6x TDD and 6x FDD cells over 2 floors, and Wi-Fi APs \\
	- 28 GHz (PtP), 60GHz (PtMP) mmWave and satellite backhauling  also supported \\
	- Core Network supports separate 4G and 5G core segments \\- Supports broadband mobile radio
		\\- Fixed core network and service platform based on software defined networking \\
		- Supports Internet of Things   \end{tabular} \\
	\hline
	\begin{tabular}{@{}c@{}@{}} 	ORBIT \\ \cite{orbit1}  \cite{orbit2} \cite{orbit3} \end{tabular}&WINLAB, Rutgers University, USA&\begin{tabular}{@{}c@{}} - Open: available for remote or on-site access\\ - Radio grid with 20x20 two-dimensional grid of programmable radio nodes \\ - Outdoor ORBIT network  provides a configurable mix of both high-speed cellular \\(WiMAX, LTE) and 802.11 wireless access \\ - SANDBOX networks used for debugging and controlled experimentation \\ - Software defined networking (SDN) resources \\- Cloud resources \end{tabular}\\
		\hline
\begin{tabular}{@{}c@{}}	PhantomNet  \\ \cite{phantom1} \cite{phantom2} \end{tabular}&Flux Group, University of Utah, USA &\begin{tabular}{@{}c@{}} - Remotely accessible and sharable \\- Mobility testbed \\  - Built on top of Emulab \\  - EPC/EPS software (OpenEPC), hardware access points (ip.access eNodeB), PC \\ nodes with mobile radios (Nexus 5 phones and SDR-based) \\ - Provides configuration directives and scripts \end{tabular}\\		
		\hline
\begin{tabular}{@{}c@{}}LuMaMi \\ \cite{lund1} \cite{lund2} \cite{lund3} \end{tabular}& Lund University, Sweden & \begin{tabular}{@{}c@{}} - Real time 128-antenna MIMO test bed \\ - National Instruments USRP RIO SDRs \\- LabVIEW system design software and PXI platforms \\ - Mobile base stations \\- Used for channel sounding, high speed data streaming, evaluation of baseband \\solutions, assessing circuit design   \\- Demonstrated mobile multi-user tests with University of Bristol \cite{lund_bristol} \end{tabular}\\		
 	\hline
	\begin{tabular}{@{}c@{}}    Firecycle \\ \cite{firecycle1} \cite{firecycle2} \end{tabular}& \begin{tabular}{@{}c@{}} Intrusion Detection Systems\\ Group, Columbia University, \\USA  \end{tabular}&  \begin{tabular}{@{}c@{}} - Scalable test bed for large-scale LTE security research \\ - Implement, test, analyze  impact of security attacks against LTE mobility network \\ - Prototyping and testing attack mitigation strategies for future cellular networks\\ - Implemented on OPNET \end{tabular}\\		
	   \hline
	   	\begin{tabular}{@{}c@{}} 	Berlin LTE-A \\ \cite{berlin1} \cite{berlin2} \cite{berlin3} \end{tabular}& \begin{tabular}{@{}c@{}@{}} Center of Berlin, operated from \\Fraunhofer HHI, Deutsche \\Telekom Laboratories and \\University of Technology, \\Berlin \end{tabular} & 	\begin{tabular}{@{}c@{}@{}} - 3 base station sites with 9 sectors \\- Incorporates LTE key features: frequency dependent scheduling in 20 MHz \\bandwidth, adaptive MIMO mode selection for 2x2 MIMO utilizing spatial\\ multiplexing, and low round-trip delay on the PHY layer of 8 m  \end{tabular}  \\
	   \hline
	\begin{tabular}{@{}c@{}}  	   CEWiT\\ LTE and 5GNR \\ \cite{india} \end{tabular}& IITMadras Research Park, Chennai, India & \begin{tabular}{@{}c@{}} - 2 types of testbeds based on: 1) CEWiT hardware 2) TI's multi-core DSPs \\ - Hardware is made using SDR radio nodes \\ 
	-  LTE PHY for UE and eNB has been developed in collaboration with IITM  \\ - Basic implementation of LTE L1 downlink and uplink chains \\ 
	- L2 MAC, RLC and a thin layer of PDCP\\
-	Both eNodeB and UE implementations\\
- End-to-end IP application flow both in DL, UL  \\ - Supports 3GPP Release 8 specifications
\\- Supports up to 10 MHz bandwidth and can be extended to 20MHz \\  - 5G NR for sub 6GHz and mm wave under development\end{tabular}\\		
	   \hline
	 \begin{tabular}{@{}c@{}}    TitanMIMO-6 \\ \cite{titan1} \cite{titan2}\end{tabular}&  Nutaq, Québec, Canada& \begin{tabular}{@{}c@{}} - Sub 6 GHz wideband Massive MIMO testbed \\ - FDD+TDD capabilities \\ - Up to 56 MHz real-time baseband processing \\- Radio tumble up to 5 GHz \\ - Nutaq's SDR systems (PicoSDR) can be combined with  TitanMIMO system \\to build up complete HetNet, MUMIMO or CRAN testbed solutions \\ - Enabling  evaluation of interoperability behavior for various deployment scenarios  \end{tabular}\\		
     \hline
     \begin{tabular}{@{}c@{}}  Aalto 5G research \\infrastructure  \\ \cite{espoo} \end{tabular}&      \begin{tabular}{@{}c@{}} Otaniemi, Espoo, \\Finland  \end{tabular}& \begin{tabular}{@{}c@{}} - Network slicing\\ - Support for NB-IOT  to be used for IoT hackathon \\ 
     	- Mobile and edge computing, VR/AR, Gaming,	Industrial Internet \\- Part of 5G TNF \end{tabular}\\		
       \hline
       
        \begin{tabular}{@{}c@{}}  University of \\Helsinki  Test \\Network \\ \cite{helsinki} \end{tabular}&      \begin{tabular}{@{}c@{}}University of Helsinki,  \\Kumpula campus,\\ (Exactum building), Finland \end{tabular}& \begin{tabular}{@{}c@{}} - 17 Nokia Flexi Zone Indoor Pico BTS (eNBs) \\
        -	Band: 2600 MHz (E-UTRA 7) FDD \\
        -  Sync: 1588v2 (PTP) / GPS / Sync-E \\
        	- 3 connections to cores through VLANs: UH core(s), Aalto core and Nokia core \\- Part of 5G TNF \end{tabular}\\		
       \hline
    \begin{tabular}{@{}c@{}}    VodaPhone \\Chair \\ \cite{vodafone1} \cite{vodafone2} \cite{vodafone3}\end{tabular}& TU Dresden, Germany & \begin{tabular}{@{}c@{}}  - Online Wireless Lab (OWL) testbed \\ - Software Defined Reconfigurable Radio Devices\\- LabVIEW/LVC in combination with USRPs\\  - Many projects and startups, e.g.,  5G Lab Germany, 5GNetMobil,
       	5G Picture, \\
       	HPE-5G-Testbed, Airrays GmbH \cite{chair_projects}\end{tabular}\\
       	\hline
       \begin{tabular}{@{}c@{}}   	CORNET \\ \cite{cornet1} \cite{cornet2} \end{tabular}& Virginia Tech University, USA & \begin{tabular}{@{}c@{}} - University-wide testbed \\- Software-defined radios, cognitive radio and dynamic spectrum access\\ - 48 indoor SDR nodes, 14 fixed outdoor nodes, 6 mobile units (O-CORNET) \\ - A few LTE-capable nodes (LTE-CORNET) \\
       	- CORNET nodes are remotely accessible \\
       	- Awarded the grant from DURIP for upgrading to LTE and LTE-A\\
       	- Outdoor network of 15 radio nodes and 2 mobile nodes\\
       \end{tabular}\\
  \hline
   \begin{tabular}{@{}c@{}} 5G Playground \\ \cite{5g_play1} \end{tabular}& Fraunhofer FOKUS and TU Berlin campus, Germany&  \begin{tabular}{@{}c@{}} - Empowers the 5G Berlin testbed \\-  Support for multi-slicing \\ - Ultra-reliable, low latency communication in Industrial IoT lab of FOKUS \\
- Automotive testbed environment in underground parking  of FOKUS building \\
   	- Coverage of dense urban areas, like portable 5G edge nodes in progress\\
  	- 3 Toolkits: Open5GCore, OpenSDNCore and Open5GMTC      \end{tabular}\\
  \hline
  
   \begin{tabular}{@{}c@{}} Tampere \\University \\Wireless Test \\Networks \\ \cite{tampere1} \cite{tampere2} \end{tabular}&Tampere University, Hervanta, Finland&  \begin{tabular}{@{}c@{}} - Part of 5G TNF\\-  FDD-LTE operating at band 1, 7, and 28 for mostly indoor coverage \\ - TDD-LTE operating at band 38 to provide campus wide outdoor test network \\ - Upcoming outdoor 5G test network in band n78 with 60 MHz channel  \\- LoRa: Digita's LoRaWAN test network in ISM band at 868 MHz
   \end{tabular}\\
  
    \hline
 \begin{tabular}{@{}c@{}} FUSECO \\Playground \\ \cite{f_play1}\end{tabular}& Fraunhofer FOKUS Institute, Berlin, Germany &  \begin{tabular}{@{}c@{}} - Open IMS Core solution\\ - Heterogeneous indoor and outdoor  radio access technologies\\ - DSL/WLAN/2G/3G/4G-LTE/LTE-A and  soon 5G\\- M2M communication, IoT, sensor networks \\- SDN/OpenFlow, NFV cloud environments \\
 	- Toolkits: Open5GCore, OpenSDNCore and Open5GMTC, OpenMTC, \\Open Source IMS Core,
 	OpenStack-based Cloud Testbed,	OpenXSP        \end{tabular}\\
   \hline
 \begin{tabular}{@{}c@{}}    5G Ready \\Trial Platform \\ \cite{5g_ready} \end{tabular} &Fraunhofer FOKUS, Berlin, Germany&  \begin{tabular}{@{}c@{}} 
 - Consolidated turn-key solution of the Fraunhofer FOKUS software components \\  - Addresses trial needs of emerging network infrastructures - \\ - Edge Instantiation: solution for micro-operators and local networks,  provides \\customized IoT connectivity for x100 devices. \\  - Data Center Instantiation:  multi-slice environment, support for multiple parallel \\instances of IoT and multimedia communication \\  - Technology Elements: Virtual Core  network, Network slicing, IoT support, \\Low delay network, Dynamic spectrum access and management\end{tabular} \\
   \hline
\begin{tabular}{@{}c@{}}  Ericsson 5G\\ \cite{Ericsson1}  \cite{Ericsson2}\end{tabular} &   Ericsson, Stockholm, Sweden&  \begin{tabular}{@{}c@{}}- Live testing of key capabilities, such as multipoint connectivity with \\ distributed MIMO and 5G-LTE dual connectivity \\- 5G devices and   base stations operate in  15 GHz band \\- TDD and OFDM \\-  Up
  	to 256 QAM modulation in  downlink and up to 64 QAM in the uplink\\- mm-Wave testbeds 15 GHz and 28 GHz \\ - Bandwidth is 80 MHz, centered at 3.5 GHz \\- Massive MIMO antenna array
  	of 128 cross-polarized antennas \end{tabular}\\
   \hline
 \begin{tabular}{@{}c@{}}   SK Telecom \\ 5G Playground  \\ \cite{SK1} \cite{SK2} \cite{SK3} \end{tabular}& \begin{tabular}{@{}c@{}}   SK Telecom  R\&D Center, \\Bundang, Korea \end{tabular}& \begin{tabular}{@{}c@{}} - Developing a centimeter-wave (cmWave) 5G radio system with Nokia\\  - 5G 3D system level simulator with Nokia and Ericsson \\ - 3D beamforming techniques with large scale array antennas with Samsung \\- Developing Anchor-Booster Cell and Massive MIMO with C-RAN with Intel \\- Achieved 19.1Gbps transmission speed over the air \\
 -	  Futuristic services including 4K live broadcast system and AR/VR \end{tabular} \\
   \hline
              \begin{tabular}{@{}c@{}} 5GTN \\(Linnanmaa) \\ \cite{oulu1} \cite{oulu2} \cite{oulu3}  \end{tabular}&      \begin{tabular}{@{}c@{}}University of Oulu and \\VTT Technical Research \\Centre of Finland \end{tabular}& \begin{tabular}{@{}c@{}} 
   	- Multi-access edge computing \\ - Core network in cloud environment \\- Cloud systems for applications \\- Secure connection to other 5G sites worldwide, 10 Gb VPN \\- Part of 5G TNF
   \end{tabular}\\		
   \hline
 \begin{tabular}{@{}c@{}}   TurboRAN \\ \cite{TurboRAN} \end{tabular}& AI4Networks Research Center, University of Oklahoma, Tulsa, USA &\begin{tabular}{@{}c@{}} - Developing first end to end programmable cellular test bed for enabling \\AI based SON research towards 5G and beyond \\
   -  Complete integrated mobile cellular network  over 300,000 $\rm m^2$ area \\
   - Tier 1:  4 outdoors macro cells on 1.2-6 GHz HF band\\
   - Tier 2: 16 small cells (programmed to pico or femto cells). 8 small cells can \\operate on the HF band, other 8 can operate on the unlicensed mmWave  \\
   -  Both tier cells  are programmable  \\
   - Both tier cells connected to EPCs and a big data processing Hadoop cluster \\
   - Hadoop cluster: 1 high performance master node, 15 slave nodes with \\high-capacity data modems \\
   - Support both high mobility and low mobility users \end{tabular}\\
      \hline
    \begin{tabular}{@{}c@{}} OAI  \\ \cite{oai1}-\nocite{oai2}\nocite{oai3}\cite{oai4} \end{tabular}&EURECOM, France & \begin{tabular}{@{}c@{}}- Open-source platform \\ - 8-node testbed, equipped 
     OAI compatible RF front-ends, UEs and VMs \\ - 4 machines that can be used for running OAI as eNodeB \\ - 4 nodes that are equipped with COTS UEs\\- 2 physical layer emulation modes \\- 64 antenna Massive MIMO testbed \end{tabular}.\\
     \hline
       \begin{tabular}{@{}c@{}}    Munich  \\ \cite{munich1} \cite{munich2} \end{tabular}&TU Munich, Munchen, Germany &  \begin{tabular}{@{}c@{}}  
      - 5G RAN with two sectors, each having carrier frequency: 3.4 GHz, \\bandwidth: 40 MHz, transmission power: 5 W antennas: up to 8\\
      - 5G Mobile Terminals with vehicular speeds up to 50 km/h,  enablingV2X \\
      - 5G Core network: HW/SW platform\\
     -  Hardware: in-house platform  of several dozen  servers representing a data centre \\
      - Software: extended network emulators, controllers, open-source and proprietary \\ switch implementations \\
      -  Testbed can deploy virtual networks with different topologies as needed\\
     - 5G Core network supporting functional split – SDN – NFV Orchestration\\
     - Distributed data centres for mobile edge computing use cases
    
    \end{tabular}
     \\
      \hline
      \begin{tabular}{@{}c@{}}    Perform Networks \\ \cite{perform1} \cite{perform2} \cite{perform3} \end{tabular} &University of Malaga, Spain&    \begin{tabular}{@{}c@{}} - T2010 conformance testing units by Keysight Technologies  \\ - LTE release 8 small cells (Pixies) by Athena Wireless working on band 7 \\ - Polaris Core Network Emulator 
    \\-  Several LTE UEs, working on different bands  \\-  ExpressMIMO2 and USRP SDR 
      cards \\ -  SIM cards from an Spanish LTE operator to be used on commercial deployments \end{tabular}  \\
      \hline
            \begin{tabular}{@{}c@{}}    Centria's \\Test Network   \\ \cite{YLIVIESKA}  \end{tabular} &      \begin{tabular}{@{}c@{}} Centria University of Applied \\ Sciences,Ylivieska, Finland \end{tabular} 
            &    \begin{tabular}{@{}c@{}} - TDD-LTE operating at band 40 and 42 for both outdoor and indoor coverage \\ - Upcoming 5G test network in band n78 with 60 MHz channel outdoor  network \\- 
            	Implementation plan of first 5G Non-Standalone during 2019 
            	\\-  Later 5G Standalone during 2020 \\ - Part of 5G TNF
             \end{tabular}  \\
      \hline
 \begin{tabular}{@{}c@{}} w-iLab.t  \\ \cite{wlab1} \cite{wlab2} \cite{wlab3} \end{tabular}&  Ghent and Zwijnaarde, Belgium&  \begin{tabular}{@{}c@{}} - w-iLab.t Office testbed: three 90 m x 18 m floors of  iMinds office  in Ghent\\ - w-iLab.t Zwijnaarde testbed: 5 km away from w-iLab.t Office in Zwijnaarde\\ - Sensor nodes, Wi-Fi based nodes, sensing platforms, and cognitive radio \\ - Heterogeneous wireless/wired experiments \\- Virtual Walls: Virtual Wall 1 and 2 containing 206 and  159 nodes respectively  \\ - OpenFlow experiments \\ - 20 programmable moving robots\end{tabular} \\
      \hline
\begin{tabular}{@{}c@{}}  5TONIC \\ \cite{5TONIC} \end{tabular}&Madrid, Spain&     \begin{tabular}{@{}c@{}}  - 9 members: Telefonica, Institute IMDEA Networks, Ericsson, Intel, \\Commscope, Universidad Carlos III de Madrid, Cohere Technologies, \\ Artesyn Embedded  Technologies and InterDigital \\
      - NFV orchestrator, implemented with Open Source MANO (OSM) \\
      - Dedicated NFVI for 5GINFIRE: 3 server computers, each with six cores, \\32GB of memory, 2TB NLSAS, network card with 4 GbE ports, DPDK support \\
      - Second NFVI: 2 high-profile servers, each equipped with eight cores in a \\NUMA architecture, 128GB RDIMM RAM, 4TB SAS and eight 10Gbps \\Ethernet optical transceivers with SR-IOV capabilities  \end{tabular}       \\
  \hline
  
  \begin{tabular}{@{}c@{}} 
  University of \\Bristol 5G \\ \cite{bristol} \end{tabular}  &University of Bristol, England&    
\begin{tabular}{@{}c@{}} 
- Multi-site network connected through a 10 km fibre \\
- Core network is located at HPN Lab at the University of Bristol \\
- Extra edge computing node is available at Watershed \\
- Access technologies are located at Millennium Square for outdoor \\
coverage and “We The Curious” science museum for indoor coverage \\
- Multi-vendor SDN enabled packet switched network \\
- SDN enabled optical (Fibre) switched network \\
- Nokia 4G and 5G NR \\
- Self-organising multipoint-to-multipoint wireless mesh network \\
- LiFi Access point, Cloud and NFV hosting \\
- 2 different NFV orchestration and management solutions: \\Open Source MANO , NOKIA CloudBand  \\
- 2 cloud/edge computing solutions:Openstack Pike, Nokia MEC \\
- 1 SDN controller: NetOS 
\end{tabular}  \\ \hline
 \begin{tabular}{@{}c@{}} 
 D-15 Labs \\\cite{D15-ercisson}  \end{tabular}  &Ericsson, Santa Clara, CA, USA&
 \begin{tabular}{@{}c@{}} 
  -  Validation and development platform for 5G use-cases, leverages  cloud edge \\support, core network, and AI-based management and orchestration \\
 \end{tabular}  \\
        \hline
        \begin{tabular}{@{}c@{}} 
ENCQOR 5G \\ \cite{ENQOR}  \end{tabular}  &
Ontario Region, Canada&
  \begin{tabular}{@{}c@{}} 
- iPaaS Services: 5G connectivity of 5 Gbps Mobile Throughput and sub 5ms \\
latency, 
cloud services of IoT Accelerator, emulation cloud, edge computing 
\\
- iPaaS Infrastructure:
5G mobile user equipment (android-based Qualcom \\ terminals operating at 3.5 GHz),  5G radio access technology \\
(NR/LTE/CAT-M1/NB-IoT), 5G transport/backhaul, distributed  core network \\ and programmable data plane \\
- Future features expected by 2021 include: 5 Gbps 5G NR, sub 5ms latency,\\ predictive analytics, federated network slicing, real time machine learning / AI \\
- Technology partners: Ericsson, Thales, CGI, IBM, Ciena
  \end{tabular}  \\
        \hline
      	\end{longtable} 
\end{small}

\onecolumn
\renewcommand*{\arraystretch}{1.3}
\begin{longtable}{|P{1.4cm}|P{1cm}|P{2.7cm}|P{3.7cm}|P{2.2cm}|P{2.2cm}|P{2.2cm}|}
\caption{Review of modeling techniques for handling scarce datasets in Radio Access Networks (RAN).} 
\label{tab:sparsity}
\\
\hline
\textbf{Reference} & \textbf{Year} & \bf Modeling technique & \textbf{Use case and data } &
 \textbf{Data type} & \textbf{Use-case type w.r.t. OSI layer} &\textbf{Use-case type w.r.t. level of analysis}\\ \hline

\textbf{\cite{aoki_few_shot_2020}} &
  2020 &
  Few-shot learning &
eNodeB performance metric analysis using cell trace data &
  Tabular data & Network & System \\ \hline
\textbf{\cite{wang2021indoor}} &
  2021 &
  Few-shot learning &
  Modeling indoor pathloss model at 28 GHz using RSS data &
  Tabular data & Physical & Link \\ \hline
\textbf{\cite{shen_lorm_2020}} &
  2020 &
  Few-shot learning + Transfer learning &
  Network power minimization in C-RAN for resource management using UE SINR data &
  Tabular data & Physical & System\\ \hline
\textbf{\cite{zappone_wireless_2019}} &
  2019 &
  Transfer learning &
  Identifying optimal deployment density of the BSs given a BS transmit power w.r.t. spectral and energy efficiency of the network using UE SINR data &
  Tabular data & Phyical & System \\ \hline
\textbf{\cite{chuai_collaborative_2019}} &
  2019 &
  Transfer learning &
  Network parameter optimization for uplink power control and user scheduling using Cell KPI/counter data & Tabular data & Application & System
  \\ \hline
\textbf{\cite{li_tact_2014}} &
  2014 & Transfer learning &
  BS ON/OFF switching for energy saving using traffic load data  &
  Tabular data & Data Link & System \\ \hline
\textbf{\cite{parera_transfer_2020}} &
  2020 &
  Transfer learning & Radio map prediction under different antenna tilt using UE RSS data &
  Tabular data & Physical & Link \\ \hline
\textbf{\cite{parera2021anticipating}} &
  2021 &
  Transfer learning &
 Cell performance prediction (CQI and Active UE count) using cell KPI/Counter data &
  Tabular data & Application & System \\ \hline
\textbf{\cite{moradi2019performance,larsson2021source}} &
  2019-2021 &
  Transfer learning &
  Network service performance prediction using testbed traces &
  Tabular data & Network & System \\ \hline
\textbf{\cite{han_two-phase_2020}} &
  2020 &
  Transfer learning + GAN &
  REM generation &
  Spatial data  & Physical & System \\ \hline
 \textbf{\cite{hughes_generative_2019}} &
  2019 &
  GAN &
  Synthetic CDR generation using CDR data (call start hour and call duration) &
  Tabular data & Network & System\\ \hline
\textbf{\cite{zhang_zipnet-gan_2017}} &
  2020 &
  ZipNet-GAN & Infer fine-grained traffic patterns from course aggregates using CDR data &
  Spatio-temporal data & Network & System \\ \hline
\textbf{\cite{zhang_generative_2020}} &
  2020 &
  GAN &
  Cell outage detection using MDT data &
  Tabular data & Application & System\\ \hline
\textbf{\cite{han_radio_2020}} &
  2020 &
  GAN &
  REM generation &
  Spatial data & Physical & System \\ \hline
 \textbf{\cite{yuan2020anomaly}} &
  2020 &
Variational autoencoder &
  Anomaly detection and root cause analysis (RCA) in RAN using KPI/KQI data &
  Tabular data & Application & System \\ \hline
 \textbf{\cite{rajendran_saife_2018}} &
  2018 &
Adversarial autoencoder &
  Detecting anomalous behavior in wireless spectrum using {power spectral density data} &
  Tabular data & { Network} & {System} \\ \hline
{\textbf{\cite{tsukamoto_highly_2018,yilmaz_location_2015,sun_simple_2010}},\textbf{\cite{pesko_indirect_2015,mezhoud_hybrid_2020}}} &
  2015-2020 &
Context-aware interpolation &
REM construction using BS location  estimated through reverse triangulation &
  Spatial data  & Physical &System \\ \hline
{\textbf{\cite{alam_clustering_2018,alam_performance_2018}},\textbf{\cite{xia_radio_2020,han_radio_2019,suchanski_radio_2019}}} &
  2018-2020 &
Kriging interpolation +  variants &
  REM generation &
  Spatial data  & Physical & System \\ \hline
\textbf{\cite{rahman_creating_2019}} &
  2019 &
  Correlation-based interpolation &
 Crowdsourced spatio-temporal REM generation &
  Spatio-temporal data  & Application & System \\ \hline
\textbf{\cite{wang_learning_2019}} &
  2019 &
Adaptive spatial  interpolation &
Uplink channel estimation in 3-D massive  MIMO systems&
  Spatial data  & Physical & Link\\ \hline
\textbf{\cite{liu_multi-criteria_2019}} &
  2019 &
Adaptive triangulation - induced interpolation &
  Multiple REM generation &
  Spatial data  &  Physical &System\\ \hline
\textbf{\cite{sato_performance_2019}} &
  2019 &
 NN-enhanced, Kriging interpolation &
  REM generation &
  Spatial data & Physical & System \\ \hline
\textbf{\cite{chen_online_2018}} &
  2018 &
  Congregate group pattern &
Signaling data (User trajectory data) for  discovering congregate group patterns &
  Spatio-temporal data & Network &System \\ \hline
 \bf \cite{qureshi2020enhanced} & 2020 &Kriging, moving average, matrix completion, IDW, nearest neighbors, natural neighbors, spline interpolation  & MDT coverage map (RSRP) construction & Spatial data & Physical & System \\ \hline
  \bf \cite{ kriging_cs_rem} & 2019 &  Kriging interpolation &  REM generation from crowdsourced data & Spatial data &Application & System \\ \hline
 \bf  \cite{comparative_REM} & 2011 & Kriging, MSM and GIDS interpolation & REM construction from total received signal power & Spatial data & Physical & System  \\ \hline
 \bf  \cite{reliability} & 2012 & IDW, adaptive IDW, MSM interpolation & REM construction  & Spatial data  & Physical &System \\ \hline
 \bf \cite{ interference_map_estimation_cognitive } & 2014 & Nearest neighbor, IDW, Kriging interpolation & Interference map estimation of MDT reports in cognitive radio networks  & Spatial data & Physical& System   \\ \hline 
 \bf \cite{delaunay} & 2012 & Nearest neighbor, natural neighbor, triangulation-based interpolation   & Interference map generation in cognitive radio networks & Spatial data & Physical &System  \\ \hline 
 \bf \cite{modeling_carrier_aggregation} & 2013 & Nearest neighbor, IDW, Kriging  & Interference maps for licensed shared
access & Spatial data & Physical &System  \\ \hline 
 \bf \cite{comparison_in_cognitive} & 2012 & Natural neighbor, kriging and spline & Interference cartography generation  in cognitive radio networks &  Spatial data & Physical &System\\ \hline 
 \bf \cite{kriging} & 2010 & Kriging  &Predict network coverage in wireless
networks &  Spatial data  &Physical &Link\\ \hline
 \bf \cite{alam_clustering_2018} &2018 & Kriging & REM construction  &  Spatial data &Physical & System\\ \hline 
 \bf \cite{mao_constructing_2018} & 2018 & Kriging & REM construction in cognitive radio networks &  Spatial data & Physical &System \\ \hline 
 \bf \cite{han_radio_2019} & 2019& Kriging, nearest neighbor, IDW & REM construction based on  RSSI mobile crowdsensing data  &  Spatial data & Application & System\\ \hline 
 \bf \cite{hosseini_tehrani_radio_2019} & 2019 & Nearest neighbor, IDW, Kriging& REM construction for spectrum sharing &  Spatial data   & Physical &System  \\ \hline 
 \bf \cite{xia_radio_2020} & 2020 & Nearest neighbor, IDW, Kriging& REM construction&  Spatial data & Physical &System   \\ \hline 
 \bf \cite{braham2014coverage} & 2014 & Kriging & REM generation for coverage mapping &  Spatial data  & Physical &System  \\ \hline 
 \bf \cite{alam_performance_2018} &2028 & Kriging & REM generation for coverage mapping &  Spatial data  & Link &System  \\ \hline 
 \bf \cite{sato_performance_2019},  \bf \cite{mezhoud_hybrid_2020} & 2019-2020 & Hybrid neural networks and Kriging interpolation &REM generation &  Spatial data &Physical & System  \\ \hline 
 \bf \cite{indoor2} &  2015 &
Kriging, splines, moving average, triangulation-based interpolation & Coverage extension and
prediction with signal strength crowdsourced measurements & Spatial data & Application & System\\ \hline 
 \bf \cite{suchanski_radio_2019}& 2019& Nearest neighbor, IDW, Kriging & REM construction for military cognitive networks & Spatial data & Physical & System\\ \hline 
 \bf \cite{indirect_methods } & 2018 & RSS and RSSD based methods & REM enrichment using RSS measurements from sensors & Spatial data & Physical & System\\ \hline 
 \bf \cite{pesko_indirect_2015} & 2015 & STM method, location estimation-based method, IDW, Kriging & REM construction  using  omnidirectional and directional transmitter antenna & Spatial data & Physical & System\\ \hline 
 \bf \cite{live} & 2015&RSS-based methods & REM construction in fading channels &  Spatial data &Physical & System\\ \hline 
 \bf \cite{tsukamoto_highly_2018} & 2018 & RSS-based method, kriging & REM construction  & Spatial data & Physical &System\\ \hline 
 \bf \cite{sun_simple_2010} & 2010 & AOA based and SNR based methods& REM construction & Spatial data  & Physical & System\\ \hline 
 \bf \cite{riaz2022deep} & 2022 & Synthetic data generation through Atoll simulator &
Cell outage detection and diagnosis using SINR-based REM maps & Tabular data  & Physical & System\\ \hline 
 \bf \cite{masood2022interpretable} & 2022 & Synthetic data generation through Atoll simulator & Modeling outdoor propagation model using RSS data & Tabular data & Physical & System \\ \hline 
 \bf \cite{farooq2022data} & 2022 & Synthetic data generation through SyntheticNet simulator & Optimization of A5 mobility parameters using RSRP, SINR, and handover success rate data  (HOSR) & Tabular data  & RSRP/SINR: Physical,  HOSR: Network & System\\ \hline 
 \bf  \cite{spatiotemporal_hasan} & 2016& Real data generation through smartphone application& Building semi-markov model based mobility prediction schemes using handover data &
 Tabular data & Network & System  \\ \hline 
 \bf  \cite{jain2020mmobile} & 2020 & Real data generation using mmWave testbed &
 Building beam tracking algorithms using mmWave channel measurement data &  Tabular data  & Physical & Link\\ \hline 
 \bf  \cite{sharakhov2014visualizing} & 2014 & Real data generation using CORNET testbed &
Evaluating  real-time radio spectrum access using RSS, packet and bit error rate data (PER/BER)&  Tabular data  & Physical & Link\\ \hline 
 \bf \cite{malkowsky2017world} & 2017 & Real data generation using LuMaMi testbed & 
Design and validation of massive MIMO research using SNR and BER data for different antenna configurations and modulation schemes &  Tabular data  & Physical & Link \\ \hline
\end{longtable}

\twocolumn

\section*{Acknowledgment}
This work was supported in part by the National Science Foundation under Grant 1923669, 1730650, the Qatar National Research Fund (QNRF) under Grant NPRP12-S 0311-190302 and in part by an unrestricted award from Ericsson Research, CA, USA.

\bibliographystyle{IEEEtran}
\bibliography{Ericsson}

\begin{IEEEbiography}
	[{\includegraphics[width=1in,height=1.25in,clip,keepaspectratio]{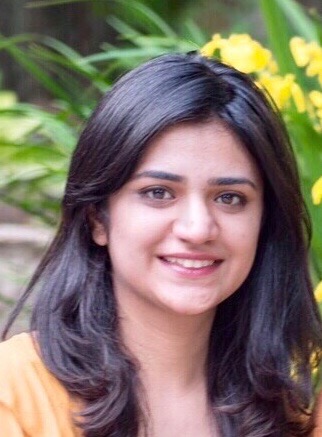}}]{HANEYA NAEEM QURESHI}
 received her BS degree in Electrical Engineering from Lahore University of Management Sciences (LUMS), Pakistan, in 2016 and M.S. and PhD degrees in Electrical and Computer Engineering from the University of Oklahoma (OU), USA in 2017 and 2021, respectively. She is currently a Postdoctoral Research Fellow at the Artificial Intelligence (AI) for Networks Research Center at OU, where she is managing and contributing to several NSF-funded projects and teaching graduate level courses.  She has also worked as an ORISE fellow with the Center for Devices and Radiological Health, U.S Food and Drug Administration (FDA), Maryland; and has significant industrial research experience in wireless communication with Ericsson Research, California, USA and in 3GPP standardization with InterDigital, Inc., New York, USA. Her other current research interests include digital smart healthcare, network automation and combination of machine learning and analytics for future cellular systems. She has also been engaged in system design of unmanned aerial vehicles deployment, channel estimation and pilot contamination problem in Massive MIMO TDD systems.
\end{IEEEbiography}

\begin{IEEEbiography}
	[{\includegraphics[width=1in,height=1.25in,clip,keepaspectratio]{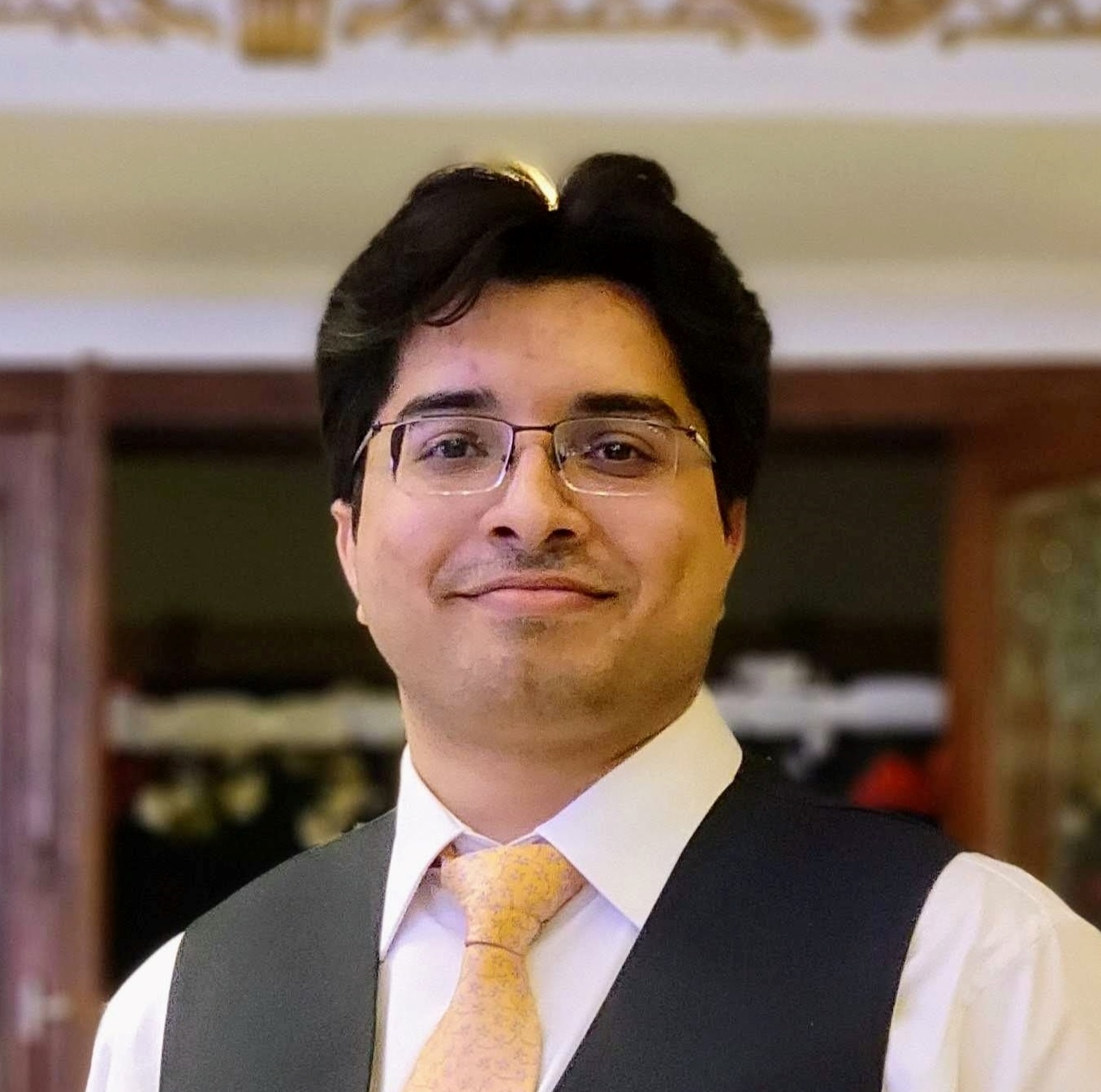}}]{USAMA MASOOD}
 is pursuing his Ph.D. degree in electrical and computer engineering at the AI4Networks Research Center, University of Oklahoma, USA, where his research focus is on designing novel Artificial Intelligence-based network modeling techniques for enabling zero touch automation in next generation networks. He is currently working at AT\&T Labs, California, where he is co-leading several projects on network analytics and optimization of AT\&T nationwide 5G network. Previously, he worked with T-Mobile USA, where he developed innovative machine learning based cloud-native applications for RAN automation use-cases.
\end{IEEEbiography}

\begin{IEEEbiography}
	[{\includegraphics[width=1in,height=1.25in,clip,keepaspectratio]{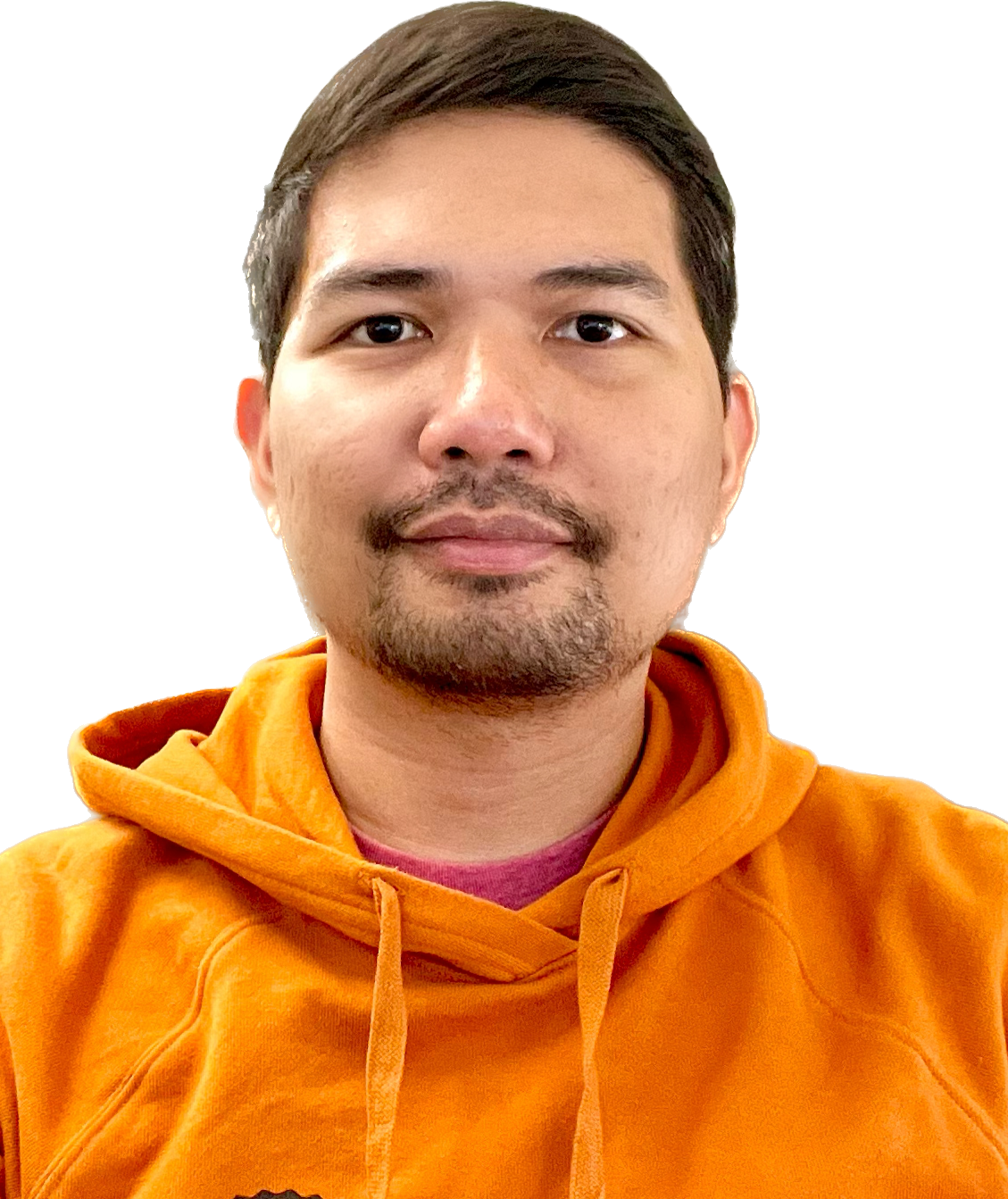}}]{MARVIN MANALASTAS}
 holds a B.S. degree in electronics and communication engineering from the Polytechnic University of the Philippines (2011), as well as an M.S. degree in electrical and computer engineering from The University of Oklahoma, USA (2020), where he is currently pursuing a Ph.D. in electrical engineering. He is also affiliated with the AI4Networks Research Center. Recently, Marvin joined Nokia Standards as a Senior RAN Architecture Research Engineer. Marvin has gained valuable industry experience in cellular network optimization through his work in the Philippines and Japan. He has also completed multiple internships in the USA, including positions as an RF Optimization Intern with Mobilecomm Professionals in TX, an AI/ML Intern with Synopsys in VA, and a Research Fellow with the U.S. FDA in MA. Marvin's research interests center on machine learning applied to optimize 5G and beyond networks.
\end{IEEEbiography}

\begin{IEEEbiography}
	[{\includegraphics[width=1in,height=1.25in,clip,keepaspectratio]{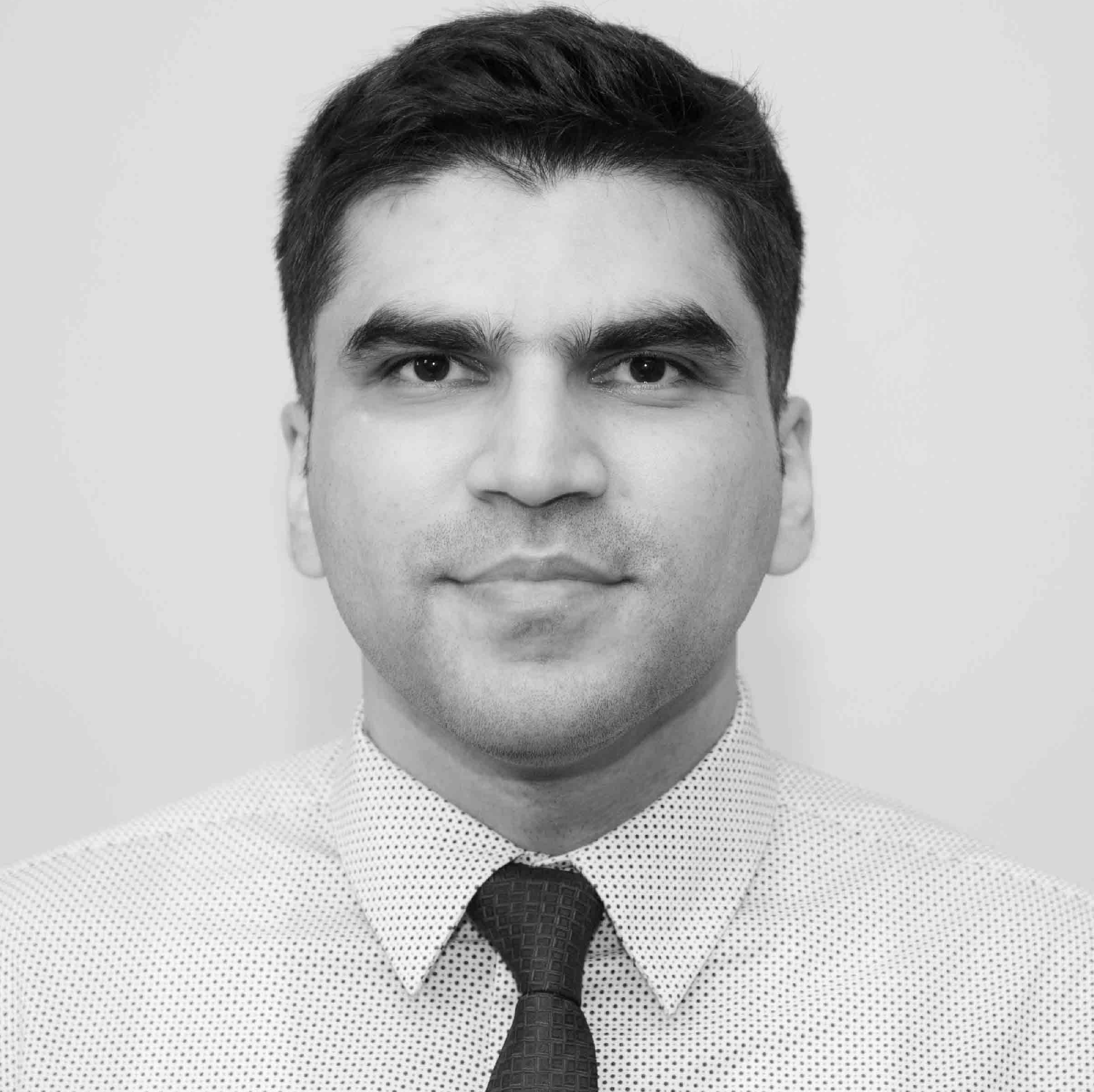}}]{SYED MUHAMMAD ASAD ZAIDI}
received the B.Sc. degree in information and communication engineering from the National University of Science and Technology (NUST), Pakistan, in 2008, MS from Ajou University, South Korea in 2013, and PhD Electrical engineering from University of Oklahoma in 2021. With almost 15 years’ experience in telecom industry, he has worked in Mobilink, Pakistan, KoreaElectronics and Technology Institute (KETI), South Korea, MOTiV Research, Japan, ATT, USA, Sprint, USA and T-Mobile, USA. Currently, he is leading 5G radio-frequency optimization team in T-Mobile networks. His research domain is mobility robustness and optimization of futuristic ultra-dense base station deployment. 
\end{IEEEbiography}

\begin{IEEEbiography}
	[{\includegraphics[width=1in,height=1.25in,clip,keepaspectratio]{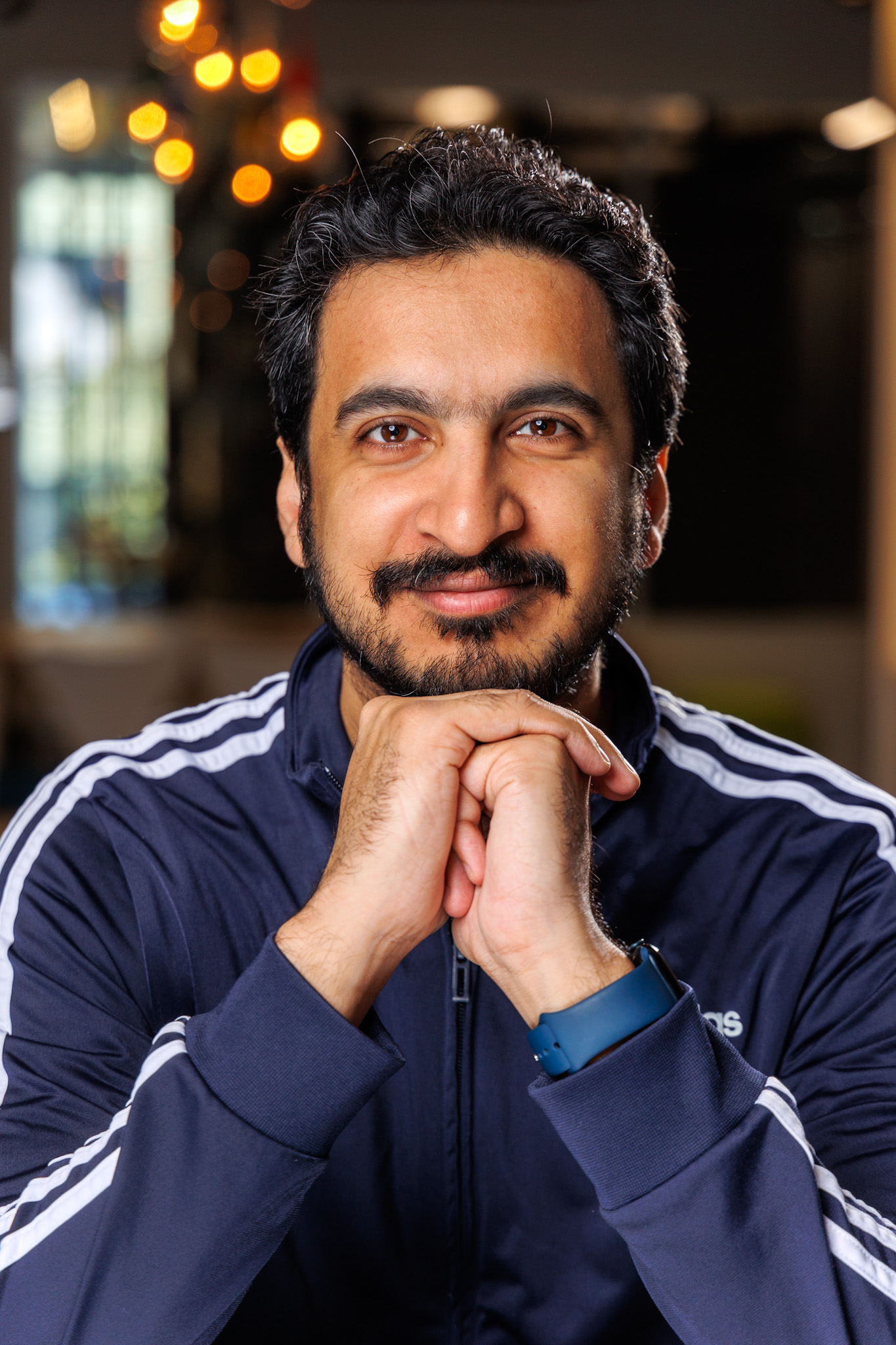}}]{HASAN FAROOQ}
is Senior AI Researcher at Ericsson Research in Santa Clara, USA. His background is AI/ML driven zero-touch automation algorithms for Radio Access Networks. He holds a B.Sc. degree in Electrical Engineering from the University of Engineering and Technology, Lahore, Pakistan, M.Sc. by Research degree in Information Technology from Universiti Teknologi PETRONAS, Malaysia, a Ph.D. degree in Electrical and Computer Engineering and Post Doc from the University of Oklahoma, USA. He has authored/co-authored over 50 publications in high impact journals, book chapters and proceedings of IEEE flagship conferences on communications. He also has patents in the area of SON algorithms.

\end{IEEEbiography}

\begin{IEEEbiography}
	[{\includegraphics[width=1in,height=1.25in,clip,keepaspectratio]{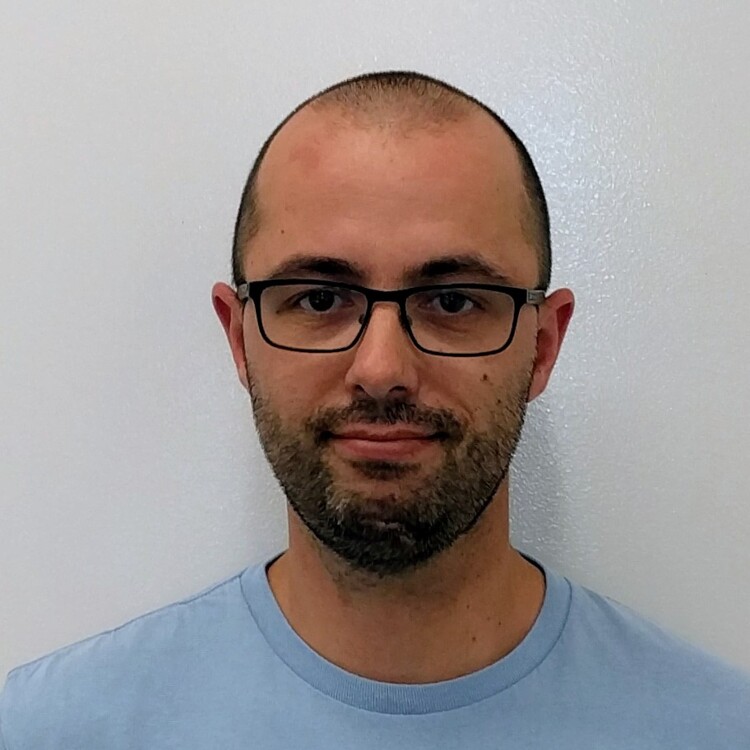}}]{JULIEN FORGEAT}
 is an artificial intelligence principal researcher at Ericsson Research. He joined Ericsson in 2010 after spending several years working on network analysis and optimization. He holds an M.Eng. in computer science from the National Institute of Applied Sciences in Lyon, France.
At Ericsson, Julien has worked on mobile learning, Internet of Things and big data analytics before specializing in machine learning and AI infrastructure. His current research focuses on the software components required to run AI and machine learning workloads on distributed infrastructures as well as the algorithmic approaches that are best suited for complex distributed and decentralized use-cases.

\end{IEEEbiography}

\begin{IEEEbiography}
	[{\includegraphics[width=1in,height=1.25in,clip,keepaspectratio]{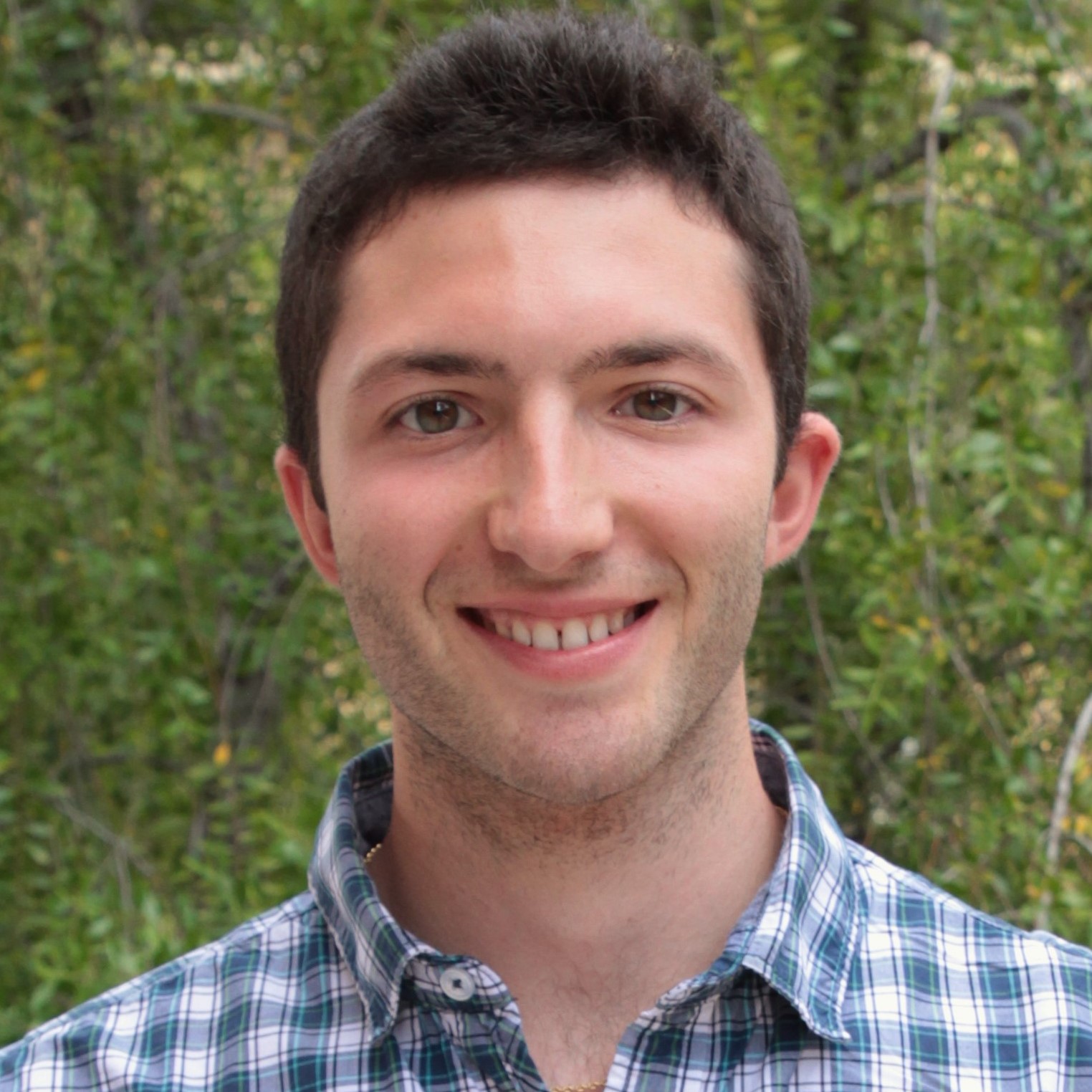}}]{MAXIME BOUTON}
 is an artificial intelligence researcher at Ericsson Research. His research interests lie in applying reinforcement learning to network optimization problems. He also works on topics related to AI safety, multi-agent systems and decision-making problems with partial observability. Maxime received his PhD from Stanford University where he worked on safety and scalability of intelligent autonomous systems. Prior to doing research, Maxime got a MS in Aeronautics and Astronautics as part of a double degree between Ecole Centrale Paris and Stanford University.
\end{IEEEbiography}

\begin{IEEEbiography}
	[{\includegraphics[width=1in,height=1.25in,clip,keepaspectratio]{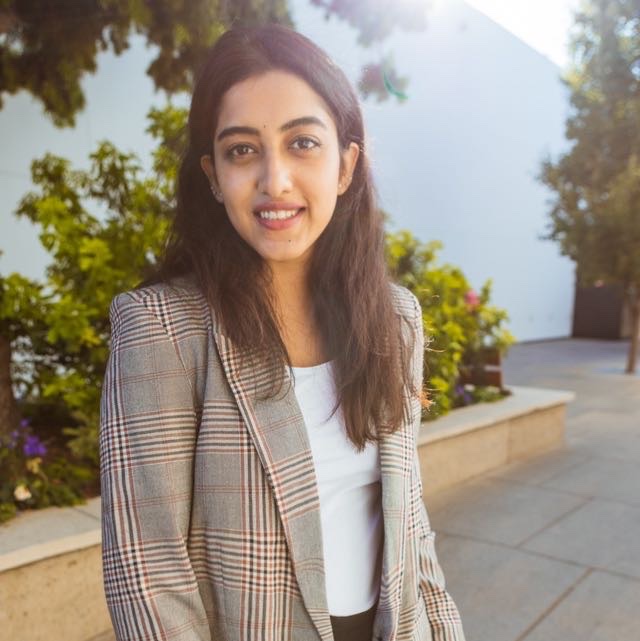}}]{SHRUTI BOTHE}
is an Artificial Intelligence Researcher at Ericsson. With several years of academic and industry experience and a proven track record of identifying issues in and achieving solutions in domains combining AI/ML to telecommunication networks, Shruti was a main contributor of Ericsson's entrepreneurial effort "Ericsson Routes" that brings autonomous and unmanned vehicles the awareness required to have consistent and reliable connectivity throughout the whole journey. Shruti also heads Ericsson Research’s multi-year collaboration with MIT CSAIL related to neuromorphic computing and Lithionics. This research is aimed to produce the next generation of more efficient algorithms and hardware that enable more efficient computing and substantial energy savings.
She has several research publications and over 18 pending patents and has recently been honored with a "Key Contributor" Award at Ericsson.

\end{IEEEbiography}

\begin{IEEEbiography}
	[{\includegraphics[width=1in,height=1.25in,clip,keepaspectratio]{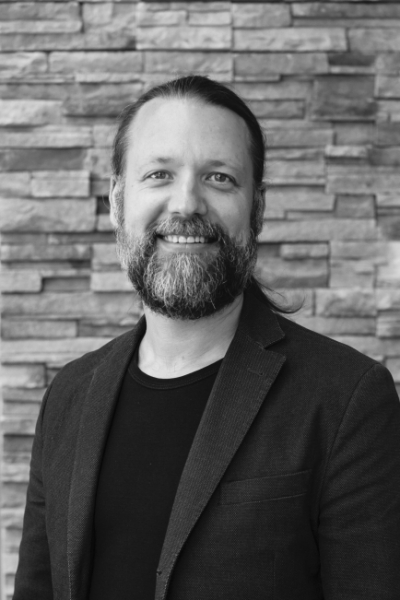}}]{PER KARLSSON}
is the Director of Media Research in Ericsson, focusing on A/V Coding, Content Analytics, and how new XR Experiences will be enabled by the rollout of 5G Networks. He is also the Director of Ericsson Research in Silicon Valley focused on the areas of Radio, AI, Networking, Media, Cloud, and Security.
The Research is performed together with Academia, Customers, Partners, and Universities. His team is currently actively engaged in collaborative projects focused on exploring new opportunities that the 5G Networks will bring to the entertainment, manufacturing, and automotive industry.
Per has been in the industry since 1993 working in the intersection of Research and Products mainly at Ericsson but also leading the Networking Research at the Swedish Research Institute Acreo.
\end{IEEEbiography}

\begin{IEEEbiography}
	[{\includegraphics[width=1in,height=1.25in,clip,keepaspectratio]{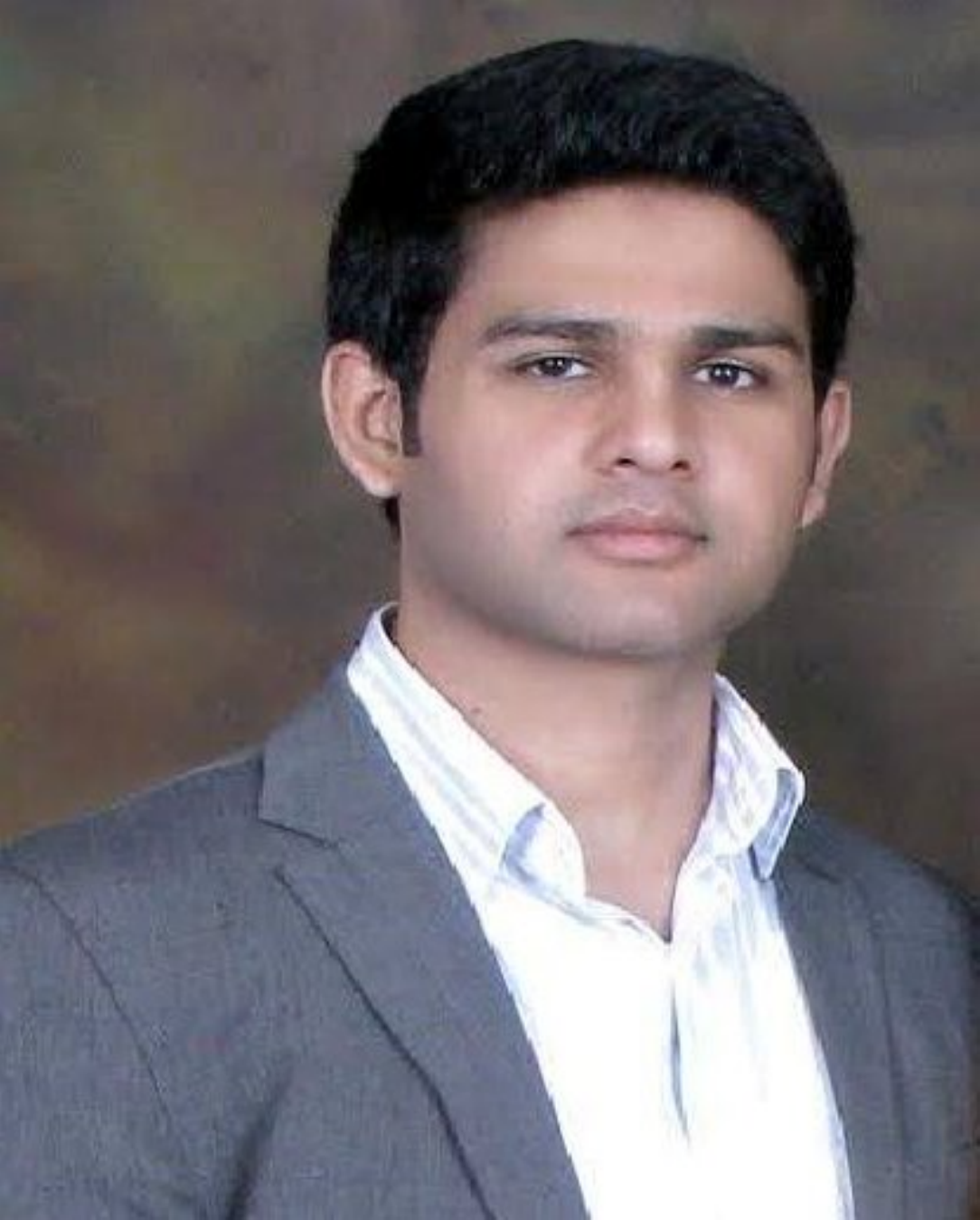}}]{ALI RIZWAN}
 obtained his Bachelor's degree in Applied and Theoretical Mathematics from Bahauddin Zakariya University, Pakistan in 2006. He then pursued an MBA-IT degree from the same institution in 2008. In 2016, he earned an M.Sc. degree in Big Data Science from Queen Mary University of London, U.K. He completed his academic journey by achieving a Ph.D. degree from the University of Glasgow, Glasgow, U.K, in 2021.
Currently, Dr. Rizwan serves as the Chief Technical Officer and Co-founder of Artificial Intelligence For Life, Pakistan. His work primarily focuses on the research and development of AI-enabled screening solutions in healthcare.
\end{IEEEbiography}

\begin{IEEEbiography}
	[{\includegraphics[width=1in,height=1.25in,clip,keepaspectratio]{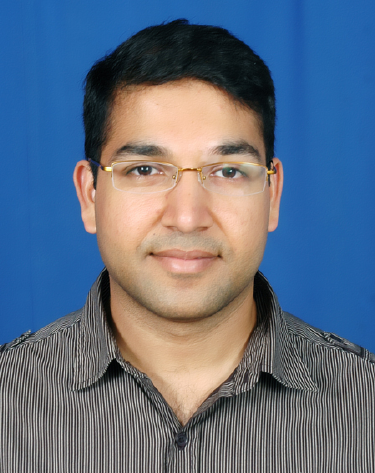}}]{ALI IMRAN}
 is Professor of Cyber Physical Systems in James Watt School of Engineering, University of Glasgow. He is currently on leave from University of Oklahoma where he is Williams Presidential Professor in ECE and the founding director of the Artificial Intelligence (AI) for Networks (AI4Networks) Research Center. His research interests include AI and its applications in wireless networks and healthcare. His work on these topics has resulted in several patents and over 150 peer-reviewed articles including some of the highly influential papers in the domain of wireless network automation. On these topics he has led numerous multinational projects, given invited talks/keynotes and tutorials at international forums and advised major public and private stakeholders and co-founded multiple start-ups. He holds a B.Sc. degree in electrical engineering from the University of Engineering and Technology Lahore, Pakistan, in 2005, and the M.Sc. degree (Hons.) in mobile and satellite communications and the PhD degree from the University of Surrey, Guildford, U.K., in 2007 and 2011, respectively. He is an Associate Fellow of the Higher Education Academy, U.K. He is also a member of the Advisory Board to the Special Technical Community on Big Data, the IEEE Computer Society.

\end{IEEEbiography}

\end{document}